\documentclass[letterpaper,showpacs,preprintnumbers,twocolumn,eqsecnum,superscriptaddress,aps,nofootinbib]{revtex4}
\usepackage{amssymb}
\usepackage[centertags]{amsmath}
\usepackage{txfonts}
\usepackage{epsfig}
\usepackage{bm}
\usepackage{color}
\usepackage{graphicx,graphics}
\usepackage{multirow}
\usepackage{float}
\usepackage[pdfstartview=FitH]{hyperref}
\hypersetup{colorlinks=true, citecolor=blue, linkcolor=blue,filecolor=black,urlcolor=blue}
\allowdisplaybreaks[2]

\usepackage{slashed}
\usepackage{ulem}
\usepackage{cancel}
\usepackage{booktabs}
\usepackage{multirow}
\usepackage{pdflscape}

\usepackage{hhtensor}
\usepackage{rotating}

\begin{document}
\title{Tensor polarization dependent fragmentation functions and $e^+e^-\to V \pi X$ at high energies}
%\date{\bf\today}

\author{Kai-bao Chen}
\affiliation{School of Physics \& Key Laboratory of Particle Physics and Particle Irradiation (MOE), Shandong University, Jinan, Shandong 250100, China}

\author{Wei-hua Yang}
\affiliation{School of Physics \& Key Laboratory of Particle Physics and Particle Irradiation (MOE), Shandong University, Jinan, Shandong 250100, China}

\author{Shu-yi Wei}
\affiliation{School of Physics \& Key Laboratory of Particle Physics and Particle Irradiation (MOE), Shandong University, Jinan, Shandong 250100, China}
\affiliation{Key Laboratory of Quark and Lepton Physics (MOE) and Institute of Particle Physics, Central China Normal University, Wuhan 430079, China}

\author{Zuo-tang Liang}
\affiliation{School of Physics \& Key Laboratory of Particle Physics and Particle Irradiation (MOE), Shandong University, Jinan, Shandong 250100, China}
%\affiliation{CAS Center for Excellence in Particle Physics (CCEPP), Box 918, Beijing 10094, China}

\begin{abstract}
We present the systematic results for three dimensional fragmentation functions of spin one hadrons defined via quark-quark correlator. 
There are totally 72 such fragmentation functions, among them 18 are twist-2, 36 are twist-3 and 18 are twist-4. 
We also present the relationships between the twist-3 parts and those defined via quark-gluon-quark correlator 
obtained from the QCD equation of motion.
We show that two particle semi-inclusive hadron production process $e^+e^-\to V\pi X$ at high energies is one of the best places  
to study the three-dimensional tensor polarization dependent fragmentation functions.  
We present the general kinematic analysis of this process and show that the cross section should be expressed in terms of 81 independent structure functions. 
After that we present parton model results for the hadronic tensor, the structure functions, the azimuthal and spin asymmetries 
in terms of these gauge invariant fragmentation functions at the leading order pQCD up to twist-3.  
\end{abstract}


\maketitle

\section{Introduction} \label{sec:introduction}
In describing high energy reactions, we need two sets of important quantities, 
the parton distribution functions (PDFs) and the fragmentation functions (FFs).
The former is used to describe the hadron structure and the latter describes the hadronization process. 
In a quantum field theoretical formulation, both PDFs and FFs are defined via the corresponding quark-quark correlator.
The quark-quark correlator is defined as a matrix in the Dirac space depending on the hadron states. 
It is then decomposed into different components expressed in terms of the basic Lorentz covariants and the scalar functions. 
These scalar functions contain the information of the hadron structure and/or hadronization mechanism and are called the corresponding PDFs or FFs. 
In many cases in literature, specific PDFs and/or FFs are introduced whenever needed, sometimes with different conventions and/or notations. 
With the development of the related studies, it is necessary and useful to make a systematic study and present a complete set of such results. 
The results for three-dimensional PDFs of the nucleon defined in this way are presented in \cite{Goeke:2005hb} in a systematical way. 
Since usually different types of hadrons with different flavors and spins are produced in a high energy reaction, 
FFs are therefore more involved and perhaps even more interesting but less studied yet.  
Specific recent discussions can also be found e.g. 
in \cite{Agashe:2014kda,Albino:2008gy,Collins:1992kk,Ji:1993vw, Mulders:1995dh,Boer:1997mf,Boer:1997nt,Boer:1997qn,Boer:1999uu,
Goeke:2003az,Bacchetta:2006tn,Boer:2008fr,Pitonyak:2013dsu,Wei:2013csa,Kanazawa:2013uia,Wei:2014pma}.
A short summary can be found in a recent unpublished note and short reviews~\cite{Chen:2015ora,Liang:2015nia,Chen:2015tca}.

In this paper, we summarize the results for three dimensional FFs defined via quark-quark correlator 
for spin one hadrons in a systematical way. 
The FFs are divided into a spin independent part, a vector polarization dependent part and a tensor polarization dependent part. 
Formerly, the spin independent part is the same as those for spin zero hadrons and the vector polarization dependent part is 
the same as those for spin-1/2 hadrons.  
They are also similar to those for PDFs presented e.g. in \cite{Goeke:2005hb} for the corresponding cases.
We will pay special attention to the tensor polarization dependent part including higher twist contributions.  
In this connection, we will in particular show also FFs defined via the quark-gluon-quark correlator and 
their relationships to those defined via quark-quark correlator obtained using Quantum Chromodynamics (QCD) equation of motion. 

The most convenient place to study the three dimensional FFs of vector mesons is perhaps $e^+e^-\to V \pi X$. 
We present the results for the general kinematical analysis of this process and calculate the 
hadronic tensor and differential cross section up to twist-3 at leading order in perturbative QCD. 
We also present the results for the tensor polarizations of $V$ in terms of the three dimensional FFs.

The rest of the paper is organized as follows. 
After this introduction, we briefly summarize the general procedure of deriving the results of FFs from the quark-quark correlator 
and present results and relationships to those defiend via quark-gluon-quark correlator at twist-3 in Sec. \ref{sec:FFs}. 
We make general kinematical analysis of $e^+e^-\to V\pi X$ in Sec.  \ref{sec:Kinematics}. 
We calculate the hadronic tensor at leading order perturbative QCD up to twist-3 in Sec. \ref{sec:HadronicTensor}.  
We present the results for the structure functions in Sec. \ref{sec:StructureFunctions} and those for azimuthal and spin asymmetries in Sec. \ref{sec:Tpol}.  
We make a summary and a discussion in Sec. \ref{sec:Summary}.
Since most of the equations are rather long, we will present the discussions in the corresponding sections  
but show most of the formulae and tables in the appendices. 

\section{Fragmentation functions defined via quark-quark correlator} \label{sec:FFs}

A systematic analysis is given in a recent unpublished note\cite{Chen:2015ora}. 
For completeness, we briefly summarize the basic ideas in this section and summarize the results in appendix \ref{FFs}. 
Similar to parton distribution and/or correlation functions, in quantum field theory, the quark fragmentation is defined via the quark-quark correlator given  by, 
\begin{align}
\hat\Xi_{ij}^{(0)}(k_F;p,S) =\frac{1}{2\pi} & \sum_X \int  d^4\xi e^{-ik_F \xi} \langle 0| \mathcal{L}^\dag (0;\infty) \psi_i(0) |p,S;X\rangle \nonumber\\
& \times \langle p,S;X|\bar\psi_j(\xi) \mathcal{L}(\xi;\infty) |0\rangle, \label{qq-correlator}
\end{align}
where $k_F$ and $p$ denote the 4-momenta of the quark and the hadron respectively, $S$ denotes the spin of the hadron; 
$\mathcal{L}(\xi;\infty)$ is the gauge link that is given by,
\begin{align}
\mathcal{L}(\xi,&\infty) =Pe^{ig\int_{\xi^-}^\infty d\eta^-A^+(\eta^-;\xi^+,\vec{\xi}_\perp)} \nonumber\\
&= 1 + ig\int_{\xi^-}^\infty d \eta^- A^+(\eta^-;\xi^+,\vec{\xi}_\perp) \nonumber\\
& + (ig)^2 \int_{\xi^-}^\infty d \eta_1^- \int_{\xi^-}^{\eta_1^-} d \eta_2^- A^+(\eta_2^-;\xi^+,\vec{\xi}_\perp)A^+(\eta_1^-;\xi^+,\vec{\xi}_\perp) \nonumber\\
& + \cdots.
\end{align}
The correlator given by Eq.~(\ref{qq-correlator}) satisfies the following constraints 
imposed by hermiticity and parity conservation,  i.e., 
\begin{align}
& \hat\Xi^{\dagger(0)}(k_F;p,S) = \gamma^0 \hat\Xi^{(0)}(k_F;p,S) \gamma^0, \label{hermiticity}\\
& \hat\Xi^{(0)}(k_F;p,S) = \gamma^0 \hat\Xi^{(0)}(k_F^{\mathcal{P}}; p^{\mathcal{P}}, S^{\mathcal{P}}) \gamma^0, \label{parity}
\end{align}
where a vector with the superscript ${\cal P}$ denotes the result after space reflection such as $p^{\cal P}_\mu=p^\mu$.
Unlike that for hadron structure, because of the presence of the gauge link and final state interactions between $h$ and $X$, 
time reversal invariance puts no such simple constraint on the correlator $\hat\Xi^{(0)}(k_F;p,S)$.

The three-dimensional or the transverse momentum dependent (TMD) FFs are defined via 
the three-dimensional quark-quark correlator $\hat\Xi^{(0)}(z,k_{F\perp};p,S)$ obtained from $\hat\Xi^{(0)}(k_{F},p,S)$ 
by integrating over $k_{F}^-$ , i.e., 
\begin{align}
\hat\Xi&^{(0)}(z,k_{F\perp};p,S) =% \int \frac{p^+dk_F^+dk_F^-}{(2\pi)^2}2\pi\delta (k_F^+-p^+/z)\hat\Xi^{(0)}(k_{F},p,S)\nonumber\\=&
\sum_X \int \frac{p^+d\xi^-}{2\pi} d^2{\xi}_\perp e^{-i(p^+\xi^-/z - \vec{k}_{F\perp} \cdot \vec{\xi}_\perp)}  \nonumber\\
&\times  \langle 0| \mathcal{L}^\dag (0;\infty) \psi(0) |p,S;X\rangle \langle p,S;X|\bar\psi(\xi) \mathcal{L}(\xi;\infty) |0\rangle, \label{Xi0}
\end{align}
where $z=p^+/k^+_F$ is the longitudinal momentum fraction defined in light cone coordinates. 
Here we use the light-cone coordinate and define the light-cone unit vectors as $\bar n = (1,0,\vec 0_\perp)$, $n = (0,1,\vec 0_\perp)$ and $n_\perp = (0,0,\vec 1_\perp)$. 
We choose the hadron's momentum as $z$-direction so that $p$ is decomposed as
$p^\mu = p^+ \bar n^\mu + (M^2/2p^+) n^\mu$.

The  FFs are obtained from $\hat\Xi^{(0)}(z,k_{F\perp};p,S)$ by decomposing it in the following two steps. 
First, we note that $\hat\Xi^{(0)}(z,k_{F\perp};p,S)$ is a matrix in Dirac space and expand it in terms of the $\Gamma$-matrices, 
$\Gamma = \bigl\{ \mathbf{I},~ i\gamma_5,~ \gamma^\alpha,~ \gamma_5\gamma^\alpha,~ i\sigma^{\alpha\beta}\gamma_5 \bigr\}$, i.e.,
\begin{align}
\hat\Xi^{(0)}(z,k_{F\perp};&p,S) =  \Xi^{(0)}(z,k_{F\perp};p,S) + i\gamma_5 \tilde\Xi^{(0)}(z,k_{F\perp};p,S) \nonumber\\
& + \gamma^\alpha \Xi_\alpha^{(0)}(z,k_{F\perp};p,S) + \gamma_5\gamma^\alpha \tilde\Xi_\alpha^{(0)}(z,k_{F\perp};p,S) \nonumber\\
& + i\sigma^{\alpha\beta}\gamma_5 \Xi_{\alpha\beta}^{(0)}(z,k_{F\perp};p,S). \label{XiExpansion}
\end{align}
The coefficient functions are given by,
\begin{align}
\Xi&^{(0)[\Gamma]}(z,k_{F\perp};p,S) =  %\frac{1}{2}{\rm{Tr}}\Bigl[\Gamma\hat\Xi^{(0)}(z,k_{F\perp};p,S)\Bigr]\nonumber\\&=
\frac{1}{4}\sum_X \int \frac{p^+d\xi^-}{2\pi} d^2\xi_\perp e^{-i(p^+\xi^-/z - \vec{k}_{F\perp} \cdot \vec{\xi}_\perp)}\nonumber\\
&\times %{\rm Tr} \Bigl[ \Gamma 
 \langle p,S;X|\bar\psi(\xi) \mathcal{L}(\xi;\infty) |0\rangle \Gamma\langle 0|\mathcal{L}^\dag (0;\infty)\psi(0) |p,S;X\rangle, \label{eq:trace}
\end{align}
where $\Xi^{(0)[\Gamma]}$ represents respectively $\Xi^{(0)}$, $\tilde\Xi^{(0)}$,  $\Xi_\alpha^{(0)}$, $\tilde\Xi_\alpha^{(0)}$ and $\Xi_{\alpha\beta}^{(0)}$ for different $\Gamma$'s.
Together with the demands imposed by the hermiticity and parity invariance [Eqs.~(\ref{hermiticity}) and (\ref{parity})],  
the Lorentz invariance demands that 
all the corresponding coefficient functions are real and are Lorentz scalar, pseudo-scalar, vector, axial-vector and tensor respectively.
Furthermore, the tensor $\Xi_{\alpha\beta}^{(0)}$  is anti-symmetric in Lorentz indices and odd under space reflection 
which implies that it can be made out of a vector and an axial vector. 

Second, we expand these coefficient functions according to their respective Lorentz transformation properties 
in terms of the basic Lorentz covariants constructed from basic variables at hand. 
They are expressed as the sum of the basic Lorentz covariants multiplied by scalar functions of $z$ and $k_{F\perp}^2$. 
These scalar functions are the three-dimensional FFs. 
We note in particular that because of the hermiticity given by Eq. (\ref{hermiticity}), 
these FFs defined via quark-quark correlator are real.  

Clearly, the basic Lorentz covariants that we can construct depend on what basic variable(s) that we have at hand. 
Besides the four-momenta $p$ and $k_F$, we have the variables describing the spin states.
Such variables are different for hadrons with different spins. 
For spin-1 hadrons, the polarization is described by a $3\times 3$ density matrix $\rho$,
which, in the rest frame of the hadron, is usually decomposed as \cite{Bacchetta:2000jk},
\begin{align}
\rho = \frac{1}{3} (\mathbf{1} + \frac{3}{2}S^i \Sigma^i + 3 T^{ij} \Sigma^{ij}), \label{eq:spin1rho}
\end{align}
where, $\Sigma^i$ is the spin operator of spin-$1$ particle, 
and $\Sigma^{ij}= \frac{1}{2} (\Sigma^i\Sigma^j + \Sigma^j \Sigma^i) - \frac{2}{3} \mathbf{1} \delta^{ij}$. 
The spin polarization tensor $T^{ij}={\rm Tr}(\rho \Sigma^{ij})$, and is parameterized as,
\begin{align}
\mathbf{T}= \frac{1}{2}
\left(
\begin{array}{ccc}
-\frac{2}{3}S_{LL} + S_{TT}^{xx} & S_{TT}^{xy} & S_{LT}^x  \\
S_{TT}^{xy}  & -\frac{2}{3} S_{LL} - S_{TT}^{xx} & S_{LT}^{y} \\
S_{LT}^x & S_{LT}^{y} & \frac{4}{3} S_{LL}
\end{array}
\right).
\label{spintensor}
\end{align}
Here, besides the polarization vector $S$, we also need a polarization tensor $T$.  
The polarization vector $S$ is similar to that for spin-1/2 hadrons and the tensor $T$ has five independent components that are given by 
a Lorentz scalar $S_{LL}$, a Lorentz vector $S_{LT}^\mu = (0, S_{LT}^x, S_{LT}^y,0)$ 
and a Lorentz tensor $S_{TT}^{\mu\nu}$ that has two nonzero independent components 
$S_{TT}^{xx} = -S_{TT}^{yy}$ and $S_{TT}^{xy} = S_{TT}^{yx}$. 
In a covariant form, the polarization vector $S$ is decomposed as,
\begin{align}
S^\mu = \lambda \frac{p^+}{M} \bar n^\mu + S_T^\mu - \lambda \frac{M}{2p^+} n^\mu,
\end{align}
where $\lambda$ denotes the helicity and $S_T = (0,0,\vec S_T)$ denotes the transverse polarization.
The tensor polarization $T^{\mu\nu}$ is expressed as~\cite{Bacchetta:2000jk},
\begin{align}
T^{\mu\nu} &= \frac{1}{2} \Bigl[ \frac{4}{3}S_{LL}\Bigl( \frac{p^+}{M} \Bigr)^2 \bar n^\mu \bar n^\nu + \frac{p^+}{M} n^{\{\mu}S_{LT}^{\nu\}} 
- \frac{2}{3}S_{LL}(\bar n^{\{\mu}n^{\nu\}} - g_\perp^{\mu\nu}) \nonumber\\
&+ S_{TT}^{\mu\nu} 
- \frac{M}{2p^+} \bar n^{\{\mu}S_{LT}^{\nu\}} + \frac{1}{3}S_{LL}\Bigl( \frac{M}{p^+} \Bigr)^2 n^\mu n^\nu \Bigr],
\end{align}
where we used the anti-commutation symbol $A^{\{\mu}B^{\nu\}} \equiv A^\mu B^\nu + A^\nu B^\mu$, 
and also in the following of this paper $A^{[\mu}B^{\nu]} \equiv A^\mu B^\nu - A^\nu B^\mu$, 
and $g_\perp^{\mu\nu} \equiv g^{\mu\nu} - \bar n^\mu n^\nu - n^{\mu} \bar n^\nu$. 

Hence, for spin-1 hadrons, the quark-quark correlator $\hat\Xi^{(0)}$ can be written as the sum of 
a polarization independent part $\hat\Xi^{U(0)}$, a vector polarization dependent part $\hat\Xi^{V(0)}$ 
and a tensor polarization dependent part $\hat\Xi^{T(0)}$, i.e, 
\begin{align}
\hat\Xi^{(0)}(z,k_{F\perp};p,S) =& \hat\Xi^{U(0)}(z,k_{F\perp};p) + \hat\Xi^{V(0)}(z,k_{F\perp};p,S) \nonumber\\
+& \hat\Xi^{T(0)}(z,k_{F\perp};p,S).
\end{align}
Since the polarization dependence is linear to the corresponding spin parameters, 
formally, the spin independent part is exactly the same as that for spin-0 hadrons, 
the vector polarization dependent part is the same as that for spin-1/2 hadrons. 
The tensor polarization dependent part is new and contributes only for  spin-$1$ hadron production.
We summarize them separately in the following. 

Before we present the results, we describe the notation system for the FFs used through out the paper.  
We will use $D$, $G$ and $H$ for unpolarized, longitudinally polarized and transversely polarized quarks. 
They correspond to those FFs obtained via decompositions of the vector, axial-vector and tensor part of the correlator. 
Those defined via the scalar and the pseudo-scalar are denoted by $E$.
A number $j$ in the subscripts specifies the twist: $j=1$ for twist-2, null (no number) for twist-3 and $j=3$ for twist-4. 
We will also use different symbols in the subscripts to denote the polarization of the produced hadron such as $L$ and $T$   
in the vector polarization case and $LL$, $LT$ or $TT$ in the tensor polarization case; 
a $\perp$ in the superscript denotes that the corresponding basic Lorentz covariant is $k_{F\perp}$-dependent.

If we decompose the quark field in Eq.~(\ref{eq:trace}) into the sum of the right- and left-handed parts, 
i.e., $\psi = \psi_R + \psi_L$ with $\psi_{R/L} \equiv \frac{1}{2}(1\pm\gamma_5)\psi$.
We see that for $\Gamma = \mathbf{I}$, $i\gamma_5$ and $i\sigma^{\alpha\beta}\gamma_5$, 
$\bar\psi_R\Gamma\psi_L$ and $\bar\psi_L\Gamma\psi_R$ are non-zero.
So the terms related to them (i.e., the $E$ and $H$-terms) 
correspond to helicity-flipped quark structure and are called chiral-odd ($\chi$-odd). 
Similarly, for $\Gamma = \gamma^\alpha$ and $\gamma^5\gamma^\alpha$, 
$\bar\psi_L\Gamma\psi_L$ and $\bar\psi_R\Gamma\psi_R$ are non-zero.
Hence, the terms related to them (i.e. the $D$'s and the $G$'s) do not flip the quark helicity and are $\chi$-even.
We also recall the properties of the fermion bilinears 
under time-reversal $\hat{\mathcal{T}}$, i.e., 
\begin{align}
\hat{\mathcal{T}}&\Bigl\{ \bar\psi \psi,~ \bar\psi i\gamma_5 \psi,~ \bar\psi \gamma_\alpha \psi, \bar\psi \gamma_5\gamma_\alpha\psi,~ \bar\psi i\sigma_{\alpha\beta}\gamma_5 \psi  \Bigr\}\nonumber\\
\Rightarrow 
&\Bigl\{ \bar\psi \psi,~ -\bar\psi i\gamma_5 \psi,~ \bar\psi \gamma^\alpha \psi,~ \bar\psi \gamma_5\gamma^\alpha\psi,~ \bar\psi i\sigma^{\alpha\beta}\gamma_5 \psi  \Bigr\}.
\end{align}
Using this, we can determine whether a component of FF defined via quark-quark correlator %given by Eqs.~(\ref{eq:XiUS}-\ref{eq:XiUT}) 
is time reversal even (T-even) or odd (T-odd) according to the time reversal behavior of the corresponding basic Lorentz covariant. 
%In this way, we find out that $G^\perp$, $H_1^\perp$, $H$ and $H_3^\perp$ are T-odd, all the others in Eqs.~(\ref{eq:XiUS}-\ref{eq:XiUT}) are T-even.
However, we should also note that they are usually referred as ``naive T-odd'' or ``naive T-even'' because the interactions between the produced hadron $h$ and the 
rest $X$ can destroy simple regularities so all of them can exist in a practical hadronization process.

\subsection{Results of the decomposition and FFs}

\subsubsection{The unpolarized part}

For the spin independent part $\hat\Xi^{U(0)}(z,k_{F\perp};p)$,  
the independent variables that can be used to construct the basic Lorentz covariants are $p_\alpha$, $k_{F\perp\alpha}$, and $n_\alpha$. 
The basic Lorentz covariants that we can construct from them are: one Lorentz scalar $p^2=M^2$, no pseudo-scalar,  
three Lorentz vectors, $p$, $k_{F\perp}$ and $n$, one axial vector $\varepsilon_{\perp\rho\alpha}k_{F\perp}^\rho \equiv \tilde k_{F\alpha}$, 
and three anti-symmetric and space reflection odd Lorentz tensors $p_{[\rho}\tilde k_{F\perp\alpha]}$, 
$\varepsilon_{\perp\rho\alpha}$ and $n_{[\rho} \tilde k_{F\perp\alpha]}$. 
Here $\varepsilon_{\perp\rho\alpha}=\varepsilon_{\mu\nu\rho\alpha}\bar n^\mu n^\nu$ and $\varepsilon_{\mu\nu\rho\alpha}$ is the anti-symmetric tensor. 
We also use the notation $\tilde a_{\perp\mu}\equiv\varepsilon_{\perp \rho\mu}a^\rho$ to denote the transverse vector perpendicular to $a_\perp$, 
and note in particular that $\tilde a_\perp\cdot b_\perp=\varepsilon_{\perp \rho\sigma}a^\rho b^\sigma=-a_\perp\cdot \tilde b_\perp$,
and $\tilde{\tilde a}_\perp=-a_\perp$.
The general decomposition of the spin independent part of the quark-quark correlator is given by Eqs.~(\ref{eq:XiUS}-\ref{eq:XiUT}) in Appendix~\ref{FFs}.
We obtain 8 unpolarized TMD FFs, 2 of them contribute at twist-2, 4 at twist-3 and the other 2 at twist-4 level. 

From Eqs.~(\ref{eq:XiUS}-\ref{eq:XiUT}), we see in particular the existence of a leading twist FF $H_1^\perp(z,k_{F\perp})$ 
that leads to azimuthal asymmetry of produced hadron in fragmentation of a transversely polarized quark. 
This was first introduced in \cite{Collins:1992kk} and is now known as Collins function.
We see also a twist-4 addendum to it described by $H_3^\perp(z,k_{F\perp})$. 

If we integrate over $d^2k_{F\perp}$,  we obtain the one dimensional results as given by Eqs.~(\ref{eq:XiUPS1}-\ref{eq:XiUT1}) in the appendix.
We see that there are only 4 left and the number density $D_1(z)$ is the only leading twist, 2 of them contribute at twist-3 and the other one at twist-4.

We note in particular the direct one to one correspondence between the results obtained in this case for FFs and those obtained in  \cite{Goeke:2005hb} for PDFs. 
The only obvious difference is the existence of the naive time reversal odd $H(z)$ due to final interaction between $h$ and $X$ 
while the corresponding term vanishes  for PDFs. 

\subsubsection{The vector polarization dependent part}

For the vector polarization dependent part, 
we have, besides $p_\alpha$, $k_{F\perp\alpha}$, and $n_\alpha$, the polarization vector $S$ to use to construct the basic Lorentz covariants. 
The results obtained are given by Eqs.~(\ref{eq:XiVS}-\ref{eq:XiVT}) in Appendix~\ref{FFs}.
We see that there are 24 vector polarization dependent TMD FFs, 6 of them contribute at twist-2, 12 at twist-3 and the other 6 at twist-4 level. 
Among them, 8 are naive T-odd ($E_T^\perp$, $E_L$, $E_T^{\prime\perp}$, 
$D_{1T}^\perp$, $D_L^\perp$, $D_T$, $D_T^{\perp}$ and $D_{3T}^\perp$), and the other 16 are T-even. 

We also note that 4 of  them ($E_L$, $G_{1L}$,  $G_{L}^\perp$, $G_{3L}$) are for longitudinal (to longitudinal) spin transfer;  
6 of them ($H_{1T}$, $H_{1T}^\perp$, $H_T^\perp$, $H_T^{\prime\perp}$, $H_{3T}$, $H_{3T}^\perp$) are for transverse (to transverse) spin transfer;
5 of them ($E^{\prime\perp}_T$, $G_{1T}^\perp$, $G_T$, $G_{T}^{\perp}$, $G_{3T}$) are for longitudinal to transverse spin transfer; 
3 of them ($H_{1L}^\perp$,  $H_L$,  $H_{3L}^\perp$) are for transverse to longitudinal spin transfer; 
and the other 6 ($E_T^\perp$,  $D_{1T}^\perp$, $D_T$, $D_L^\perp$, $D_T^{\perp}$, $D_{3T}^\perp$) are for induced polarizations 
which leads to hadron polarizations in fragmentation of unpolarized quark. 
At leading twist, we have a $D_{1T}^\perp$ for induced polarization, 
a longitudinal ($G_{1L}$), two transverse ($H_{1T}$, $H_{1T}^\perp$),  
a longitudinal to transverse ($G_{1T}^\perp$) and 
a transverse to longitudinal ($H_{1L}^\perp$) spin transfer. 

We note in particular the induced polarization terms described by $E_T^\perp$ and the $D$'s in fragmentation of an unpolarized quark. 
At leading twist, there is a Sivers type~\cite{Sivers:1989cc} FF $D_{1T}^\perp$ describing polarization transverse to the production plane 
and corresponding to the transverse hyperon polarizations observed in high energy hadron-hadron and hadron-nucleus collisions~\cite{tran-exp}. 
Other higher twist FFs describe polarizations in longitudinal as well as two transverse directions.    

If we integrate over $d^2k_{F\perp}$, we obtain the results given by Eqs.~(\ref{eq:XiVS1}-\ref{eq:XiVT1}) in Appendix~\ref{FFs}.
We see that only 8 terms survive, which means that, in the one-dimensional case, for the vector polarization dependent part, we have totally 8 FFs. 
We see also that 2 of them are leading twist, they are the longitudinal spin transfer $G_{1L}(z)$ and the transverse spin transfer $H_{1T}(z)$. 
We also have 4 twist-3 FFs that lead to induced polarization of hadron and 2 twist-4 FFs that are addenda to the 
longitudinal and transverse spin transfer respectively. 
We also see that in this case induced polarization in the transverse direction exists at twist-3.  

We note again the direct one to one correspondence between the results obtained in this case and those obtained in \cite{Goeke:2005hb} for PDFs. 
The difference is the existence of the naive time reversal odd $E_L(z)$ and $D_T(z)$ due to final state interactions between $h$ and $X$ 
while the corresponding term vanishes  for PDFs. 
While $E_L(z)$ is an addendum to $G_{1L}(z)$, $D_T(z)$ leads to transverse polarization in fragmentation of unpolarized quark. 
Both of them contribute at twist-3. 

\subsubsection{The tensor polarization dependent part}

The general decomposition of the tensor polarization dependent part is given 
by Eqs.~(\ref{eq:XiTS}-\ref{eq:XiTT}) in Appendix~\ref{FFs} which is obtained by 
constructing basic Lorentz covariants by using, 
besides $p$, $k_{F\perp}$ and $n$, the Lorentz scalar $S_{LL}$, Lorentz vector $S_{LT}$, 
and Lorentz tensor $S_{TT}$. 
We see that there are totally 40 tensor polarization dependent TMD FFs, 
10 contribute at twist-2, 20 at twist-3 and the other 10 at twist-4. 
Among them, 24 (those related to $\tilde \Xi_\alpha^{T(0)}$ and $\Xi_{\rho\alpha}^{T(0)}$) are T-odd and the other 16 are T-even. 

We emphasize in particular the similarities between the tensor polarization dependent terms given by Eqs.~(\ref{eq:XiTS}-\ref{eq:XiTT}) in Appendix~\ref{FFs}  
and those unpolarized and vector polarization dependent terms given by Eqs.~(\ref{eq:XiUS}-\ref{eq:XiVT}) in Appendix~\ref{FFs}. 

(1) Since $S_{LL}$ is a Lorentz scalar thus has no influence on the basic Lorentz covariants, 
the $S_{LL}$-dependent terms have exactly one to one correspondence to the unpolarized terms. 

(2) For the $S_{LT}$-dependent terms, because $S_{LT}$ and $S$ behave differently under space reflection, 
the $S_{LT}$-dependent terms are different from the $S$-dependent terms. 
Since $S_{LT}$ has only two independent transverse components, 
we have one to one correspondence for $S_{LT}$- to $S_T$-terms with the replacement of $S_{T\alpha}$ by $\tilde S_{LT\alpha}$. 

(3) Although there is no counterpart for the $S_{TT}$-dependent terms in other cases, however, there is no direct $S_{TT\rho\alpha}$-term contributing because 
$S_{TT\rho\alpha}=S_{TT\alpha\rho}$ is symmetric while $\Xi_{\rho\alpha}^{T(0)}=-\Xi_{\alpha\rho}^{T(0)}$ is anti-symmetric.  
All the independent $S_{TT}$-terms are in the form of $S_{TT\alpha\sigma}k_{F\perp}^{\sigma}$ which is denoted by $S_{TT\alpha}^{k_F}$. 
Because $S_{TT\alpha}^{k_F}$ has exactly the same Lorentz  and space reflection behaviors as $S_{LT\alpha}$,  
we obtain a direct one to one correspondence between $S_{LT}$- and $S_{TT}$-dependent terms 
with the replacement of $S_{LT\alpha}$ by $S_{TT\alpha}^{k_F}$.

We note again the induced polarizations in the fragmentation of an unpolarized quark. 
We see that at leading twist an $S_{LL}$-dependent term exist and is described by $D_{1LL}$. 
There exist also terms depend on the other components of the tensor polarization at higher twists. 
We emphasize that, since they are independent of the polarization of the fragmenting quark, 
they might be much easier to study in experiments since no polarization in the initial state is needed.   

We integrate over $d^2k_{F\perp}$ and obtain Eqs.~(\ref{eq:XiTS1}-\ref{eq:XiTT1}) in Appendix~\ref{FFs}.
We have totally 8 terms, 4 of them are $S_{LL}$-dependent and the other 4 are $S_{LT}$-dependent.  
They have exact one to one correspondence to the unpolarized and $S_T$-dependent parts. 
We see that there is completely no $S_{TT}$-dependent terms exist in the one-dimensional case.
This means that no $S_{TT}$-dependent one-dimensional FF can be defined via quark-quark correlator. 
The $S_{TT}$-dependent one dimensional FFs can only be higher twists.

We list those twist-2 FFs in table \ref{tab:TMDFF1}, and those twist-3 FFs in table \ref{tab:TMDFF2}. 
The twist-4 FFs have the same structure of those at twist-2, so we will not make a separate table for them.
We also list them according to chiral and time-reversal properties in table \ref{tab:TMDFFChiralTime}.  

We note in particular the $S_{LL}$-dependent terms exist also in the one-dimensional case.
We see that the leading twist contribution $D_{1LL}$-term survives the integration over $k_{F\perp}$ 
also the higher twist addenda such as $E_{LL}$ and $D_{3LL}$. 
This means that it can be studied even in inclusive high energy reactions. 
In the case that the leading twist effect dominates, the results should be not very much dependent  of energy. 
The energy dependence can be used as a sensitive test of higher twist contributions.  

\subsection{Relation to those defined via quark-gluon-quark correlator at twist-3}\label{subsec:qgq}

Higher twist PDFs and FFs can also be defined via the corresponding quark-$j$-gluon-quark correlators ($j=1,2,...$ represents the number of gluons) 
too~\cite{Mulders:1995dh,Boer:1997mf,Boer:1997nt,Boer:1997qn,Boer:1999uu,
Goeke:2003az,Bacchetta:2006tn,Boer:2008fr,Pitonyak:2013dsu,Wei:2013csa,Kanazawa:2013uia,Wei:2014pma}. 
However, because of QCD equation of motion $\gamma \cdot D(y)\psi(y)=0$, 
the higher twist PDFs and  FFs defined via these quark-$j$-gluon-quark correlators are often not independent.  
They are related to those defined via  the quark-quark correlator by a set of equations derived using the equation of motion 
and can often be replaced by using these relationships when calculating the cross sections and other measurable quantities for different high energy reactions.
In this section, we take twist-3 as an example to illustrate the results for FFs defined via quark-$j$-gluon-quark correlator and their relationships to those defined via quark-quark correlator. 

Up to twist-3, we need to consider the quark-gluon-quark correlator defined as, 
\begin{align}
\hat\Xi_{\rho,ij}^{(1)}(k_F;p,S)=&\frac{1}{2\pi}  \sum_X \int  d^4\xi e^{-ik_F \xi}  \langle p,S;X|\bar\psi_j(\xi) \mathcal{L}(\xi;\infty) |0\rangle \nonumber\\
& \times \langle 0| \mathcal{L}^\dag (0;\infty) D_\rho(0)\psi_i(0) |p,S;X\rangle,  \label{eq:qgq-correlator}
\end{align}
where $D_\rho(y)\equiv -i\partial_\rho+gA_\rho(y)$ and $A_\rho(y)$ denotes the gluon field. 
Similar to the quark-quark correlator $\hat\Xi^{(0)}$, we decompose it as, 
\begin{align}
\hat\Xi^{(1)}_\rho(z,k_{F\perp};&p,S) =  \Xi^{(1)}_\rho(z,k_{F\perp};p,S) + i\gamma_5 \tilde\Xi^{(1)}_\rho(z,k_{F\perp};p,S) \nonumber\\
& + \gamma^\alpha \Xi_{\rho\alpha}^{(1)}(z,k_{F\perp};p,S) + \gamma_5\gamma^\alpha \tilde\Xi_{\rho\alpha}^{(1)}(z,k_{F\perp};p,S) \nonumber\\
& + i\sigma^{\alpha\beta}\gamma_5 \Xi_{\rho\alpha\beta}^{(1)}(z,k_{F\perp};p,S) . \label{Xi1Expansion}
\end{align}

Twist-3 components are the leading twist contributions that we obtain from $\hat\Xi^{(1)}_\rho$. 
There has to be one $\bar n$ involved in the basic Lorentz covariants and the other(s) are from the transverse components. 
Since the $\bar n$ component of gluon field goes into the gauge link, we only have the other three components for $D_\rho$ 
thus no $\bar n_\rho$-component exists in the basic Lorentz covariants. 
We therefore do not have twist-3 contributions from $\Xi_\rho^{(1)}$ or $\tilde\Xi_\rho^{(1)}$. 
The twist-3 contributions are obtained from $\Xi^{(1)}_{\rho\alpha}$,  $\tilde\Xi^{(1)}_{\rho\alpha}$ and $\Xi^{(1)}_{\rho\alpha\beta}$ 
and are given by Eqs.~(\ref{eq:Xi1UV}-\ref{eq:Xi1TT}) in Appendix~\ref{FFs}.
Here, we use a subscript $d$ to specify that they are defined via quark-gluon-quark correlator. 
A prime in the superscript before the $\perp$ denotes different polarization situation, that after the $\perp$ specifies 
different FFs for the same polarization situation. 
We see that we have totally 36 FFs at twist-3 defined via quark-gluon-quark correlator.  
This is just the same as what we obtained from the quark-quark correlator. 
Among them, 18 are $\chi$-even and the other 18 are $\chi$-odd; 
4 contribute to unpolarized part, 12 to vector polarized part and 20 to the tensor polarized part. 
We note in particular that the hermiticity in this case does not demand that the FFs defined via quark-gluon-quark correlator are real. 
They can have both real and imaginary parts. 

For the 18 chiral even FFs (the $D_d$'s and $G_d$'s), QCD equation of motion leads to rather simple relationships. 
They can be written in the following unified form, i.e.,
\begin{align}
D^K_{dS}(z,k_{\perp}) +G^K_{dS}(z,k_{\perp})&=\frac{1}{z} \Bigl[ D^K_S(z,k_{\perp})+iG^K_S(z,k_{\perp})\Bigr] , \label{eq:qgq2qq1}
\end{align}
where the superscript $K$ can be null (no superscript), a ``$\perp$" or a ``$\prime$$\perp$";
the subscript $S$ specifies the polarization of hadron and can be null (unpolarized), $L$, $T$, $LL$, $LT$ or $TT$.  
There are in fact totally 9 such equations with the following combinations of $K$ and $S$: 
 $K=$ null and $S=T$ or $LT$; $K=\perp$ and $S=$ null, $L$, $T$, $LL$, $LT$ or $TT$;  $K=\prime$$\perp$ and $S=TT$.  
For the 18 chiral odd FFs, we have also 9 equations in the form, 
\begin{align}
H^K_{dS}(z,k_{\perp}) + \frac{k_{\perp}^2}{2M^2} H^{K'}_{dS}(z,k_{\perp})& =\frac{1}{2z} \Bigl[ H^K_S(z,k_{\perp}) + \frac{i}{2}E^K_S(z,k_{\perp})\Bigr],\label{eq:qgq2qq2}
\end{align}
with the following combinations of $K$, $K'$ and $S$: 
$(K,K')$=(null, $\perp$) and $S$=null, $L$ or $LL$;
$(K,K')$=($\perp, \perp$$\prime$) or $(\prime$$\perp, \prime$$\perp$$\prime$) and $S=T$, $LT$, or $TT$. 
We note in particular that these 18 equations in fact represent 36 real equations which imply that 
all the 36 twist-3 FFs defined via quark-quark correlator are given either by the real or imaginary part 
of those defined via quark-gluon-quark correlator. 
We note also that there are of course different choices for the basic Lorentz covariants used here in defining 
these FFs via quark-quark and/or quark-gluon-quark correlators.  
We choose them in the way so the defined FFs satisfy the relationships given by Eqs.~(\ref{eq:qgq2qq1}) and (\ref{eq:qgq2qq2}).

These relationships reveal the physical essences of these FFs and also help us to choose correct conventions in defining FFs.
It is also very interesting to observe that, although not generally proved, the final results obtained for the physical observables 
up to twist-3 are all expressed in terms of FFs defined via quark-quark correlator~\cite{Mulders:1995dh,Boer:1997mf,Boer:1997nt,Boer:1997qn,Boer:1999uu,
Goeke:2003az,Bacchetta:2006tn,Boer:2008fr,Pitonyak:2013dsu,Wei:2013csa,Kanazawa:2013uia,Wei:2014pma}.
The contributions from the quark-gluon-quark correlator can be replaced by using the relations given by Eqs.~(\ref{eq:qgq2qq1}) and (\ref{eq:qgq2qq2}).

\section{Kinematic analysis of $e^+e^-\to V\pi X$}\label{sec:Kinematics}

\begin{figure}[!ht]
\centering \includegraphics[width=0.3\textwidth]{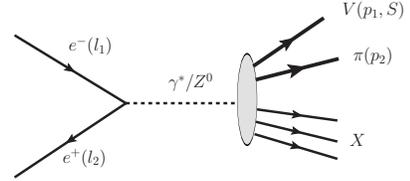}
\caption{Illustrating diagram for $e^+e^-\to V\pi X$.}
\label{fig:ff1}
\end{figure}

As mentioned in the introduction, among all different high energy reactions, 
$e^+e^-$-annihilation is most suitable for studying FFs. 
For one-dimensional FFs, the inclusive hadron production process $e^+e^-\to VX$ is the simplest case to study. 
In order to study transverse momentum dependence, we need at least two hadrons in the final state.
Hence $e^+e^-\to V\pi X$ as illustrated in Fig.~\ref{fig:ff1} is most suitable for 
studying tensor polarization dependent part of the three dimensional FFs. 
We now concentrate on this reaction and present the results for cross sections in this and next sections. 

For explicitness, we take $e^+e^-\to Z^0\to V\pi X$ as an example.  
The differential cross section is given by,
\begin{align}
\frac{2E_1E_2 d\sigma}{d^3p_1 d^3p_2} = \frac{\alpha^2 \chi}{sQ^4} L_{\mu\nu}(l_1,l_2) W^{\mu\nu}(q,p_1,S,p_2).
\end{align}
Here we use the same notations as illustrated in Fig.~\ref{fig:ff1}; 
$\alpha = e^2/4\pi$, $\chi = Q^4/[(Q^2-M_Z^2)^2 + \Gamma_Z^2 M_Z^2]\sin^4 2\theta_W$, $Q^2 = s = q^2$, $\theta_W$ is the Weinberg angle, 
$M_Z$ is the $Z$-boson's mass and $\Gamma_Z$ is the decay width. 
The leptonic tensor is well known and is given by, 
\begin{align}
L_{\mu\nu}(l_1,l_2) =& c_1^e\left[l_{1\mu} l_{2\nu}+l_{1\nu} l_{2\mu}-(l_1\cdot l_2)g_{\mu\nu}\right] + ic_3^e\varepsilon_{\mu\nu\rho\sigma}l_{1}^{\rho}l_{2}^{\sigma},  %\nonumber\\=& c_1^e L_{\mu\nu}^{(S)}(l_1,l_2) + ic_3^e L_{\mu\nu}^{(A)}(l_1,l_2). 
\label{eq:leptonictensor}
\end{align}
where $c_1^e = (c_V^e)^2 + (c_A^e)^2$ and $c_3^e = 2 c_V^e c_A^e$; 
$c_V^e$ and $c_A^e$ are defined in the weak interaction current $ \bar \psi \gamma^\mu (c_V^e - c_A^e \gamma^5) \psi$. 
Similar notations are also used for quarks.
The hadronic tensors is defined as,
\begin{align}
W_{\mu\nu} &(q,p_1,S,p_2) = \frac{1}{(2\pi)^4}  \sum_X (2\pi)^4 \delta^4 (q-p_1-p_2- p_X) \nonumber\\
& ~~~\times \langle 0| J_\nu (0) |p_1,S,p_2,X\rangle \langle p_1,S,p_2,X |J_\mu (0)|0\rangle,
\end{align}
where $S$ denotes the polarization of the hadron and for vector meson it includes both the vector and tensor polarization parts, 
$J_\mu (x)=\bar \psi(x)\Gamma_\mu\psi(x)$ and $\Gamma_\mu= \gamma^\mu (c_V^q - c_A^q \gamma^5)$. 
 
Besides the Lorentz covariance, the hadronic tensor $W^{\mu\nu}$ satisfies the general constraints imposed by Hermiticity, current conservation, 
and parity conservation in electromagnetic process, i.e.,
\begin{align}
&W^{*\mu\nu}(q,p_1,S,p_2) = W^{\nu\mu}(q,p_1,S,p_2), \label{eq:HC}\\
&q_\mu W^{\mu\nu}(q,p_1,S,p_2) = q_\nu W^{\mu\nu}(q,p_1,S,p_2) = 0, \label{eq:CC} \\
&W^{\mu\nu}(q,p_1,S,p_2) = W_{\mu\nu}(q^{\cal P}, p_1^{\cal P}, S^{\mathcal{P}},  p_2^{\cal P}). \label{eq:PC}
\end{align}
We emphasize that parity conservation is not valid in weak process via $Z$-exchange. 

\subsection{The general structure of $W^{\mu\nu}(q,p_1,S,p_2)$}\label{subsec:Wmunu}

A systematic analysis of the hadronic tensor $W_{\mu\nu}$ for $e^+e^-\to h_1h_2X$ for the case that both $h_1$ and $h_2$ are spin-1/2 hadrons 
are presented in \cite{Pitonyak:2013dsu}. 
Here, we extend the analysis to $e^+e^-\to V\pi X$ including parity conserving as well as violating contributions. 
We present the results for the basic Lorentz tensors, the cross section and structure functions in the Lorentz invariant form 
as well as in the form of azimuthal angular dependences in a particular Lorentz frame.   

\subsubsection{The basic Lorentz tensors for $W^{\mu\nu}(q,p_1,S,p_2)$}

For the spin-independent and vector polarization dependent parts, the results can just be taken from \cite{Pitonyak:2013dsu}.  
We list them here for completeness and also for unification of notations that are more convenient to extend to including tensor polarization dependent parts. 

First, the spin-independent (or unpolarized) part, we take the notations as,  
\begin{align}
h_{Ui}^{S\mu\nu}=& \Bigl\{ g^{\mu\nu}-\frac{q^\mu q^\nu}{q^2}, ~ p_{1q}^{\mu} p_{1q}^{\nu}, ~ p_{1q}^{\{\mu} p_{2q}^{\nu\}}, ~ p_{2q}^{\mu} p_{2q}^{\nu}  \Bigr\}, \\
\tilde h_{Ui}^{S\mu\nu}=& \Bigl\{ \varepsilon^{\{\mu qp_1p_2}\bigl(p_{1q}, ~ p_{2q}\bigr)^{\nu\}} \Bigr\}, \\
h_U^{A\mu\nu}=& p_{1q}^{[\mu} p_{2q}^{\nu]}, \\
\tilde h_{Ui}^{A\mu\nu}=& \Bigl\{ \varepsilon^{\mu\nu qp_1},  ~\varepsilon^{\mu\nu qp_2} \Bigr\}, 
\end{align}
where $h$ represents the parity conserved (space reflection $P$-even) tensors i.e. those satisfying Eq.~(\ref{eq:PC}) or more precisely 
$h^{\mu\nu}(q^{\cal P}, p_1^{\cal P}, S^{\mathcal{P}},  p_2^{\cal P})= h_{\mu\nu}(q, p_1, S,  p_2)$
while $\tilde h$ represents those parity non-conserved ($P$-odd)
i.e. satisfying $\tilde h^{\mu\nu}(q^{\cal P}, p_1^{\cal P}, S^{\mathcal{P}},  p_2^{\cal P})=-\tilde h_{\mu\nu}(q, p_1, S,  p_2)$;  
the superscript $S$ or $A$ denotes symmetric or anti-symmetric under exchange of ($\mu\leftrightarrow\nu$), 
the subscript $U$ denotes the unpolarized part~\cite{ffTR}.
A 4-momentum $p$ with a subscript $q$ denotes $p_q \equiv p-q(p\cdot q)/q^2$ satisfying $p_q\cdot q=0$. 
We use the short-handed notations to make the expressions more concise 
such as  $\varepsilon^{\mu qp_1p_2}\equiv \varepsilon^{\mu\alpha\rho\sigma}q_\alpha p_{1\rho}p_{2\sigma}$, 
and $\varepsilon^{\{\mu qp_1p_2}\bigl(p_{1q},~p_{2q}\bigr)^{\nu\}}$ means 
$\varepsilon^{\{\mu qp_1p_2}p_{1q}^{\nu\}}$ and $\varepsilon^{\{\mu qp_1p_2}p_{2q}^{\nu\}}$.
We see that there are totally 9 such basic tensors in the unpolarized case. 

For the vector polarization dependent part, we have, 
\begin{align}
h_{Vi}^{S\mu\nu}=& \Bigl\{ \bigl[ (q\cdot S),~(p_2\cdot S)\bigr] \tilde h_{Ui}^{S\mu\nu},~\varepsilon^{Sqp_1p_2}h_{Uj}^{S\mu\nu} \Bigr\},  \label{eq:hViS}\\
\tilde h_{Vi}^{S\mu\nu}=& \Bigl\{ \bigl[ (q\cdot S),~(p_2\cdot S)\bigr] h_{Ui}^{S\mu\nu},~\varepsilon^{Sqp_1p_2} \tilde h_{Uj}^{S\mu\nu} \Bigr\}, \label{eq:thViS}\\
h_{Vi}^{A\mu\nu}=& \Bigl\{ \bigl[ (q\cdot S),~(p_2\cdot S)\bigr] \tilde h_{Ui}^{A\mu\nu},~\varepsilon^{Sqp_1p_2}h_U^{A\mu\nu} \Bigr\}, \label{eq:hViA}\\
\tilde h_{Vi}^{A\mu\nu}=&  \Bigl\{ \bigl[ (q\cdot S),~(p_2\cdot S)\bigr] h_{U}^{A\mu\nu},~\varepsilon^{Sqp_1p_2} \tilde h_{Uj}^{A\mu\nu} \Bigr\}. \label{eq:thViA}
\end{align}
There are totally 27 such $S$-dependent basic tensors, 3 times as many as those for the unpolarized part, corresponding to 3 independent 
vector polarization modes.
 
For the tensor polarization dependent part, after some lengthy algebra, we find out that 
if we consider $S_{LL}$-, $S_{LT}$- and $S_{TT}$-dependent parts separately, we obtained the following nice symmetric forms.   

(1) {\it The $S_{LL}$-dependent part}. 
Since $S_{LL}$ is a scalar, the $S_{LL}$-dependent part is very simple. 
The $S_{LL}$-dependent basic tensors are just given by the corresponding spin-independent tensors multiplied by $S_{LL}$ 
such as $h_{LLi}^{S\mu\nu}=S_{LL}h_{Ui}^{S\mu\nu}$ and so on. 
We have therefore 9 such tensors in this case.  

(2) {\it The $S_{LT}$-dependent part}. 
In contrast to the axial-vector $S$, $S_{LT}$ is a vector satisfying the constraint $S_{LT}\cdot p_1=0$, 
the $S_{LT}$-dependent part is thus different from the $S$-dependent part. 
Furthermore, both $S_{LT}$ and $S_{TT}$ each has only two independent transverse components in the rest frame of the vector meson, 
this is guaranteed by demanding a further constraint $S_{LT}\cdot q=0$ for $S_{LT}$. 
The basic $S_{LT}$-dependent Lorentz tensors are given by, 
\begin{align}
h_{LTi}^{S\mu\nu}=& \Bigl\{ (p_2\cdot S_{LT}) h_{Ui}^{S\mu\nu},~\varepsilon^{S_{LT}qp_1p_2} \tilde h_{Uj}^{S\mu\nu} \Bigr\}, \\ 
\tilde h_{LTi}^{S\mu\nu}=&\Bigl\{ (p_2\cdot S_{LT}) \tilde h_{Ui}^{S\mu\nu},~\varepsilon^{S_{LT}qp_1p_2} h_{Uj}^{S\mu\nu} \Bigr\}, \\
h_{LTi}^{A\mu\nu}=& \Bigl\{ (p_2\cdot S_{LT}) h_{U}^{A\mu\nu},~\varepsilon^{S_{LT}qp_1p_2} \tilde h_{Uj}^{A\mu\nu} \Bigr\}, \\
\tilde h_{LTi}^{A\mu\nu}=&  \Bigl\{ (p_2\cdot S_{LT}) \tilde h_{Ui}^{A\mu\nu},~\varepsilon^{S_{LT}qp_1p_2}h_U^{A\mu\nu} \Bigr\}.
\end{align}
There are totally 18 such tensors, corresponding to the two independent $S_{LT}$-components.

(3) {\it The $S_{TT}$-dependent part}. 
$S_{TT}^{\alpha\beta}$ is a tensor satisfying the constraints, $S_{TT}^{\alpha\beta}=S_{TT}^{\beta\alpha}$, 
$g_{\alpha\beta}S_{TT}^{\alpha\beta}=0$, 
$S_{TT}^{p_1\beta}(\equiv S_{TT}^{\alpha\beta}p_{1\alpha})=0$ and $S_{TT}^{q\beta}=0$. 
We have, the $S_{TT}$-dependent basic Lorentz tensors as given by,  
\begin{align}
h_{TTi}^{S\mu\nu}=& \Bigl\{S_{TT}^{p_2p_2} h_{Ui}^{S\mu\nu}, ~\varepsilon^{S_{TT}^{p_2}qp_1p_2} \tilde h_{Uj}^{S\mu\nu} \Bigr\}, \\
\tilde h_{TTi}^{S\mu\nu}=&\Bigl\{ S_{TT}^{p_2p_2} \tilde h_{Ui}^{S\mu\nu},~\varepsilon^{S_{TT}^{p_2}qp_1p_2} h_{Uj}^{S\mu\nu}\Bigr\}, \\
h_{TTi}^{A\mu\nu}=& \Bigl\{ S_{TT}^{p_2p_2}h_{U}^{A\mu\nu}, ~\varepsilon^{S_{TT}^{p_2}qp_1p_2} \tilde h_{Uj}^{A\mu\nu} \Bigr\}, \\
\tilde h_{TTi}^{A\mu\nu}=&  \Bigl\{ S_{TT}^{p_2p_2} \tilde h_{Ui}^{A\mu\nu},~\varepsilon^{S_{TT}^{p_2}qp_1p_2} h_U^{A\mu\nu}\Bigr\}.
\end{align}
%We note in particular that $S_{TT}^{\mu\nu}$ itself is not independent from those given here since, 
%\begin{align}
%p_{2T}^4 S_{TT}^{\mu\nu} =& S_{TT}^{p_2p_2} \left( 2p_{2T}^{\mu} p_{2T}^{\nu} - \frac{|p_{2T}|^2}{\tilde p_1^2} \tilde p_1^{\mu} \tilde p_1^{\nu} + |p_{2T}|^2 h_{u1}^{\mu\nu} \right) \nonumber\\
%& - \frac{1}{q^2 \tilde p_1^2} \varepsilon^{S_{TT}^{p_2} q p_1 p_2} \varepsilon^{\{\mu q p_1 p_2} p_{2T}^{\nu\}}, \\
%p_{2T} =& \tilde p_2 - \frac{\tilde p_1 \cdot p_2}{\tilde p_1^2} \tilde p_1.
%\end{align} 
There are also totally 18 $S_{TT}$-dependent basic Lorentz tensors. 
For $W_{\mu\nu}(q,p_1,S,p_2)$, we have totally 81 basic Lorentz tensors, 41 of them are space reflection even and 40 are odd.

\subsubsection{General form of $W^{\mu\nu}(q,p_1,S,p_2)$}

The hadronic tensor $W^{\mu\nu}(q,p_1,S,p_2)$ is in general expressed as a sum of all these basic Lorentz tensors multiplied by corresponding coefficients. 
The coefficients are real and functions of the Lorentz scalars $q^2$, $q\cdot p_1$, $q\cdot p_2$ and $p_1\cdot p_2$, 
which can be replaced by $s=q^2$, $\xi_1=2q\cdot p_1/q^2$, $\xi_2=2q\cdot p_2/q^2$ and $\xi_{12}=s_{12}/s=(p_1+p_2)^2/s$. 
More precisely, we have, 
\begin{align}
W^{\mu\nu}(q,p_1,S,p_2)=&W^{S\mu\nu}(q,p_1,S,p_2)+iW^{A\mu\nu}(q,p_1,S,p_2),\label{eq:WeqWsWa}\\
W^{S\mu\nu}(q,p_1,S,p_2)=&\sum_{\sigma,i} W^S_{\sigma i}(s,\xi_1,\xi_2,\xi_{12}) h_{\sigma i}^{S\mu\nu}\nonumber\\
&+\sum_{\sigma,j} \tilde W^S_{\sigma j}(s,\xi_1,\xi_2,\xi_{12}) \tilde h_{\sigma j}^{S\mu\nu},\\
W^{A\mu\nu}(q,p_1,S,p_2)=&\sum_{\sigma,i} W^A_{\sigma i}(s,\xi_1,\xi_2,\xi_{12}) h_{\sigma i}^{A\mu\nu}\nonumber\\
&+\sum_{\sigma,j} \tilde W^A_{\sigma j}(s,\xi_1,\xi_2,\xi_{12}) \tilde h_{\sigma j}^{A\mu\nu},
\end{align}
where the subscript $\sigma$ denotes $U$, $V$, $LL$, $LT$ and $TT$ for different polarizations; 
all the coefficients $W$'s  are scalar functions of the Lorentz scalars $s$, $\xi_1$, $\xi_2$ and $\xi_{12}$.

\subsection{The general structure for the cross section}

Since the number of independent structure functions is rather large, in practice, it is often more convenient to write the cross section directly. 

\subsubsection{The Lorentz invariant form}
Making the Lorentz contraction of $W^{\mu\nu}(q,p_1,S,p_2)$ with $L_{\mu\nu}(l_1,l_2)$, we obtain the 
general form of the cross section. 
For the unpolarized part, this is given by, 
\begin{align}
 \frac{2 E_1 E_2 d\sigma^U}{d^3p_1 d^3p_2} =\frac{\alpha^2 \chi}{s^2}\bigl[ &{\cal F}_U(s,\xi_1,\xi_2,\xi_{12};y_1,y_2)\nonumber\\
 &+\tilde{\cal F}_U(s,\xi_1,\xi_2,\xi_{12};y_1,y_2,\tilde y)\bigr], 
 \end{align}
where ${\cal F}_U$and $\tilde{\cal F}_U$ represent the space reflection even and odd parts respectively and they have the structures as given by, 
\begin{align}
 {\cal F}_U =& F_U^0 + F_U^1 y_1 + F_U^2 y_2 + F_U^{11} y_1^2 + F_U^{22} y_2^2 + F_U^{12} y_1 y_2, \label{eq:calFU}\\
 \tilde{\cal F}_U =&  \tilde y(\tilde F_U^0  + \tilde F_U^1 y_1 + \tilde F_U^2  y_2),  \label{eq:tcalFU}
\end{align}
where besides $\xi_1$, $\xi_2$ and $\xi_{12}$ defined before, we introduced two new 
Lorentz scalars $y_1 = 2 p_1 \cdot l_1 / q^2$, $y_2 = 2 p_2 \cdot l_1 / q^2$ and one pseudo-scalar $\tilde y = \varepsilon^{l_1qp_1p_2} / q^4$. 
The ``structure functions" $F$'s are all scalar functions depending on $(s,\xi_1,\xi_2,\xi_{12})$. 
We see also clearly that the six $F$'s describe the parity conserved contributions while the three $\tilde F$'s represent the parity violated part.
They are related to the $W$'s by, 
\begin{align}
&F_U^0= -\frac{1}{2}c_1^e \Bigl[ 2W_{U1}^S + (m_1^2 W_{U2}^S + m_2^2 W_{U3}^S) + \nonumber\\ 
&~~~ -(s\xi_{12} - m_1^2 - m_2^2) W_{U4}^S \Bigr] +  \frac{1}{2}sc_3^e ( \xi_1 \tilde W_{U1}^A + \xi_2 \tilde W_{U2}^A ), \label{eq:FU0}\\
&F_U^1 =   \frac{1}{2} c_1^e s (\xi_1 W_{U2}^S + \xi_2 W_{U4}^S) - c_3^e s \tilde W_{U1}^A,\label{eq:FU1}\\
&F_U^2 = \frac{1}{2} c_1^e  s(\xi_2 W_{U3}^S + \xi_1 W_{U4}^S) - c_3^e  s\tilde W_{U2}^A, \label{eq:FU2}\\
&F_U^{11} = -\frac{1}{2} c_1^e s W_{U2}^S, \label{eq:FU11}\\
&F_U^{22} =-\frac{1}{2} c_1^e  s W_{U3}^S, \label{eq:FU22}\\
&F_U^{12} =-c_1^e s W_{U4}^S,\label{eq:FU12}\\
&\tilde F_U^0= c_1^e s^2 (\xi_1 \tilde W_{U1}^S + \xi_2 \tilde W_{U2}^S) - 2 c_3^e  s W_U^A,\\
&\tilde F_U^1 =-2c_1^e s^2 \tilde W_{U1}^S,\\
&\tilde F_U^2 =-2c_1^e s^2 \tilde W_{U2}^S. \label{eq:tFU2}
\end{align}
We see here that although the $F_{Ui}$'s and $\tilde F_{Ui}$'s are all functions of $s,\xi_1,\xi_2,\xi_{12}$, they contain already information from the 
leptonic tensor due to the coefficient $c_1^e$ and $c_3^e$. 
We also see that the parity conserved parts come from parity conserved hadronic tensor terms (characterized by $W$'s) contracted with 
parity conserved leptonic tensor terms (characterized by $c_1^e$) or 
parity violated hadronic tensor terms (characterized by $\tilde W$'s) contracted with 
the parity violated leptonic tensor term (characterized by $c_3^e$). 
We have six such $F_{Ui}$'s. 
Similarly we have three $\tilde F_{Ui}$'s for the parity violated parts obtained from Lorentz contractions of 
parity conserved leptonic tensor terms with parity violated hadronic tensor terms or 
parity violated leptonic tensor term with parity conserved tensor terms.   

The polarization dependent part has completely the same structure. 
For the vector polarization dependent part, from Eqs.~(\ref{eq:hViS}-\ref{eq:thViA}), we obtain immediately that,  
\begin{align}
&\frac{2E_1 E_2 d\sigma^V}{d^3p_1 d^3p_2} = \frac{\alpha^2}{s^2} \chi \Bigl\{ (q \cdot S)({\cal F}_{V1}+\tilde {\cal F}_{V1})  \nonumber\\
&~~~~~~~~~ +(p_2 \cdot S)({\cal F}_{V2}+\tilde {\cal F}_{V2})+\varepsilon^{Sqp_1p_2} ({\cal F}_{V3}+\tilde {\cal F}_{V3}) \Bigr\}. \label{eq:CsVpol}
\end{align}
Here, we note that since $q\cdot S$ and $p_2\cdot S$ are space reflection odd hence the parity conserved parts ${\cal F}_{V1}$ and ${\cal F}_{V2}$ 
take exactly the same form as $\tilde {\cal F}_U$ given by Eq. (\ref{eq:tcalFU}), 
while the parity violated  parts $\tilde {\cal F}_{V1}$ and $\tilde {\cal F}_{V2}$ 
take the same form as ${\cal F}_U$ given by Eq. (\ref{eq:calFU}) with the subscript $U$ replaced by $V1$ or $V2$. 
Since $\varepsilon^{Sqp_1p_2}$ is a scalar,  ${\cal F}_{V3}$ and $\tilde{\cal F}_{V3}$ take exactly the same form as ${\cal F}_{U}$ and $\tilde{\cal F}_{U}$ 
given by Eqs. (\ref{eq:calFU}-\ref{eq:tFU2}) respectively with the subscript $U$ replaced by $V3$. 
We have three set of $F_{Vi}$ and ${\tilde F}_{Vi}$ because there are three independent components of vector polarization. 

For the tensor polarization dependent part, we have, 
\begin{align}
\frac{2E_1 E_2 d\sigma^{LL}}{d^3p_1 d^3p_2} =  \frac{\alpha^2}{s^2} \chi  & S_{LL} ({\cal F}_{LL}+\tilde{\cal F}_{LL}), \\
\frac{2E_1 E_2 d\sigma^{LT}}{d^3p_1 d^3p_2}  =  \frac{\alpha^2}{s^2} \chi & \Bigl\{ (p_2 \cdot S_{LT})  ( {\cal F}_{LT1} + \tilde {\cal F}_{LT1} ) \nonumber\\
& + \varepsilon^{S_{LT}qp_1p_2} ( {\cal F}_{LT2} + \tilde {\cal F}_{LT2} ) \Bigr\},\\
\frac{2E_1 E_2 d\sigma^{TT}}{d^3p_1 d^3p_2} =  \frac{\alpha^2}{s^2} \chi & \Bigl\{ S_{TT}^{p_2p_2} ( {\cal F}_{TT1} + \tilde {\cal F}_{TT1} ) \nonumber\\
& + \varepsilon^{S_{TT}^{p_2}qp_1p_2} ( {\cal F}_{TT2} + \tilde {\cal F}_{TT2} )\Bigr\}.
\end{align}
Here $S_{LL}$, $p_2\cdot S_{LT}$ and $S_{TT}^{p_2p_2}$ are scalars, $\varepsilon^{S_{LT}qp_1p_2}$ and $\varepsilon^{S_{TT}^{p_2}qp_1p_2}$ are pseudo-scalars. 
Hence, ${\cal F}_{LL}$, ${\cal F}_{LT1}$, ${\cal F}_{TT1}$, $\tilde {\cal F}_{LT2}$ and $\tilde {\cal F}_{TT2}$ take exactly the same form as ${\cal F}_U$ given by Eq. (\ref{eq:calFU}),  
while $\tilde {\cal F}_{LL}$, $\tilde {\cal F}_{LT1}$, $\tilde {\cal F}_{TT1}$, ${\cal F}_{LT2}$ and ${\cal F}_{TT2}$ take exactly the same form as $\tilde {\cal F}_U$ given by Eq. (\ref{eq:tcalFU}). 

\subsubsection{In the Helicity-GJ-frame}\label{Subsec:FFs_HGJ}

Going into a special reference frame, we can express the cross section in terms of angular dependences. 
The polarization of high energy particles are described and/or studied  most conveniently in the helicity frame, 
i.e. where we choose the direction of motion of the particle as $z$-direction. 
Hence, to study polarization dependent FFs for $V$ in $e^+e^-\to V\pi X$, we suggest to choose the following frame. 
We choose center of mass frame of the $e^+e^-$-system, and direction of motion of $V$ i.e. $\vec p_1$  as $z$-direction, 
and the lepton-hadron (vector meson) plane as $Oxz$ plane. 
This is a particular Gottfried-Jackson frame~\cite{Gottfried:1964nx} which we will refer to as ``Helicity-GJ frame'' in the following of this paper. 
In this frame, we have, % the leptonic polar angle $\theta$ and the azimuthal angle $\varphi$ of $p_2$,
\begin{align}
& p_1 = (E_1, 0, 0, p_{1z}),\\
& p_2 = (E_2, |\vec p_{2T}|\cos\varphi, |\vec p_{2T}|\sin\varphi, p_{2z}),\\
& l_1 = \frac{Q}{2} (1, \sin\theta, 0, \cos\theta),\\
& l_2 = \frac{Q}{2} (1, -\sin\theta, 0, -\cos\theta),\\
& q = l_1 + l_2 = (Q, 0,0,0),
\end{align}
and we choose $\xi_1$, $\xi_2$, $|\vec p_{2T}|$, $\theta$ or $y=l_2 \cdot p_1/q\cdot p_1\approx (1+\cos\theta)/2$ and $\varphi$ as the independent variable set. 
The other variables are replaced.  
%such as $y_2=-|\vec p_{2T}|\sin\theta\cos\varphi/Q$, and 
%$\tilde y=\xi_1(|\vec p_{2T}|\sin\theta\cos\varphi/Q)(1-4M_1^2/s\xi_1^2)^{1/2}$. 
% and $\xi_{12}=\xi_1\xi_2[1+(1-4M_1^2/s\xi_1^2)^{1/2}(1-4E_{2T}^2)/s\xi_2^2)^{1/2}]/2+(m_1^2+M_2^2)/s$. 
The basic volume element transforms as,   
\begin{align}
\frac{d^3p_1 d^3p_2}{E_1E_2} =&\frac{\pi \xi_1}{ \xi_2} s(1- 4{M_{2T}^2}/{s\xi_2^2})^{-1/2} d\xi_1 d\xi_2 dy d^2p_{2T},
%\nonumber\\
%\approx& \frac{\pi \xi_1}{2 \xi_2} s d\xi_1 d\xi_2 d\vec p_{2T}^2 d\varphi dy.
\end{align}
where $M_{2T}^2=M_2^2+\vec p_{2T}^2$ and $d^2p_{2T}=d\vec p_{2T}^2d\varphi/2$.\\

(i) {\it The structure functions}\\

For the unpolarized part, we have, 
\begin{align}
{\cal F}_U=& (1+\cos^2\theta) F_{1U}+  \sin^2\theta F_{2U} + \cos\theta F_{3U} \nonumber\\
&+ \cos\varphi \bigl[ \sin\theta F_{1U}^{\cos\varphi} +  \sin2\theta F_{2U}^{\cos\varphi}\bigr] \nonumber\\
&+\cos2\varphi \sin^2\theta F_U^{\cos2\varphi}, \\
\tilde{\cal F}_U=& \sin\varphi  \bigl[ \sin\theta \tilde F_{1U}^{\sin\varphi} + \sin2\theta \tilde F_{2U}^{\sin\varphi} \bigr] \nonumber\\
&+ \sin2\varphi\sin^2\theta \tilde F_U^{\sin2\varphi},
\end{align} %%%%
where $F_{Ui}$ and $\tilde F_{Ui}$ are all scalar functions of $s$, $\xi_1, \xi_2$ and $p_{2T}^2$.
We see also clearly that we have totally 9 independent structure functions in the unpolarized case, 
6 of them are denoted by $F_{U}$'s and correspond to parity conserving terms and the other 3 are $\tilde F_{U}$'s describing parity odd part of the cross section. 
This is just the same as those shown by Eqs.~(\ref{eq:FU0}-\ref{eq:tFU2}). 
We note in particular that the structure functions $F_U$'s and $\tilde F_U$'s themselves are scalar functions of $s$, $\xi_1, \xi_2$ and $p_{2T}^2$ 
and are invariant under space reflection. But the angular dependent coefficients have the corresponding space reflection properties. 
The different basic Lorentz tensors $h_{Ui}^{\mu\nu}$'s and $\tilde h_{Ui}^{\mu\nu}$'s are transformed to different angular dependences. 
We also see that there are 3 azimuthal angle independent structure functions, 
3 parity conserving and 3 parity violating azimuthal angle dependent structure functions. 
They correspond to $\cos$ or $\sin$ asymmetries and are parity conserving and violating respectively. 

Here we take the following conventions for the notations of structure functions, i.e.,  
the superscript to denote the corresponding azimuthal angle $\varphi$-dependence, the capital letter in the subscripts to denote the polarization, 
the digital number in front of the capital letter to specify if we have more than one such structure functions corresponding to the same 
azimuthal angle $\varphi$-dependence but different $\theta$- or $y$-dependences~\cite{fn:diff}. 
We also note that to replace $\theta$ by $y$ we have,
\begin{align}  
&1+\cos^2\theta\approx 1+(2y-1)^2=2A(y),\label{eq:A(y)} \\
&\cos\theta\approx -1+2y=-B(y), \label{eq:B(y)} \\
&\sin^2\theta\approx 1-(1-2y)^2=4y(1-y)=C(y), \label{eq:C(y)} %\\
%&\sin\theta\approx 2\sqrt{y(1-y)}=D(y), \label{eq:D(y)}
\end{align}
that appear frequently in the expressions of the cross section.

For the vector polarized part, we note that, 
\begin{align}
&S = (\lambda\frac{p_{1z}}{M_1}, |\vec{S}_T|\cos\varphi_S, |\vec{S}_T|\sin\varphi_S, \lambda\frac{E_1}{M_1}). \label{eq:Sdecom}
\end{align}
The $(q\cdot S)$- and $\varepsilon^{Sqp_1p_2}$-terms in Eq.~(\ref{eq:CsVpol}) contribute to longitudinal and transverse polarization separately, while the $(p_2\cdot S)$-terms contribute to both cases. 
The contributions to transverse polarization from $(p_2\cdot S)$- and $\varepsilon^{Sqp_1p_2}$-terms
are characterized by additional $\cos(\varphi_S-\varphi)$- and $\sin(\varphi_S-\varphi)$-dependence. 
We absorb the different kinematic factors into ${\cal F}$ and $\tilde{\cal F}$ and write the cross section as, 
\begin{align}
\frac{2E_1 E_2 d\sigma^V}{d^3p_1 d^3p_2} &= \frac{\alpha^2}{s^2} \chi \Bigl\{ \lambda ({\cal F}_{L}+\tilde {\cal F}_{L})  +|\vec S_T| ({\cal F}_{T}+\tilde {\cal F}_{T})\Bigr\}.
\end{align}
Since $\lambda$ changes sign under space reflection, the parity conserving ${\cal F}_{L}$ and parity violating $\tilde {\cal F}_{L}$ 
take exactly the same form as $\tilde{\cal F}_{U}$ and ${\cal F}_{U}$ respectively. 
We have 3 $F_{iL}$'s that have one to one correspondence to $\tilde F_{iU}$'s and 
6 $\tilde F_{iL}$'s that have one to one correspondence to $F_{iU}$'s. 

For the transverse (vector) polarization dependent part, due to $\varphi_S$-dependence, the structure looks a bit different, they are given by,  
%\begin{align}
%\frac{2E_1 E_2 d\sigma^T}{d^3p_1 d^3p_2} &=\frac{\alpha^2 \chi}{(2\pi)^6s^2} |\vec{S}_T| \Big\{{\cal F}_{T} + \tilde {\cal F}_{T} \Bigr\},
%\end{align}
%
\begin{align}
{\cal F}_T =&
\sin\varphi_S \bigl[ \sin\theta F_{1T}^{\sin\varphi_S} + \sin2\theta F_{2T}^{\sin\varphi_S} \bigr] \nonumber \\
& + \sin(\varphi_S + \varphi) \sin^2\theta F_{T}^{\sin(\varphi_S + \varphi)} \nonumber\\
& + \sin(\varphi_S - \varphi) \bigl[ (1+\cos^2\theta) F_{1T}^{\sin(\varphi_S - \varphi)} \nonumber\\
&\qquad\qquad\qquad + \sin^2\theta F_{2T}^{\sin(\varphi_S - \varphi)} + \cos\theta F_{3T}^{\sin(\varphi_S - \varphi)} \bigr] \nonumber\\
& + \sin(\varphi_S - 2\varphi) \bigl[ \sin\theta F_{1T}^{\sin(\varphi_S - 2\varphi)} + \sin2\theta F_{2T}^{\sin(\varphi_S - 2\varphi)} \bigr] \nonumber \\
& + \sin(\varphi_S - 3\varphi) \sin^2\theta F_{T}^{\sin(\varphi_S - 3\varphi)},\label{eq:FT}\\
%\end{align}
%\begin{align}
\tilde{\cal F}_T =&
\cos\varphi_S \bigl[ \sin\theta \tilde F_{1T}^{\cos\varphi_S} + \sin2\theta \tilde F_{2T}^{\cos\varphi_S} \bigr] \nonumber \\
& + \cos(\varphi_S + \varphi) \sin^2\theta \tilde F_{T}^{\cos(\varphi_S + \varphi)} \nonumber\\
& + \cos(\varphi_S - \varphi) \bigl[ (1+\cos^2\theta) \tilde F_{1T}^{\cos(\varphi_S - \varphi)}  \nonumber\\
&\qquad\qquad\qquad + \sin^2\theta \tilde F_{2T}^{\cos(\varphi_S - \varphi)} + \cos\theta \tilde F_{3T}^{\cos(\varphi_S - \varphi)} \bigr] \nonumber\\
& + \cos(\varphi_S - 2\varphi) \bigl[ \sin\theta \tilde F_{1T}^{\cos(\varphi_S - 2\varphi)} + \sin2\theta \tilde F_{2T}^{\cos(\varphi_S - 2\varphi)} \bigr] \nonumber \\
& + \cos(\varphi_S - 3\varphi) \sin^2\theta \tilde F_{T}^{\cos(\varphi_S - 3\varphi)}.\label{eq:tFT}
\end{align}
There are 18 such transverse polarization dependent structure functions, 
9 of them are space reflection even and 9 are space reflection odd. 
Totally we have 27 vector polarization dependent structure functions corresponding to 
the 27 independent basic Lorentz tensors $h_{Vi}^{\mu\nu}$'s for the hadronic tensor. 
Among them, 12 contribute to space reflection even terms in the cross section, the other 15 to space reflection odd terms. 
We note in particular the $\sin\varphi_S$- and $\cos\varphi_S$-terms correspond to single transverse spin asymmetries 
in deep-inelastic lepton-nucleon scattering $e^-h\to e^-X$ with respect to the leptonic plane. 
They are either parity or time reversal odd and do not exist in $e^-h\to e^-X$.
In $e^+e^-$-annihilation, they describe the transverse polarization in or transverse to the lepton-hadron plane. 

The $S_{LL}$-dependent part is again completely the same as that for the unpolarized case, i.e., 
we have a one to one correspondence of $F_{LL}$ to $F_U$ and $\tilde F_{LL}$ to $\tilde F_U$. 

For the $S_{LT}$-dependent part, we define, 
\begin{align}
& S_{LT}^x = |S_{LT}| \cos\varphi_{LT}, \\
& S_{LT}^y = |S_{LT}| \sin\varphi_{LT}, \\
& |S_{LT}| = \sqrt{(S_{LT}^x)^2 + (S_{LT}^y)^2},
\end{align}
and we have, 
\begin{align}
\frac{2E_1 E_2 d\sigma^{LT}}{d^3p_1 d^3p_2} &= \frac{\alpha^2}{s^2} \chi |{S}_{LT}| \Big\{{\cal F}_{LT} + \tilde {\cal F}_{LT} \Bigr\},
\end{align}
Because $S_{LT}$ behaves differently from $S_T$ under space reflection, 
we obtain that ${\cal F}_{LT} $ takes exactly the same form as $\tilde{\cal F}_{T}$ 
and $\tilde{\cal F}_{LT}$ behaves in the same way as ${\cal F}_{T}$. 
More precisely, we obtain the results for ${\cal F}_{LT}$ by replacing $\varphi_S$ with $\varphi_{LT}$ and $\tilde{F}_{jT}$ with ${F}_{jLT}$ 
in Eq.~(\ref{eq:tFT}), and those for $\tilde {\cal F}_{LT}$ by replacing $\varphi_S$ with $\varphi_{LT}$ and ${F}_{jT}$ with $\tilde {F}_{jLT}$ 
in Eq.~(\ref{eq:FT}). We have exactly one to one correspondence here.

For the $S_{TT}$-dependent part, we take,
\begin{align}
& S_{TT}^{xx} = |S_{TT}| \cos2\varphi_{TT}, \\
& S_{TT}^{xy} = |S_{TT}| \sin2\varphi_{TT}, \\
& |S_{TT}| = \sqrt{(S_{TT}^{xx})^2 + (S_{TT}^{xy})^2},
\end{align}
so that $S_{TT}^{p_2p_2}$ and $\varepsilon^{S_{TT}^{p_2} q p_1 p_2}$ will contribute $\cos(2\varphi_{TT} - 2\varphi)$ and $\sin(2\varphi_{TT} - 2\varphi)$ terms.
Compare with the $S_T$ part, by changing $\varphi_S \to 2\varphi_{TT} - \varphi$, the $S_{TT}$-dependent part is classified into $\cos2\varphi_{TT}$-, $\cos(2\varphi_{TT}-\varphi)$-,  $\cos(2\varphi_{TT}-2\varphi)$-, $\cos(2\varphi_{TT}-3\varphi)$-, $\cos(2\varphi_{TT}-4\varphi)$- and the corresponding $\sin$ terms.
More precisely, they are given by,
\begin{align}
\frac{2E_1 E_2 d\sigma^{TT}}{d^3p_1 d^3p_2} &= \frac{\alpha^2}{s^2} \chi |{S}_{TT}| \Big\{{\cal F}_{TT} + \tilde {\cal F}_{TT} \Bigr\},
\end{align}
%%%
%%%
\begin{align}
{\cal F}_{TT} =& \cos2\varphi_{TT} \sin^2\theta F_{TT}^{\cos2\varphi_{TT}} \nonumber\\
&+ \cos(2\varphi_{TT}-\varphi) \bigl[ \sin\theta F_{1TT}^{\cos(2\varphi_{TT}-\varphi)} + \sin2\theta F_{2TT}^{\cos(2\varphi_{TT}-\varphi)} \bigr] \nonumber \\
& + \cos(2\varphi_{TT}-2\varphi) \bigl[ (1+\cos^2\theta) F_{1TT}^{\cos(2\varphi_{TT}-2\varphi)}  \nonumber\\
&\qquad\qquad\qquad + \sin^2\theta F_{2TT}^{\cos(2\varphi_{TT}-2\varphi)} + \cos\theta F_{3TT}^{\cos(2\varphi_{TT}-2\varphi)} \bigr] \nonumber\\
& + \cos(2\varphi_{TT} - 3\varphi) \bigl[ \sin\theta F_{1TT}^{\cos(2\varphi_{TT} - 3\varphi)} + \sin2\theta F_{2TT}^{\cos(2\varphi_{TT} - 3\varphi)} \bigr] \nonumber \\
& + \cos(2\varphi_{TT} - 4\varphi) \sin^2\theta F_{TT}^{\cos(2\varphi_{TT} - 4\varphi)}, \\
%%%
%%%
\tilde{\cal F}_{TT} =& \sin2\varphi_{TT} \sin^2\theta \tilde F_{TT}^{\sin2\varphi_{TT}} \nonumber\\
&+ \sin(2\varphi_{TT}-\varphi) \bigl( \sin\theta \tilde F_{1TT}^{\sin(2\varphi_{TT}-\varphi)} + \sin2\theta \tilde F_{2TT}^{\sin(2\varphi_{TT}-\varphi)} \bigr) \nonumber \\
& + \sin(2\varphi_{TT} - 2\varphi) [ (1+\cos^2\theta) \tilde F_{1TT}^{\sin(2\varphi_{TT} - 2\varphi)}  \nonumber\\
&\qquad\qquad\qquad + \sin^2\theta \tilde F_{2TT}^{\sin(2\varphi_{TT} - 2\varphi)} + \cos\theta \tilde F_{3TT}^{\sin(2\varphi_{TT} - 2\varphi)} ] \nonumber\\
& + \sin(2\varphi_{TT} - 3\varphi) \bigl( \sin\theta \tilde F_{1TT}^{\sin(2\varphi_{TT} - 3\varphi)} + \sin2\theta \tilde F_{2TT}^{\sin(2\varphi_{TT} - 3\varphi)} \bigr) \nonumber \\
& + \sin(2\varphi_{TT} - 4\varphi) \sin^2\theta \tilde F_{TT}^{\sin(2\varphi_{TT} - 4\varphi)}.
\end{align}

To show the regularities we list all the 81 structure functions together with the leading twist parton model results in a table. 
See table~\ref{tab:StructureFunctions} in Sec.~\ref{sec:StructureFunctions}.   \\

(ii) {\it The azimuthal asymmetries}\\

From these equations, we can calculate the azimuthal asymmetries and 
different components of hadron polarization in a straightforward way. 
E.g., 
\begin{align}
&\langle \cos\varphi\rangle_U =(\sin\theta F_{1U}^{\cos\varphi} + \sin2\theta F_{2U}^{\cos\varphi})/{2F_{Ut}}, \label{eq:Acos}\\
&\langle \cos2\varphi\rangle_U={\sin^2\theta F_{U}^{\cos2\varphi}}/{2F_{Ut}}, \label{eq:Acos2}\\
&\langle \sin\varphi\rangle_U =(\sin\theta \tilde F_{1U}^{\sin\varphi} + \sin2\theta \tilde F_{2U}^{\sin\varphi})/{2F_{Ut}}, \label{eq:Asin}\\
&\langle \sin2\varphi\rangle_U={\sin^2\theta \tilde F_{U}^{\sin2\varphi}}/{2F_{Ut}}, \label{eq:Asin2}
%
%&\langle \cos(\varphi_s-\varphi)\rangle=\frac{1}{2F_{Ut}} \bigl[
%(1+\cos^2\theta) \tilde F_{1T}^{\cos(\varphi_S - \varphi)}  \nonumber\\ 
%&~~~~~~~~~~~~~~~~~~~~~ + \sin^2\theta \tilde F_{2T}^{\cos(\varphi_S - \varphi)} + \cos\theta \tilde F_{3T}^{\cos(\varphi_S - \varphi)} \bigr],
\end{align}
where $F_{Ut}$ denotes the result of ${\cal F}_U+\tilde{\cal F}_U$ averaging over $\varphi$, i.e, 
\begin{align}
F_{Ut}&(s,\xi_1,\xi_2,p_{2T},\theta)\equiv\int \frac{d\varphi}{2\pi} ({\cal F}_U+\tilde{\cal F}_U) \nonumber\\
&= (1+\cos^2\theta) F_{1U}+  \sin^2\theta F_{2U} + \cos\theta F_{3U}.
\end{align} 
We see that these azimuthal asymmetries just equal to the corresponding structure functions divided by the azimuthal angle independent part. 
We also see that the $\cos$-asymmetries correspond to parity conserving part and  the $\sin$-asymmetries correspond to parity violating part
of the cross section so the latter vanish in parity conserving processes. \\

(iii) {\it The polarization of the vector meson $V$} \\

The average value of each component of the polarization is obtained from their correspondences to the probability differences in different polarization 
such as $\bar S_{LL}=[1-3{\cal P}(0;0,0)]/2$ where ${\cal P}(m;\theta_n,\phi_n)$ is 
the probability for $V$ to be in the eigenstate of $\Sigma^n$ with the eigenvalue $m$~\cite{Bacchetta:2000jk}.
For the five components describing the tensor polarization,  we obtain,
\begin{align}
\bar S_{LL}=\frac{1}{2}\frac{{\cal F}_{LL} + \tilde {\cal F}_{LL}}{{\cal F}_U +  \tilde {\cal F}_U}, \label{eq:barSLL}\\
\bar S_{LT}^i= \frac{2}{3} \frac{{\cal F}_{LT}^i+\tilde{\cal F}_{LT}^i}{{\cal F}_U +  \tilde {\cal F}_U},  \label{eq:barSLT}\\
\bar S_{TT}^{xi}=\frac{2}{3}\frac{{\cal F}_{TT}^{xi}+\tilde{\cal F}_{TT}^{xi}}{{\cal F}_U +  \tilde {\cal F}_U},  \label{eq:barSTT}
\end{align}
where $i=x$ or $y$ denotes different components of the polarization tensor. 
It is also interesting to see that the numerator  ${\cal F}_{LT}^x$ and ${\cal F}_{LT}^y$ 
are equal to the $\cos\varphi_{LT}$ and $\sin\varphi_{LT}$-terms of ${\cal F}_{LT}$ respectively. 
They can be obtained as follows,
\begin{align}
{\cal F}_{LT}^x =&\int \frac{d\varphi_{LT}}{\pi}\cos\varphi_{LT}{\cal F}_{LT}, \label{eq:FLTx}\\ 
{\cal F}_{LT}^y =&\int \frac{d\varphi_{LT}}{\pi}\sin\varphi_{LT}{\cal F}_{LT}, \label{eq:FLTy}
\end{align}
and similar for $\tilde{\cal F}_{LT}^i$.  For ${\cal F}_{TT}^{xi}$, we have,  
\begin{align}
{\cal F}_{TT}^{xx} =&\int \frac{d\varphi_{TT}}{\pi}  \cos2\varphi_{TT}{\cal F}_{TT}, \label{eq:FTTxx}\\
{\cal F}_{TT}^{xy} =& \int \frac{d\varphi_{TT}}{\pi}  \sin2\varphi_{TT}{\cal F}_{TT}, \label{eq:FTTxy}
\end{align}
and similar for $\tilde{\cal F}_{TT}^{xi}$. 
The explicit expressions can be obtained easily from those for the corresponding ${\cal F}_{\sigma}$ or $\tilde{\cal F}_{\sigma}$. 
We omit them here but simply emphasize that they are in general dependent 
on the variables $s$, $\xi_1$, $\xi_2$, $p_{2T}$, $\theta$ and $\varphi$.

If we average over $\varphi$, we see that only the $\varphi$ independent terms in the expressions of ${\cal F}$'s and $\tilde{\cal F}$'s survive. 
We denote them as, 
\begin{align}
\langle {\cal F}_\sigma\rangle=\int\frac{d\varphi}{2\pi}  {\cal F}_\sigma, 
\end{align}
and we obtain, 
%%%%
\begin{align}
&\langle{\cal F}_{U}\rangle =(1+\cos^2\theta) F_{1U}+  \sin^2\theta F_{2U} + \cos\theta F_{3U},\label{eq:FUint} \\
&\langle{\tilde{\cal F}}_{U}\rangle =0,\label{eq:tFUint} \\
&\langle{\cal F}_{L}\rangle =0,\label{eq:FLint} \\
&\langle\tilde{\cal F}_{L}\rangle = (1+\cos^2\theta) \tilde F_{1L}+  \sin^2\theta \tilde F_{2L} + \cos\theta \tilde F_{3L}, \label{eq:tFLint} \\
&\langle{\cal F}_{T}\rangle= \sin\varphi_S \Bigl( \sin\theta F_{1T}^{\sin\varphi_S} + \sin2\theta F_{2T}^{\sin\varphi_S} \Bigr), \label{eq:FTint}\\
&\langle\tilde{\cal F}_{T}\rangle= \cos\varphi_S \Bigl( \sin\theta \tilde F_{1T}^{\cos\varphi_S} + \sin2\theta \tilde F_{2T}^{\cos\varphi_S} \Bigr), \label{eq:tFTint} \\
&\langle{\cal F}_{LT}\rangle= \cos\varphi_{LT} \Bigl( \sin\theta F_{1LT}^{\cos\varphi_{LT}} + \sin2\theta F_{2LT}^{\cos\varphi_{LT}} \Bigr), \label{eq:FLTint}\\
&\langle\tilde{\cal F}_{LT}\rangle = \sin\varphi_{LT} \Bigl( \sin\theta \tilde F_{1LT}^{\sin\varphi_{LT}} + \sin2\theta \tilde F_{2LT}^{\sin\varphi_{LT}} \Bigr), \label{eq:tFLTint} \\
&\langle{\cal F}_{LL}\rangle= (1+\cos^2\theta) F_{1LL}+  \sin^2\theta F_{2LL} + \cos\theta F_{3LL}, \label{eq:FLLint} \\
&\langle\tilde{\cal F}_{LL}\rangle=0, \label{eq:tFLLint}\\
&\langle{\cal F}_{TT}\rangle= \cos2\varphi_{TT} \sin^2\theta F_{TT}^{\cos2\varphi_{TT}}, \label{eq:FTTint} \\
&\langle\tilde{\cal F}_{TT}\rangle= \sin2\varphi_{TT} \sin^2\theta \tilde F_{TT}^{\sin2\varphi_{TT}}. \label{eq:tFTTint}
\end{align}
We see the similarities between different components and also the $\cos\varphi_\sigma$ or $\sin\varphi_\sigma$-term 
corresponding to $x$ or $y$-component of the polarization.  
More precisely, in this case, we obtain, 
\begin{align}
&\langle \lambda \rangle=\frac{2}{3F_{Ut}} \Bigl( (1+\cos^2\theta) \tilde F_{1L}+  \sin^2\theta \tilde F_{2L} + \cos\theta \tilde F_{3L}\Bigr), \label{eq:Alambda}\\
&\langle S_{LL}\rangle= \frac{1}{2F_{Ut}} \Bigl({(1+\cos^2\theta) F_{1LL}+  \sin^2\theta F_{2LL} + \cos\theta F_{3LL}}\Bigr), \label{eq:ASll}\\
&\langle S_{T}^x\rangle=\frac{2}{3F_{Ut}} \Bigl( \sin\theta \tilde F_{1T}^{\cos\varphi_S} + \sin2\theta \tilde F_{2T}^{\cos\varphi_S}\Bigr),\label{eq:AStx}\\
&\langle S_{T}^y\rangle=\frac{2}{3F_{Ut}}\Bigl( \sin\theta F_{1T}^{\sin\varphi_S} + \sin2\theta F_{2T}^{\sin\varphi_S} \Bigr), \label{eq:ASty}\\
&\langle S_{LT}^x\rangle=\frac{2}{3F_{Ut}} \Bigl( \sin\theta F_{1LT}^{\cos\varphi_{LT}} + \sin2\theta F_{2LT}^{\cos\varphi_{LT}}\Bigr),\label{eq:ASltx}\\
&\langle S_{LT}^y\rangle=\frac{2}{3F_{Ut}}\Bigl( \sin\theta \tilde F_{1LT}^{\sin\varphi_{LT}} + \sin2\theta \tilde F_{2LT}^{\sin\varphi_{LT}} \Bigr), \label{eq:ASlty}\\
&\langle S_{TT}^{xx}\rangle=\frac{2}{3F_{Ut}}  \sin^2\theta F_{TT}^{\cos2\varphi_{TT}},\label{eq:ASttxx}\\
&\langle S_{TT}^{xy}\rangle= \frac{2}{3F_{Ut}} \sin^2\theta \tilde F_{TT}^{\sin2\varphi_{TT}}. \label{eq:ASttxy} 
\end{align}
We see that in this way we just pick up the corresponding $\varphi$-independent and, in the transverse polarization case, 
the $\cos\varphi_\sigma$ or $\sin\varphi_\sigma$-terms.
These results are much simpler and can be used to study the corresponding components of the structure functions more conveniently. 
We also note that $\langle S_{LL}\rangle$, $\langle S_{T}^y\rangle$,  
$\langle S_{LT}^x\rangle$ and $\langle S_{TT}^{xx}\rangle$ are parity conserving,  
and the other components such as $\langle \lambda \rangle$, $\langle S_{T}^x\rangle$,  
$\langle S_{LT}^y\rangle$ and $\langle S_{TT}^{xy}\rangle$ are parity violating.  
This implies that if we consider parity conserving reactions, only the $F$-terms survive and the $\tilde F_i$'s have to vanish. 
In this case we see that we have only $\langle S_{LL}\rangle$, $\langle S_{T}^y\rangle$,  
$\langle S_{LT}^x\rangle$ and $\langle S_{TT}^{xx}\rangle$ are non-zero. 
Other components such as $\langle \lambda \rangle$, $\langle S_{T}^x\rangle$,  
$\langle S_{LT}^y\rangle$ and $\langle S_{TT}^{xy}\rangle$ have to vanish.  
 
In the case where transverse components are concerned, 
it is often useful to study different components with respect to the two transverse directions $\vec e_n$ and $\vec e_t$ 
defined as $\vec e_n=\vec p_1\times\vec p_2/ |\vec p_1\times\vec p_2| = (-\sin\varphi, \cos\varphi)$ and $\vec e_t=\vec p_{2T}/|\vec p_{2T}| = (\cos\varphi, \sin\varphi)$, 
i.e. the normal and tangent of the hadron-hadron plane respectively.
The corresponding components of the polarization are given by exactly the same equations such as 
Eqs.~(\ref{eq:barSLT})-(\ref{eq:barSTT}) with $i = n$ or $t$. 
It can easily be shown that such components can also be obtained from Eqs. (\ref{eq:barSLT}) and (\ref{eq:barSTT}) 
with $\varphi_\sigma$ being replaced by $\varphi_\sigma-\varphi$ in the integrations given in Eqs. (\ref{eq:FLTx}-\ref{eq:FTTxy}), e.g.,   
\begin{align}
&{\cal F}_{T}^n =\int \frac{d\varphi_{S}}{\pi}\sin(\varphi_{S}-\varphi){\cal F}_{T},\\
&{\cal F}_{T}^t =\int \frac{d\varphi_{S}}{\pi}\cos(\varphi_{S}-\varphi){\cal F}_{T},\\
&{\cal F}_{LT}^n =\int \frac{d\varphi_{LT}}{\pi}\sin(\varphi_{LT}-\varphi){\cal F}_{LT},\\
&{\cal F}_{LT}^t =\int \frac{d\varphi_{LT}}{\pi}\cos(\varphi_{LT}-\varphi){\cal F}_{LT},\\
&{\cal F}_{TT}^{nn} =-\int \frac{d\varphi_{TT}}{\pi}\cos(2\varphi_{TT}-2\varphi){\cal F}_{TT},\\
&{\cal F}_{TT}^{nt} =\int \frac{d\varphi_{TT}}{\pi}\sin(2\varphi_{TT}-2\varphi){\cal F}_{TT}.
\end{align}
It will be also interesting to see the results after integrating over $\varphi$,  we just pick the corresponding 
$\cos(\varphi_\sigma-\varphi)$- or $\sin(\varphi_\sigma-\varphi)$-terms. 
More precisely, we have, 
\begin{align}
\langle S_{T}^n\rangle=& \frac{2}{3F_{Ut}} \Bigl[ (1+\cos^2\theta) F_{1T}^{\sin(\varphi_S - \varphi)} \nonumber\\
&+ \sin^2\theta F_{2T}^{\sin(\varphi_S - \varphi)} + \cos\theta F_{3T}^{\sin(\varphi_S - \varphi)} \Bigr], \label{eq:AStn}\\
\langle S_{T}^t\rangle=&\frac{2}{3F_{Ut}} \Bigl[ (1+\cos^2\theta) \tilde F_{1T}^{\cos(\varphi_S - \varphi)} \nonumber\\
& + \sin^2\theta \tilde F_{2T}^{\cos(\varphi_S - \varphi)} + \cos\theta \tilde F_{3T}^{\cos(\varphi_S - \varphi)}\Bigr], \label{eq:AStt}\\
\langle S_{LT}^n\rangle=& \frac{2}{3F_{Ut}} \Bigl[ (1+\cos^2\theta) \tilde F_{1LT}^{\sin(\varphi_{LT} - \varphi)} \nonumber\\
&+ \sin^2\theta \tilde F_{2LT}^{\sin(\varphi_{LT} - \varphi)} + \cos\theta \tilde F_{3LT}^{\sin(\varphi_{LT} - \varphi)} \Bigr],\label{eq:ASltn}\\
\langle S_{LT}^t\rangle=&\frac{2}{3F_{Ut}}\Bigl[ (1+\cos^2\theta) F_{1LT}^{\cos(\varphi_{LT} - \varphi)} \nonumber\\
& + \sin^2\theta F_{2LT}^{\cos(\varphi_{LT} - \varphi)} + \cos\theta F_{3LT}^{\cos(\varphi_{LT} - \varphi)} \Bigr], \label{eq:ASltt}\\
\langle S_{TT}^{nn}\rangle=& \frac{-2}{3F_{Ut}} \Bigl[ (1+\cos^2\theta) F_{1TT}^{\cos(2\varphi_{TT} - 2\varphi)}  \nonumber\\
& + \sin^2\theta F_{2TT}^{\cos(2\varphi_{TT} - 2\varphi)} + \cos\theta F_{3TT}^{\cos(2\varphi_{TT} - 2\varphi)} \Bigr], \label{eq:ASttnn}\\
\langle S_{TT}^{nt}\rangle=&\frac{2}{3F_{Ut}} \Bigl[ (1+\cos^2\theta) \tilde F_{1TT}^{\sin(2\varphi_{TT} - 2\varphi)}  \nonumber\\
& + \sin^2\theta \tilde F_{2TT}^{\sin(2\varphi_{TT} - 2\varphi)} + \cos\theta \tilde F_{3TT}^{\sin(2\varphi_{TT} - 2\varphi)} \Bigr]. \label{eq:ASttnt}
\end{align}
It is interesting to see that all the average transverse polarizations w.r.t. the hadron-hadron plane take similar form in terms of the corresponding structure functions. 
We also see that in this case $\langle S_{T}^n\rangle$, $\langle S_{LT}^t\rangle$ and $\langle S_{TT}^{nn}\rangle$ are parity conserving 
while $\langle S_{T}^t\rangle$, $\langle S_{LT}^n\rangle$ and $\langle S_{TT}^{nt}\rangle$ are parity violating. 

In experiments, it is usually very difficult to study azimuthal dependence and hadron polarization simultaneously. 
From the kinematic analysis given above, we see that we can either study the azimuthal asymmetries given by Eqs.~(\ref{eq:Acos}-\ref{eq:Asin2}) in the unpolarized case, 
or study the longitudinal hadron polarization in the helicity frame and transverse polarizations w.r.t. lepton-hadron plane 
or the hadron-hadron plane averaged over the azimuthal angle $\varphi$  
to study the corresponding structure functions as given by Eqs.~(\ref{eq:AStx}-\ref{eq:ASttxy}) or Eqs.~(\ref{eq:AStn}-\ref{eq:ASttnt}).  

\subsection{Reduce to $e^+e^-\to VX$}

It is also clear that if we consider the inclusive process $e^+e^-\to VX$, we should integrate over $p_2$, 
i.e. carrying out the integration $\int d^3p_2/(2E_2)$, to obtain the corresponding hadronic tensor and/or cross section.
In this case, we obtain 3 for unpolarized, 3 for $\lambda$-, 3 for $S_{LL}$-, 4 for $S_T$-, 4 for $S_{LT}$- and 2 for $S_{TT}$-dependent part.
The basic Lorentz tensors for the hadronic tensor obtained in this case are given by,
\begin{align}
& h_{Ui,in}^{S\mu\nu} = \Bigl\{ g^{\mu\nu}-\frac{q^\mu q^\nu}{q^2}, ~~ p_{1q}^{\mu} p_{1q}^{\nu}  \Bigr\}, \\
& \tilde h_{U,in}^{A\mu\nu} = \varepsilon^{\mu\nu qp_1}, \\
%%%
& h_{V,in}^{S\mu\nu} = \varepsilon^{\{\mu q p_1 S} p_{1q}^{\nu\}}, \\
& \tilde h_{Vi,in}^{S\mu\nu} = \Bigl\{ (q \cdot S) h_{Uj,in}^{S\mu\nu}, ~~S_q^{\{\mu} p_{1q}^{\nu\}} \Bigr\}, \label{eq:thVS} \\
& h_{Vi,in}^{A\mu\nu} = \Bigl\{ (q \cdot S) \tilde h_{U,in}^{A\mu\nu}, ~~\varepsilon^{[\mu q p_1 S} p_{1q}^{\nu]} \Bigr\}, \\
& \tilde h_{V,in}^{A\mu\nu} = S_q^{[\mu} p_{1q}^{\nu]}, \\
%%%
& h_{LLi,in}^{S\mu\nu} =S_{LL} h_{Ui,in}^{S\mu\nu}, \\
& \tilde h_{LL,in}^{A\mu\nu} = S_{LL} \tilde h_{U,in}^{A\mu\nu}, \\
%%%
& h_{LT,in}^{S\mu\nu} = S_{LT}^{\{\mu} p_{1q}^{\nu\}}, \\
& \tilde h_{LT,in}^{S\mu\nu} = \varepsilon^{\{\mu q p_1 S_{LT}} p_{1q}^{\nu\}}, \\
& h_{LT,in}^{A\mu\nu} = S_{LT}^{[\mu} p_{1q}^{\nu]}\\
& \tilde h_{LT,in}^{A\mu\nu} = \varepsilon^{[\mu q p_1 S_{LT}} p_{1q}^{\nu]}, \\
%%%
& h_{TTi,in}^{S\mu\nu} = S_{TT}^{\mu\nu}, \\
& \tilde h_{TT,in}^{S\mu\nu} = \varepsilon^{\{\mu \alpha q p_1} S_{TT\alpha}^{\nu\}}. 
\end{align}
There are totally 19 such independent basic Lorentz tensors, 10 of them are space reflection even and 9 of them are space reflection odd.  
We note in particular the spin-dependent time reversal odd term $h_{V,in}^{S\mu\nu} = \varepsilon^{\{\mu q p_1 S} p_{1q}^{\nu\}}$ 
novel to deep-inelastic lepton-nucleon scattering (DIS) as discussed in \cite{Lu:1995rp}.  
This corresponds to single transverse polarization of $V$ w.r.t. the lepton-hadron plane. 
There could be also parity violating transverse polarization in the lepton-hadron plane described by the last one in 
Eq.~(\ref{eq:thVS}) i.e.  $\tilde h_{V3,in}^{S\mu\nu} =S_q^{\{\mu} p_{1q}^{\nu\}}$. 

The inclusive process $e^+e^-\to VX$ can also be studied in the Helicity-GJ frame. 
Formerly, the differential cross section for $e^+e^-\to VX$ takes exactly the same form as that for $e^+e^-\to V\pi X$ integrated over $\varphi$.  
The corresponding inclusive structure functions just have one to one correspondence to those given by Eqs.~(\ref{eq:FUint}-\ref{eq:tFTTint}).
They are just equal to the counterparts in Eqs.~(\ref{eq:FUint}-\ref{eq:tFTTint}) integrated over $\xi_2$ and $p_{2T}^2$.  
In this case, we can study the longitudinal polarization and the transverse polarization with respect to the lepton-hadron plane 
that have similar expressions in terms of the structure functions as those given by Eqs.~(\ref{eq:Alambda}-\ref{eq:ASttxy}).

\section{Hadronic tensor in terms of FFs}\label{sec:HadronicTensor}

We now calculate the hadronic tensor and differential cross section in the partonic picture at leading order in pQCD but with leading and twist-3 contributions. 
In this section we present the results obtained for the hadronic tensor. 

\begin{figure}[!ht]
\centering \includegraphics[width=0.23\textwidth]{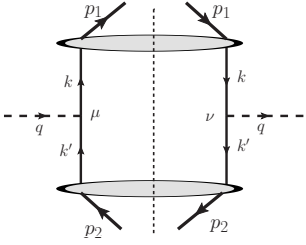}
\caption{Feynman diagram for $Z\to V\pi X$ without gluon exchange that contributes at leading and higher twists.}
\label{fig:ff2}
\end{figure}

\begin{figure}[!ht]
\centering \includegraphics[width=0.45\textwidth]{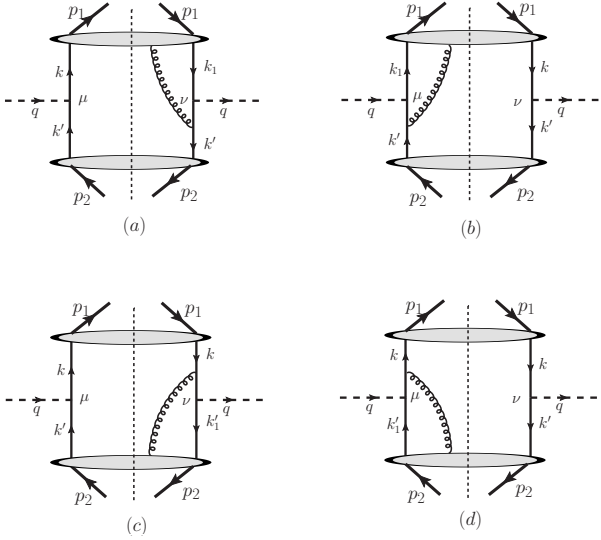}
\caption{Feynman diagrams for $Z\to V\pi X$ with one gluon exchange that contribute at twist-3 and higher twists.}
\label{fig:ff3}
\end{figure}

In the partonic picture at the leading order in pQCD,  
we need to consider the contributions from the diagrams shown in Figs.~\ref{fig:ff2} and \ref{fig:ff3} just as in \cite{Boer:1997mf} 
where spin-1/2 hadrons are considered. 
We need to perform the collinear expansion and pick up the results up to the order $1/Q$ 
in order to get the twist-3 contributions. 
Collinear expansion was first proposed for inclusive process~\cite{Ellis:1982wd,Qiu:1990xxa} 
and has now been applied to all processes where one hadron is explicitly involved~\cite{Liang:2006wp,Wei:2013csa,Wei:2014pma}.  
Systematic derivations have been given for such processes (for a recent short summary see e.g. \cite{Chen:2015tca}). 
However, for processes with no less than two hadrons are involved, systematic derivation for collinear expansion is still lacking. 
Usually, one just picks up terms up to $1/Q$ from these diagrams~\cite{Mulders:1995dh,Boer:1997mf,Boer:1997nt,Boer:1997qn,Boer:1999uu,
Goeke:2003az,Bacchetta:2006tn,Boer:2008fr,Kanazawa:2013uia}.
We do it in the same way in the following of this paper. 
 
\subsection{Hadronic tensor in the collinear frame}

The leading power contribution from Fig.~\ref{fig:ff2} gives us the leading twist contribution where no transverse gluon exchange is involved. 
The longitudinal gluon exchanges lead to the gauge link that is needed to keep the quark-quark correlator gauge invariant. 
Up to twist-3, we need the next to the leading power contribution from Fig.~\ref{fig:ff2} and also the leading power contributions from Fig.~\ref{fig:ff3}, 
where the quark-gluon-quark correlator is involved. 
We use the definition of the quark-gluon-quark correlator as given in Eq. (\ref{eq:qgq-correlator}) i.e. to use the covariant derivative $D$ instead of $A$. 
This is not only to use the simple relationships as given by Eqs. (\ref{eq:qgq2qq1}) and (\ref{eq:qgq2qq2}) 
but also to be consistent to the cases of $e^+e^-\to V\bar q X$ and $e^+e^-\to VX$ 
where collinear expansion has already been systematically proven~\cite{Wei:2013csa,Wei:2014pma}.  
To do so, we need to pick up the corresponding $k_\perp$-terms from Fig.~\ref{fig:ff2} and add them to those from Fig.~\ref{fig:ff3}. 
In this way, we obtain $W_{\mu\nu}=\tilde W_{\mu\nu}^{(0)} +\tilde W_{\mu\nu}^{(1)}-\Delta\tilde W_{\mu\nu}^{(0)}$. 
For the contribution $\tilde W_{\mu\nu}^{(0)}$ from Fig.~\ref{fig:ff2}, we have, 
\begin{align}
\tilde W_{\mu\nu}^{(0)}=&\frac{1}{p_1^+ p_2^-} \int \frac{d^2k_\perp}{(2\pi)^2} \frac{d^2k^\prime_\perp}{(2\pi)^2} \delta^2(k_\perp + k^\prime_\perp - q_\perp) \nonumber\\ 
&~\times {\rm Tr} \bigl[ \Xi^{(0)}(z_1,k_\perp,p_1,S) \Gamma_\mu \bar\Xi^{(0)}(z_2,k^\prime_\perp,p_2) \Gamma_\nu \bigr].  \label{eq:HT0}
\end{align}
Corresponding to Fig.~\ref{fig:ff3}a, we have, 
\begin{align}
\tilde W_{\mu\nu}^{(1a)} =& \frac{-1}{\sqrt{2}Qp_1^+ p_2^-}  \int \frac{d^2k_\perp}{(2\pi)^2} \frac{d^2k^\prime_\perp}{(2\pi)^2} \delta^2(k_\perp + k^\prime_\perp - q_\perp)  \nonumber\\
 \times & {\rm Tr} \bigl[ \Gamma_\mu \bar\Xi^{(0)}(z_2,k^\prime_\perp,p_2) \gamma_\rho \slashed{\bar n} \Gamma_\nu \Xi^{(1)\rho}(z_1,k_\perp,p_1,S) \bigr],  \label{eq:HT1}
\end{align}
\begin{align}
\Delta\tilde W_{\mu\nu}^{(0a)}&=
\frac{1}{\sqrt{2}Q p_1^+ p_2^-} \int \frac{d^2k_\perp}{(2\pi)^2} \frac{d^2k^\prime_\perp}{(2\pi)^2} \delta^2(k_\perp + k^\prime_\perp - q_\perp) \nonumber\\
\times & k_\perp^\rho{\rm Tr} \bigl[ \Gamma_\mu \bar\Xi^{(0)}(z_2,k^\prime_\perp,p_2) \gamma_\rho \slashed{\bar n} \Gamma_\nu \Xi^{(0)}(z_1,k_\perp,p_1,S) \bigr],\label{eq:dHT1}
%%%%%%%%%
\end{align}
and similar for those from Figs.~\ref{fig:ff3}(b-d). 
The transverse momentum dependent quark-quark or quark-gluon-quark correlator, $\Xi^{(0)}(z,k_\perp,p,S)$ or $\Xi^{(1)\rho}(z,k_\perp,p,S)$ are given by 
Eq.~(\ref{Xi0}) or (\ref{eq:qgq-correlator}) respectively. 
We use $\bar\Xi$ to denote that for anti-quark fragmentation that differs from the corresponding one for quark by exchanging $\psi$ and $\bar\psi$ in the definition. 
Here as well as in the following of this paper, for explicitness, we consider only $q\to VX$ and $\bar q\to\pi X$. 
The complete results should be the sum of these contributions and those from $\bar q\to VX$ and $q\to\pi X$. 
The latter are just obtained simply by changing the $\Xi$'s to the corresponding $\bar\Xi$'s and $\bar\Xi$'s to the corresponding $\Xi$'s.  
Also a summation over the flavor of $q$ is implicit. 
  
We emphasize in particular that these expressions (\ref{eq:HT0}-\ref{eq:dHT1}) are obtained from Figs.~(\ref{fig:ff2}-\ref{fig:ff3}) and 
they are also straightforward extensions of the results obtained for $e^+e^-\to V\bar qX$ which is a special case 
by setting $|p_2,X\rangle$ as an anti-quark final state $|k^\prime\rangle$. 
In the latter case $\tilde W_{\mu\nu}^{(0)}-\Delta\tilde W_{\mu\nu}^{(0)}$ together reduces to the corresponding results of $\tilde W_{\mu\nu}^{(0)}$ 
while $\tilde W_{\mu\nu}^{(1)}$ reduces to the corresponding result directly. 

To obtain the corresponding results for the hadronic tensors, we need to substitute 
the Lorentz decompositions of the quark-quark and quark-gluon-quark correlators 
as given by the equations in Appendix A into the above Eqs.(\ref{eq:HT0}-\ref{eq:dHT1}) and carry out the traces.  
We note that all the decompositions of the quark-quark and quark-gluon-quark correlators are given in the collinear frame of the corresponding hadron, 
i.e. the direction of motion of the hadron is taken as the longitudinal direction. 
Hence, the most convenient frame to carry out the calculations of the hadronic tensor is the collinear frame of the hadron. 
Fortunately, in the case we discuss here, we have only two hadrons and we can make a Lorentz transformation into a frame where the two hadrons moving in the opposite directions. 
We call it the collinear frame of the two hadrons. 
We first present the results of the hadronic tensor in this frame and then transform them into the Helicity-GJ-frame.  

\subsubsection{Hadronic tensor at twist-2}

The leading twist contribution to the hadronic tensor comes solely from $\tilde W_{\mu\nu}^{(0)}$ given by Eq.~(\ref{eq:HT0}). 
To obtain the results, we insert the leading twist parts for the quark-quark correlator given in Appendix A. 
The unpolarized part and the vector polarization dependent parts are the same as those for spin-$1/2$ hadrons and 
can be found e.g. in \cite{Pitonyak:2013dsu}. 
We present here for completeness and for unification of notations. 
First of all, the simplest case, i.e. the unpolarized part is given by, 
\begin{align}
W^{(0)U}_{\mu\nu}&(q,p_1,p_2) =%\frac{4}{(2\pi)^2 z_1z_2} \int d^2k_\perp d^2k^\prime_\perp \delta^2(\vec k_\perp + \vec k^\prime_\perp - \vec q_\perp)   \nonumber\\
\frac{4}{ z_1z_2} \int\frac{d^2k_\perp}{(2\pi)^2} \frac{d^2k^\prime_\perp}{(2\pi)^2}  \delta^2(k_\perp + k^\prime_\perp - q_\perp)   \nonumber\\
&\times  \biggl\{ -( c_1^q g_{\perp\mu\nu} + ic_3^q \varepsilon_{\perp\mu\nu} ) D_1(z_1,k_\perp) \bar D_1(z_2,k^\prime_\perp) \nonumber\\
&+  \frac{4c_2^q}{M_1M_2} ( k_{\perp\{\mu} k'_{\perp\nu\}} - k_\perp \cdot k'_\perp g_{\perp\mu\nu} ) 
H_1^\perp(z_1,k_\perp) \bar H_1^\perp(z_2,k^\prime_\perp)  \biggr\}, \label{eq:W0U}
\end{align}
where $c_2^q = (c_V^q)^2 - (c_A^q)^2$; $z_1\approx \xi_1$ and $z_2\approx \xi_2$ up to $1/Q$.
To make the results look more concise and explicit, we introduce the basic Lorentz tensors similar to those defined in \cite{Wei:2014pma}, i.e., 
\begin{align}
&c_{\perp\mu\nu}=c_1^q g_{\perp\mu\nu} + ic_3^q \varepsilon_{\perp\mu\nu}, \\
&\tilde c_{\perp\mu\nu}=c_3^q g_{\perp\mu\nu} + ic_1^q \varepsilon_{\perp\mu\nu},\\
&\alpha_{\perp\mu\nu}(a,b)=a_{\perp\{\mu} b_{\perp\nu\}} - (a_{\perp} \cdot b_{\perp}) g_{\perp\mu\nu},
\end{align}
for two Lorentz vectors $a$ and $b$. 
We will also omit the arguments of FFs in the expressions in the following of this paper.  
Since we are considering only the case of $q\to VX$ and $\bar q\to\pi X$, this omission will not cause any ambiguity. 
The FFs defined via $\Xi$'s, i.e. $D$'s, $G$'s, $E$'s and $H$'s, are for $q\to VX$ and have the arguments $(z_1,k_\perp)$, 
while those defined via $\bar\Xi$'s, i.e. $\bar D$'s, $\bar G$'s, $\bar E$'s and $\bar H$'s, are for $\bar q\to\pi X$ and have the arguments $(z_2,k'_\perp)$. 
With such simplified notations, we have, 
\begin{align}
W^{(0)U}_{\mu\nu}=& \frac{4}{ z_1z_2} \int\frac{d^2k_\perp}{(2\pi)^2} \frac{d^2k^\prime_\perp}{(2\pi)^2} \delta^2(k_\perp + k^\prime_\perp - q_\perp)  \nonumber\\
\times & \biggl\{ -c_{\perp\mu\nu} D_1 \bar D_1 + \frac{4c_2^q}{M_1M_2} \alpha_{\perp\mu\nu}(k,k^\prime) H_1^\perp \bar H_1^\perp \biggr\}. \label{eq:W0Usimple}
\end{align}
We see that for the unpolarized part at twist-2, we have chiral even contribution from $D_1$ convoluted with $\bar D_1$ and 
chiral odd contribution from $H_1^\perp$ convoluted with $\bar H_1^\perp$. 
We also note that for the chiral even contribution, there is a symmetric and an anti-symmetric part. 
However for the chiral odd contribution, there is only a symmetric part.

For the vector polarization dependent part, we write the longitudinally and transversely polarized parts separately. 
For the longitudinally polarized part, we have
\begin{align}
W^{(0)L}_{\mu\nu} &= \frac{4\lambda}{z_1z_2} \int\frac{d^2k_\perp}{(2\pi)^2} \frac{d^2k^\prime_\perp}{(2\pi)^2} \delta^2(k_\perp + k^\prime_\perp - q_\perp)  \nonumber\\
 \times &\biggl\{ \tilde c_{\perp\mu\nu} G_{1L} \bar D_1 
 +\frac{4c_2^q}{M_1M_2}  \alpha_{\perp\mu\nu}( \tilde k^\prime, k) H_{1L}^\perp \bar H_1^\perp \biggr\}. \label{eq:W0L}
\end{align}
We see that besides the helicity $\lambda$ factor, this takes quite similar form as that for the unpolarized part.  
Here, we have contributions from $G_{1L}$ convoluted with $\bar D_1$ and from $H_{1L}^\perp$ with $\bar H_{1}^\perp$.
For the transverse polarization dependent part, we have, 
\begin{align}
W^{(0)T}_{\mu\nu} &= \frac{4}{z_1z_2} \int\frac{d^2k_\perp}{(2\pi)^2} \frac{d^2k^\prime_\perp}{(2\pi)^2} \delta^2(k_\perp + k^\prime_\perp - q_\perp)  \nonumber\\
 \times &\biggl\{\frac{k_\perp \cdot S_T}{M_1}  \Bigl[  \tilde c_{\perp\mu\nu} G_{1T}^\perp \bar D_1
+ \frac{4c_2^q}{M_1M_2}   \alpha_{\perp\mu\nu}( \tilde k^\prime, k)  H_{1T}^\perp \bar H_1^\perp \Bigr] \nonumber\\
-& \frac{\tilde k_\perp\cdot S_\perp}{M_1} c_{\perp\mu\nu} D_{1T}^\perp  \bar D_1  
+ \frac{4c_2^q}{M_2}  \alpha_{\perp\mu\nu}(\tilde k^{\prime},S) H_{1T} \bar H_1^\perp \biggr\}. \label{eq:W0T}
\end{align}
Because there are two transverse directions, this part looks more complicated. 
We see clearly that we have both contributions in $k_\perp$ or transverse to $k_\perp$ (i.e. in $\tilde k_\perp$) directions. 

%%%%%%%%%
The $S_{LL}$-dependent part looks very much the same as the unpolarized part, i.e., 
\begin{align}
W^{(0)LL}_{\mu\nu} &=  \frac{4S_{LL}}{ z_1z_2} \int\frac{d^2k_\perp}{(2\pi)^2} \frac{d^2k^\prime_\perp}{(2\pi)^2}  \delta^2(k_\perp + k^\prime_\perp - q_\perp) \nonumber\\
\times & \biggl\{ -c_{\perp\mu\nu} D_{1LL}\bar D_1 + \frac{4c_2^q}{M_1M_2} \alpha_{\perp\mu\nu}( k, k^{\prime}) H_{1LL}^\perp \bar H_1^\perp \biggr\}, \label{eq:W0LL}
\end{align}
where we have the chiral even contribution from $D_{1LL}$ convoluted with $\bar D_1$ and chiral odd part from $H_{1LL}^\perp$ with $\bar H_1$. 

For the $S_{LT}$- and $S_{TT}$-dependent part, we have, 
\begin{align}
W^{(0)LT}_{\mu\nu} &= \frac{4}{ z_1z_2} \int\frac{d^2k_\perp}{(2\pi)^2} \frac{d^2k^\prime_\perp}{(2\pi)^2} \delta^2(k_\perp + k^\prime_\perp - q_\perp)   \nonumber\\
\times & \biggl\{ \frac{k_\perp\cdot S_{LT}}{M_1}  \Bigl[ -c_{\perp\mu\nu} D_{1LT}^\perp\bar D_1
+ \frac{4c_2^q}{M_1M_2}   \alpha_{\perp\mu\nu}(k,k') H_{1LT}^\perp\bar H_1^\perp \Bigr]\nonumber\\
+&\frac{\tilde k_\perp\cdot S_{LT}}{M_1}\tilde c_{\perp\mu\nu} G_{1LT}^\perp  \bar D_1  
+ \frac{4c_2^q}{M_2}  \alpha_{\perp\mu\nu}(k',S_{LT}) H_{1LT} \bar H_1^\perp \biggr\}, \label{eq:W0LT}\\
%\end{align}
%\begin{align}
W^{(0)TT}_{\mu\nu} & =\frac{4}{ z_1z_2} \int\frac{d^2k_\perp}{(2\pi)^2} \frac{d^2k^\prime_\perp}{(2\pi)^2} \delta^2(k_\perp + k^\prime_\perp - q_\perp)  \nonumber\\
\times & \biggl\{ \frac{S_{TT}^{kk}}{M_1^2} \Bigl[-c_{\perp\mu\nu} D_{1TT}^\perp \bar D_1 
+ \frac{4c_2^q}{M_1M_2}  \alpha_{\perp\mu\nu}(k,k')  H_{1TT}^\perp \bar H_1^\perp \Bigr] \nonumber\\
+&  \frac{S_{TT}^{\tilde kk}}{M_1^2} \tilde c_{\perp\mu\nu}  G_{1TT}^\perp \bar D_1
+ \frac{4c_2^q}{M_1M_2} \alpha_{\perp\mu\nu}(k',S_{TT}^k) H_{1TT}^{\prime\perp} \bar H_1^\perp \biggr\}. \label{eq:W0TT}
\end{align}
We see clearly the similarities and differences between them and the transverse polarization dependent part. 
We note once more that the chiral even contributions contain a symmetric and an anti-symmetric part 
given by the basic tensor $c_{\perp\mu\nu}$ or $\tilde c_{\perp\mu\nu}$
while the chiral odd contributions are always characterized by $c_2^q$ and have only symmetric tensor $\alpha_{\perp\mu\nu}$.

\subsubsection{Hadronic tensor at twist-3}
The twist-3 contribution to the hadronic tensor comes from both Eq. (\ref{eq:HT0}) and (\ref{eq:HT1}). In Eq. (\ref{eq:HT0}), 
we either expand $\Xi^{(0)}$ to leading twist and $\bar\Xi^{(0)}$ to twist-3 or $\bar\Xi^{(0)}$ to leading twist and $\Xi^{(0)}$ to twist-3. 
In Eq. (\ref{eq:HT1}), we expand all the $\Xi$'s to their leading twist contribution.
The equations are a bit longer than those at leading twist, we present as examples the results for the unpolarized and $S_{LL}$-dependent parts here but 
other parts in the appendix, 
\begin{align}
W^{(1)U}_{\mu\nu} & = \frac{4}{ z_1z_2} \int\frac{d^2k_\perp}{(2\pi)^2} \frac{d^2k^\prime_\perp}{(2\pi)^2} \delta^2(k_\perp + k^\prime_\perp - q_\perp) \nonumber\\
 \times \biggl\{ &  \frac{1}{p_1^+} \Bigl[ \omega_{\mu\nu}(k)   D^\perp  +  \tilde\omega_{\mu\nu}(\tilde k) G^\perp\Bigr] \bar D_1 \nonumber\\
 & - \frac{1}{p_2^-} D_1 \Bigl[ \omega_{\mu\nu}(k') \bar D^\perp + \tilde\omega_{\mu\nu}(\tilde k^\prime) \bar G^\perp ] \nonumber\\
 & - \frac{2c_2^qM_2}{M_1 p_2^-} H_1^\perp \Bigl[ 2(k_n - k_{\bar n})_{\{\mu\nu\}}  \bar H  + i (k_n - k_{\bar n})_{[\mu\nu]}  \bar E \Bigr]  \nonumber \\
 & + \frac{2c_2^qM_1}{M_2p_1^+} \Bigl[ 2(k'_n -k'_{ \bar n})_{\{\mu\nu\}} H  + i(k'_n -k'_{ \bar n})_{[\mu\nu]} E \Bigr] \bar H_1^\perp \nonumber\\
%%%%%%%%%%%%%%%%%
&+ \frac{\sqrt{2}}{Q} \Bigl[ \omega_{\mu\nu}(k',k)  D_1  \bar D_1
 - \frac{4c_2^q}{M_1M_2} \omega_{\mu\nu}^{(n)}(k,k^\prime) H_1^\perp \bar H_1^\perp \Bigr] \biggr\}, \label{eq:W1U}
\end{align}
where we introduce the short handed notations defined as, 
\begin{align}
& a_{n\{\mu\nu\}}\equiv a_{\perp\{\mu} n_{\nu\}}, ~~~~~~a_{n[\mu\nu]}\equiv a_{\perp[\mu} n_{\nu]}, \\
&\omega_{\mu\nu}(a,b)=c_1^q ( a_{n} + b_{\bar n})_{\{\mu\nu\}}  - ic_3^q ( \tilde a_{n} + \tilde b_{\bar n})_{[\mu\nu]},\\
&\tilde \omega_{\mu\nu}(a,b) = c_3^q ( a_{n} + b_{\bar n})_{\{\mu\nu\}} - ic_1^q ( \tilde a_{n} + \tilde b_{\bar n})_{[\mu\nu]}, \\
%
%&\omega_{\mu\nu}(a)\equiv\omega_{\mu\nu}(a,-a)=c_1^q a_{(n-\bar n)\{\mu\nu\}}  - ic_3^q \tilde a_{(n-\bar n)[\mu\nu]},\\
%&\tilde \omega_{\mu\nu}(a) \equiv\tilde \omega_{\mu\nu}(a,-a)= c_3^q a_{(n-\bar n)\{\mu\nu\}} - ic_1^q \tilde a_{(n-\bar n)[\mu\nu]}, \\
%
& \omega_{\mu\nu}^{(n)}(a,b) = (a_\perp^2 b_{\bar n} + b_\perp^2 a_{n})_{\{\mu\nu\}}, 
\end{align}
and $\omega_{\mu\nu}(a)\equiv\omega_{\mu\nu}(a,-a)$, $\tilde\omega_{\mu\nu}(a)\equiv\tilde\omega_{\mu\nu}(a,-a)$.

The $S_{LL}$-dependent part looks very much similar, i.e., 
\begin{align}
W^{(1)LL}_{\mu\nu} & = \frac{4S_{LL}}{z_1z_2} \int\frac{d^2k_\perp}{(2\pi)^2} \frac{d^2k^\prime_\perp}{(2\pi)^2} \delta^2(k_\perp + k^\prime_\perp - q_\perp)  \nonumber\\
 \times \biggl\{ &  \frac{1}{p_1^+} \Bigl[ \omega_{\mu\nu}(k)  D_{LL}^\perp  + \tilde\omega_{\mu\nu}(\tilde k) G_{LL}^\perp\Bigr] \bar D_1 \nonumber\\
 & - \frac{1}{p_2^-} D_{1LL} \Bigl[\omega_{\mu\nu}(k') \bar D^\perp + \tilde\omega_{\mu\nu}(\tilde k^\prime) \bar G^\perp ] \nonumber\\
 & - \frac{2c_2^qM_2}{M_1 p_2^-} H_{1LL}^\perp \Bigl[ 2(k_n -k_{ \bar n})_{\{\mu\nu\}}  \bar H  + i(k_n -k_{ \bar n})_{[\mu\nu]}  \bar E \Bigr]  \nonumber \\
 & + \frac{2c_2^qM_1}{M_2p_1^+} \Bigl[ 2(k'_n -k'_{ \bar n})_{\{\mu\nu\}} H_{LL}  + i(k'_n -k'_{ \bar n})_{[\mu\nu]}  E_{LL} \Bigr] \bar H_1^\perp \nonumber\\
 %%%%%%%%%%%%%%%%%
&+ \frac{\sqrt{2}}{Q} \Bigl[ \omega_{\mu\nu}(k',k) D_{1LL}  \bar D_1 
 - \frac{4c_2^q}{M_1M_2}\omega_{\mu\nu}^{(n)}(k,k^\prime) H_{1LL}^\perp \bar H_1^\perp \Bigr] \biggr\}. \label{eq:W1LL}
\end{align}

\subsection{Transform into the Helicity-GJ-frame}

We now transform the hadronic tensor into Helicity-GJ-frame as described in Sec.~\ref{Subsec:FFs_HGJ}. 
Since our goal is to express the hadronic tensor by FFs that are usually defined in the collinear way, 
we should just keep the FFs defined this way and transform the coefficients into the Helicity-GJ-frame of the vector meson $V$. 
This is achieved by replacing the vectors and tensors in the hadronic tensor by their expressions in the Helicity-GJ-frame. 
Up to $1/Q$, we have~\cite{Boer:1997mf,Boer:2008fr}, 
\begin{align}
& \bigl( k_{\perp\mu} \bigr)_{coll}= k_{\perp\mu} - {\sqrt{2}q_\perp \cdot k_\perp} \bar n_\mu/Q+\cdots , \\
& \bigl( g_{\perp\mu\nu} \bigr)_{coll} = g_{\perp\mu\nu} - {\sqrt{2}} q_{\bar n\{\mu\nu\}}/Q+\cdots , \\
& \bigl( \varepsilon_{\perp\mu\nu} \bigr)_{coll} = \varepsilon_{\perp\mu\nu} + {\sqrt{2}}\tilde{q}_{\bar n[\mu\nu]}/Q+\cdots ,
\end{align}
and $q_\perp=-p_{2T}/z_2+\cdots$, where $\cdots$ are higher power suppressed terms. 
We see that the differences are all higher twist. 
It implies that the leading twist part is unchanged but there are additional twist-3 terms generated by transforming the twist-2 parts. 
E.g., for the unpolarized part, we have,
\begin{align}
\delta W^{(1)U}_{\mu\nu} &= \frac{4\sqrt{2}}{z_1z_2 Q} \int\frac{d^2k_\perp}{(2\pi)^2} \frac{d^2k^\prime_\perp}{(2\pi)^2}  \delta^2(k_\perp + k^\prime_\perp - q_\perp)  \nonumber\\
\times & \Bigl\{ -\bigl( c_1^q q_{\bar n\{\mu\nu\}} - ic_3^q \tilde q_{\bar n[\mu\nu]} \bigr) D_1 \bar D_1 \nonumber\\
& + \frac{4c_2^q}{M_1M_2} \bigl(  k_\perp^2 k^\prime_{\bar n} + k_\perp^{\prime2} k_{\bar n})_{\{\mu\nu\}}  H_1^\perp \bar H_1^\perp \Bigr\}. \label{eq:dW1U}
\end{align}
Others are given in the appendix \ref{sec:twist3Ws}.

\section{Structure functions in terms of FFs}\label{sec:StructureFunctions}

Making Lorentz contraction with the leptonic tensor, we obtain the cross section and the structure functions. 
The parton model results for the structure functions are given as convolution of the gauge invariant TMD FFs in the form,
\begin{align}
\mathcal{C} [ w D \bar D ] = \frac{1}{z_1z_2}\int & \frac{d^2k_\perp}{(2\pi)^2} \frac{d^2k^\prime_\perp}{(2\pi)^2} \delta^2(k_\perp +k_\perp^\prime - q_\perp)\nonumber\\
& \times w(k_\perp,k_\perp^\prime) D(z_1,k_\perp) \bar D(z_2,k_\perp^\prime).
\end{align}
The weight $w$ is a scalar function of $k_\perp$ and $k'_\perp$. 
As in \cite{Pitonyak:2013dsu}, we introduce the following dimensionless scalars, 
\begin{align}
& w_0 = -{{k}_\perp^2}/{M_1^2},\label{eq:w0} \\
& \bar w_0 =- {{k}_\perp^{\prime 2}}/{M_2^2}, \label{eq:bw0} \\
& w_1 = -{p_{2T} \cdot k_\perp}/{M_1|\vec p_{2T}|},\label{eq:w1} \\
& \bar w_1 =- {p_{2T} \cdot k'_\perp}/{M_2|\vec p_{2T}|}, \label{eq:bw1}\\
& w_2 = -{{k}_\perp \cdot {k}_\perp^\prime}/{M_1M_2}. \label{eq:w2}
\end{align}
Others are just functions of them and are given when needed.

\subsection{Structure functions at twist-2}

We note that the twist-2 results presented here are for leading order in pQCD. 
Formally they just correspond to the results obtained from the naive or intuitive parton model.  

We introduce a second digital in the subscript to specify the contributions at twist level, e.g $F_{jTi}^{\sin(\varphi_S-\varphi)}$, 
and $i=1,2,3,...$ to specify the twist-$(i+1)$ contributions. 
The unpolarized and vector polarization dependent parts can be derived from those given in e.g. \cite{Pitonyak:2013dsu}. 
We list them here for completeness and comparison.  
We list only those non-zero structure functions. 
Those are not listed are zero at twist-2. 

For the unpolarized part, we have,
\begin{align}
& F_{1U1} ={2c_1^e c_1^q} \mathcal{C} [ D_1 \bar D_1 ], \label{eq:F1U1} \\
%& F_{2U1} = 0, \\
& F_{3U1} = {4c_3^e c_3^q} \mathcal{C} [ D_1 \bar D_1 ],  \label{eq:F3U1} \\
%& F_{1U1}^{\cos \varphi} =0, \\
%& F_{2U1}^{\cos \varphi} =0, \\
& F_{U1}^{\cos 2\varphi} = -{8c_1^e c_2^q} \mathcal{C} [ w_{hh} H_1^\perp \bar H_1^\perp ], \label{eq:FU1cos2phi}
%& \tilde F_{1U1}^{\sin \varphi} =0, \\
%& \tilde F_{2U1}^{\sin \varphi} =0, \\
%& \tilde F_{U1}^{\sin 2\varphi} =0, \\ 
\end{align}
where $w_{hh}=2w_1 \bar w_1 - w_2$.
The other 6 $F_U$'s are zero at twist-2. 
We note in particular that there is a twist-2 contribution to $\cos2\varphi$ due to Collins function~\cite{Collins:1992kk} 
but no such contribution to $\cos\varphi$ or $\sin\varphi$. 

The longitudinal polarization dependent part is very much the same as the unpolarized part. 
There are 3 non-zero $F_L$'s at twist-2, they are given by, 
\begin{align}
& \tilde F_{1L1} = -{2c_1^e c_3^q} \mathcal{C} [ G_{1L} \bar D_1 ], \label{eq:tF1L1} \\
& \tilde F_{3L1} = -{4c_3^e c_1^q} \mathcal{C} [ G_{1L} \bar D_1 ], \label{eq:tF3L1} \\
& F_{L1}^{\sin2\varphi} = -{8c_1^e c_2^q} \mathcal{C} [ w_{hh} H_{1L}^\perp \bar H_1^\perp ]. \label{eq:FLsin2phi}
\end{align}
We see a one to one correspondence to the unpolarized terms. 
More precisely we have $\tilde F_{jL1}$ just corresponds to $F_{jU1}$ upon exchange of $D_1$ to $G_{1L}$ 
and $F_{jL1}^{\sin2\varphi}$ just corresponds to $F_{U1}^{\cos2\varphi}$ upon exchange of $H_1^\perp$ to $H_{1L}^\perp$. 

For the transverse polarization dependent part, we have,  
\begin{align}
& F_{1T1}^{\sin(\varphi_S-\varphi)} ={2c_1^e c_1^q} \mathcal{C} [ w_1 D_{1T}^\perp \bar D_1 ], \label{eq:F1T1} \\
& F_{3T1}^{\sin(\varphi_S-\varphi)} = {4c_3^e c_3^q} \mathcal{C} [ w_1 D_{1T}^\perp \bar D_1 ], \label{eq:F3T1}\\
& \tilde F_{1T1}^{\cos(\varphi_S-\varphi)} ={2c_1^e c_3^q} \mathcal{C} [ w_1 G_{1T}^\perp \bar D_1 ], \label{eq:tF1T1} \\
& \tilde F_{3T1}^{\cos(\varphi_S-\varphi)} = {4c_3^e c_1^q} \mathcal{C} [ w_1 G_{1T}^\perp \bar D_1 ], \label{eq:tF3T1} \\
& F_{T1}^{\sin(\varphi_S+\varphi)} = -{8c_1^e c_2^q} \mathcal{C} [ \bar w_1 {\mathcal H}_{1T}^{\perp} \bar H_1^\perp ], \\
& F_{T1}^{\sin(\varphi_S-3\varphi)} = -{8c_1^e c_2^q} \mathcal{C} [ w_{hh}^t H_{1T}^\perp \bar H_1^\perp ],
\end{align}
where $w_{hh}^t=w_1w_{hh} - w_0 \bar w_1/2$, ${\mathcal H}_{1T}^{\perp}$ is defined by Eq.~(\ref{eq:calK}). 
%\equiv   H_{1\sigma}+k_\perp^2 H_{1\sigma}^\perp/2M_1^2, \label{eq:calH}
%\end{align}
%for  $\sigma=T$ and $LT$ used here and in the following of this paper.  
We see that there are 6 non-zero transverse polarization dependent structure functions ($F_T$ or $\tilde F_T$) 
at twist-2, 4 of them are parity conserving and the other 2 are parity violating. 

We see that among the 36 spin-independent and vector polarization dependent structure functions, 
12 of them have twist-2 contributions while the other 24 are zero at twist -2. 
Among these 12 non-zero $F$'s, 8 are parity conserving 4 are parity violating, 8 of them correspond to azimuthal asymmetries. 

For the tensor polarization dependent part, the results are much similar. 
First the $S_{LL}$-dependent part looks very much the same as the unpolarized part. 
There are only three non-zero $F_{LL}$'s at twist-2, they are given by,
\begin{align}
& F_{1LL1} = {2c_1^e c_1^q}\mathcal{C} [ D_{1LL} \bar D_1 ], \\
& F_{3LL1} = {4c_3^e c_3^q} \mathcal{C} [ D_{1LL} \bar D_1 ], \\
& F_{LL1}^{\cos 2\varphi} = -{8c_1^e c_2^q} \mathcal{C} [ w_{hh} H_{1LL}^\perp \bar H_1^\perp ].
\end{align}

The $S_{LT}$-dependent part is very much similar to the $S_T$-part. 
The 6 non-zeros are given by,
\begin{align}
& F_{1LT1}^{\cos(\varphi_{LT}-\varphi)} = -{2c_1^e c_1^q} \mathcal{C} [ w_1 D_{1LT}^\perp \bar D_1 ], \\
& F_{3LT1}^{\cos(\varphi_{LT}-\varphi)} = -{4c_3^e c_3^q} \mathcal{C} [ w_1 D_{1LT}^\perp \bar D_1 ], \\
& \tilde F_{1LT1}^{\sin(\varphi_{LT}-\varphi)} = -{2c_1^e c_3^q} \mathcal{C} [ w_1 G_{1LT}^\perp \bar D_1 ], \\
& \tilde F_{3LT1}^{\sin(\varphi_{LT}-\varphi)} = -{4c_3^e c_1^q} \mathcal{C} [ w_1 G_{1LT}^\perp \bar D_1 ], \\
& F_{LT1}^{\cos(\varphi_{LT}+\varphi)} = -{8c_1^e c_2^q} \mathcal{C} [ \bar w_1 {\cal H}_{1LT}^\perp \bar H_1^\perp ], \\
& F_{LT1}^{\cos(\varphi_{LT}-3\varphi)} = {8c_1^e c_2^q} \mathcal{C} [ w_{hh}^{t}H_{1LT}^\perp \bar H_1^\perp ]. 
\end{align}
%where ${\cal H}_{1LT}^{\perp} \equiv  H_{1LT} + k_\perp^2 H_{1LT}^\perp/2M_1^2$.

The $S_{TT}$-dependent part is similar to the $S_T$-part but the weights are different. 
\begin{align}
& F_{1TT1}^{\cos(2\varphi_{TT}-2\varphi)} = {2c_1^e c_1^q} \mathcal{C} [ w_{dd}^{tt} D_{1TT}^\perp \bar D_1 ], \\
& F_{3TT1}^{\cos(2\varphi_{TT}-2\varphi)} = {4c_3^e c_3^q} \mathcal{C} [ w_{dd}^{tt} D_{1TT}^\perp \bar D_1 ], \\
& \tilde F_{1TT1}^{\sin(2\varphi_{TT}-2\varphi)} = {2c_1^e c_3^q} \mathcal{C} [ w_{dd}^{tt} G_{1TT}^\perp \bar D_1 ], \\
& \tilde F_{3TT1}^{\sin(2\varphi_{TT}-2\varphi)} = {4c_3^e c_1^q} \mathcal{C} [ w_{dd}^{tt} G_{1TT}^\perp \bar D_1 ], \\
& F_{TT1}^{\cos(2\varphi_{TT}-4\varphi)} = -{4c_1^e c_2^q} \mathcal{C} [ w_{hh}^{tt} H_{1TT}^\perp \bar H_1^\perp ], \\
& F_{TT1}^{\cos2\varphi_{TT}} = {8c_1^e c_2^q} \mathcal{C} [ w_2 {H}_{1TT}^{\perp\prime}  \bar H_1^\perp ],
\end{align}
where $w_{dd}^{tt} = 2w_1 - w_0$, $w_{hh}^{tt}=w_0 w_2 - 4w_0 w_1 \bar w_1 + 4w_1^2 w_2+ 8w_1^3 \bar w_1$, 
and $H_{1TT}^{\perp\prime}\equiv %H_{1TT}^{\prime\perp} -\frac{1}{2}(w_0 + 8w_1^2) H_{1TT}^\perp=
H_{1TT}^{\prime\perp}+[k_\perp^2+8(k_\perp\cdot p_{2T})^2/p_{2T}^2]H_{1TT}^\perp/2M_1^2$. 

\begin{widetext}

%\begin{landscape}
%\begin{sidewaystable}[ht]
\begin{table}[!ht]
\caption{The 81 structure functions and their twist-2 results.
The capital letter U, L, LL and so on in the first column denote the polarization of $V$. 
The $\theta$-dependence is common for all lines so we list it only in the first line. 
For the U, L and LL polarized parts, the space reflection behaviors of first three columns are different from the other six. 
For the others, i.e. T, LT and TT polarized parts, the whole line has the same reflection behavior and we use 
PC and PV to denote party conserved and parity violated respectively.} \label{tab:StructureFunctions}
\begin{tabular}{c|cc|c|ccc|cc|c}\hline
$\theta$-dep. & $\sin\theta$ & $\sin2\theta$ & \multicolumn{1}{c||}{$\sin^2\theta$} & $1+\cos^2\theta$ & $\sin^2\theta$ & $\cos\theta$ & $\sin\theta$ & $\sin2\theta$ & $\sin^2\theta$  \rule[-0.20cm]{0mm}{0.6cm} \\ \hline\hline
$\varphi$-dep. & \multicolumn{2}{c|} {$\sin\varphi$} &  \multicolumn{1}{c||}{$\sin2\varphi$}  & \multicolumn{3}{c|} {$1$} & \multicolumn{2}{c|} {$\cos\varphi$} & $\cos2\varphi$ \rule[-0.20cm]{0mm}{0.6cm} \\ \hline
U &$\tilde F_{1U}^{\sin\varphi}$&$\tilde F_{2U}^{\sin\varphi}$& \multicolumn{1}{c||}{$\tilde F_{U}^{\sin2\varphi}$} 
& $F_{1U}/2c_1^ec_1^q$ & $F_{2U}$ & $F_{3U}/4c_3^ec_3^q$ & $F_{1U}^{\cos\varphi}$ & $F_{2U}^{\cos\varphi}$ & $F_{U}^{\cos2\varphi}/8c_1^ec_2^q$ \rule[-0.20cm]{0mm}{0.6cm} \\ 
%$\theta$-dep. & $\sin\theta$ & $\sin2\theta$ & $\sin^2\theta$ & $1+\cos^2\theta$& $\sin^2\theta$ & $\cos\theta$ & $\sin\theta$ & $\sin2\theta$ & $sin^2\theta$ \rule[-0.20cm]{0mm}{0.6cm} \\ \hline
twist-2 & 0 & 0 &   \multicolumn{1}{c||}{0} & $\mathcal{C} [ D_1 \bar D_1 ]$ & 0 & $\mathcal{C} [ D_1 \bar D_1 ]$ & 0 & 0 & $- \mathcal{C} [ w_{hh}  H_1^\perp \bar H_1^\perp ]$  \rule[-0.20cm]{0mm}{0.6cm} \\ \hline%\hline
%%%%%%%
L &$F_{1L}^{\sin\varphi}$&$F_{2L}^{\sin\varphi}$& \multicolumn{1}{c||}{$F_{L}^{\sin2\varphi}/8c_1^ec_2^q$} & $\tilde F_{1L}/2c_1^ec_3^q$ & $\tilde F_{2L}$ & $\tilde F_{3L}/4c_3^ec_1^q$ & $\tilde F_{1L}^{\cos\varphi}$ & $\tilde F_{2L}^{\cos\varphi}$ & $\tilde F_{L}^{\cos2\varphi}$  \rule[-0.20cm]{0mm}{0.6cm} \\ %\hline
%$\varphi$-dep. & \multicolumn{2}{c|} {$\sin\varphi$} &  \multicolumn{1}{c||}{$\sin2\varphi$} & \multicolumn{3}{c|} {$1$} & \multicolumn{2}{c|} {$\cos\varphi$} & $\cos2\varphi$  \rule[-0.20cm]{0mm}{0.6cm} \\ %\hline
%$\theta$-dep. & $\sin\theta$ & $\sin2\theta$ & $\sin^2\theta$ & $1+\cos^2\theta$& $\sin^2\theta$ & $\cos\theta$ & $\sin\theta$ & $\sin2\theta$ & $sin^2\theta$  \rule[-0.20cm]{0mm}{0.6cm} \\ \hline
twist-2 & 0 & 0 &  \multicolumn{1}{c||}{$- \mathcal{C} [ w_{hh}  H_{1L}^\perp \bar H_1^\perp ]$} & $ - \mathcal{C} [ G_{1L} \bar D_1 ]$ & 0 & $-\mathcal{C} [ G_{1L} \bar D_1 ]$ & 0 & 0 
& 0  \rule[-0.20cm]{0mm}{0.6cm} \\ \hline%\hline
%%%%%%%%
LL &$\tilde F_{1LL}^{\sin\varphi}$&$\tilde F_{2LL}^{\sin\varphi}$& \multicolumn{1}{c||}{$\tilde F_{LL}^{\sin2\varphi}$} & $F_{1LL}/2c_1^ec_1^q$ & $F_{2LL}$ & $F_{3LL}/4c_3^ec_3^q$ & $F_{1LL}^{\cos\varphi}$ & $F_{2LL}^{\cos\varphi}$ & $F_{LL}^{\cos2\varphi}/8c_1^ec_2^q$  \rule[-0.20cm]{0mm}{0.6cm} \\ %\hline
%$\varphi$-dep. & \multicolumn{2}{c|} {$\sin\varphi$} &  \multicolumn{1}{c||}{$\sin2\varphi$} & \multicolumn{3}{c|} {$1$} & \multicolumn{2}{c|} {$\cos\varphi$} & $\cos2\varphi$  \rule[-0.20cm]{0mm}{0.6cm} \\ %\hline
%$\theta$-dep. & $\sin\theta$ & $\sin2\theta$ & $\sin^2\theta$ & $1+\cos^2\theta$& $\sin^2\theta$ & $\cos\theta$ & $\sin\theta$ & $\sin2\theta$ & $sin^2\theta$  \rule[-0.20cm]{0mm}{0.6cm} \\ \hline
twist-2 & 0 & 0 &  \multicolumn{1}{c||}{0} & $\mathcal{C} [ D_{1LL} \bar D_1 ]$ & 0 & $ \mathcal{C} [ D_{1LL} \bar D_1 ]$ & 0 & 0 
& $- \mathcal{C} [ w_{hh} H_{1LL}^\perp \bar H_1^\perp ]$  \rule[-0.20cm]{0mm}{0.6cm} \\ \hline\hline
%%%%%%%
%\end{tabular}
%\end{table}
%
%\begin{table}[!ht]
%\caption{The 9 structure functions for T-dependent cases}\label{tab:FFunpol}
%\begin{tabular}{c|cc|c|ccc|cc|c}\hline
%\color{red}$\theta$-dep. & \color{red}$\sin\theta$ & \color{red}$\sin2\theta$ & \color{red}$\sin^2\theta$ & \color{red}$1+\cos^2\theta$& \color{red}$\sin^2\theta$ & \color{red}$\cos\theta$ & \color{red}$\sin\theta$ & \color{red}$\sin2\theta$ & \color{red}$\sin^2\theta$ \rule[-0.20cm]{0mm}{0.6cm} \\ \hline%\hline
%$\theta$-dep. & $\sin\theta$ & $\sin2\theta$ & $\sin^2\theta$ & $1+\cos^2\theta$& $\sin^2\theta$ & $\cos\theta$ & $\sin\theta$ & $\sin2\theta$ & $\sin^2\theta$  \rule[-0.20cm]{0mm}{0.6cm} \\ \hline
T-PC & $F_{1T}^{\sin\varphi_S}$ & $F_{2T}^{\sin\varphi_S}$ &$F_{T}^{\sin(\varphi_S+\varphi)}/{8c_1^e c_2^q} $ & $F_{1T}^{\sin(\varphi_S-\varphi)}/2c_1^ec_1^q$ &$F_{2T}^{\sin(\varphi_S-\varphi)}$ &$F_{3T}^{\sin(\varphi_S-\varphi)}/4c_3^ec_3^q$ & 
$F_{1T}^{\sin(\varphi_S-2\varphi)}$ &$F_{2T}^{\sin(\varphi_S-2\varphi)}$ &$F_{T}^{\sin(\varphi_S-3\varphi)}/8c_1^ec_2^q$ \rule[-0.20cm]{0mm}{0.6cm} \\ %\hline
$\varphi$-dep. & \multicolumn{2}{c|} {$\sin\varphi_S$} & $\sin(\varphi_S+\varphi)$ & \multicolumn{3}{c|}{$\sin(\varphi_S-\varphi)$} & \multicolumn{2}{c|}{$\sin(\varphi_S-2\varphi)$}  & $\sin(\varphi_S-3\varphi)$ \rule[-0.20cm]{0mm}{0.6cm} \\ %\hline
twist-2 & 0 & 0 & $- \mathcal{C} [ \bar w_1 {\cal H}_{1T}^{\perp} \bar H_1^\perp ]$ & $ \mathcal{C} [ w_1 D_{1T}^\perp \bar D_1 ]$ 
    & 0 & $\mathcal{C} [ w_1 D_{1T}^\perp \bar D_1 ]$ & 0 & 0 & $- \mathcal{C} [ w_{hh}^t H_{1T}^\perp \bar H_1^\perp ]$ \rule[-0.20cm]{0mm}{0.6cm}  \\ \hline%\hline
%%%%%%%
T-PV & $\tilde F_{1T}^{\cos\varphi_S}$ & $\tilde F_{2T}^{\cos\varphi_S}$ &$\tilde F_{T}^{\cos(\varphi_S+\varphi)}$ & $\tilde F_{1T}^{\cos(\varphi_S-\varphi)}/2c_1^ec_3^q$ &$\tilde F_{2T}^{\cos(\varphi_S-\varphi)}$ &$\tilde F_{3T}^{\cos(\varphi_S-\varphi)}/4c_3^ec_1^q$ & 
$\tilde F_{1T}^{\cos(\varphi_S-2\varphi)}$ &$\tilde F_{2T}^{\cos(\varphi_S-2\varphi)}$ &$\tilde F_{T}^{\cos(\varphi_S-3\varphi)}$ \rule[-0.20cm]{0mm}{0.6cm} \\ %\hline
$\varphi$-dep. & \multicolumn{2}{c|} {$\cos\varphi_S$} & $\cos(\varphi_S+\varphi)$ & \multicolumn{3}{c|}{$\cos(\varphi_S-\varphi)$} & \multicolumn{2}{c|}{$\cos(\varphi_S-2\varphi)$}  & $\cos(\varphi_S-3\varphi)$ \rule[-0.20cm]{0mm}{0.6cm} \\ %\hline
%$\theta$-dep. & $\sin\theta$ & $\sin2\theta$ & $\sin^2\theta$ & $1+\cos^2\theta$& $\sin^2\theta$ & $\cos\theta$ & $\sin\theta$ & $\sin2\theta$ & $sin^2\theta$ \rule[-0.20cm]{0mm}{0.6cm} \\ \hline
%%
twist-2 & 0 & 0 & 0 & $\mathcal{C} [ w_1 G_{1T}^\perp \bar D_1 ] $ & 0 & $ \mathcal{C} [ w_1 G_{1T}^\perp \bar D_1 ]$ & 0 & 0 &0 \rule[-0.20cm]{0mm}{0.6cm} \\ \hline%\hline
%\end{tabular}
%\end{table} 
%
%\begin{table}[!ht]
%\caption{The 9 structure functions $w_1 H_{1LT}^{\perp\prime}=\frac{1}{2} w_0 \bar w_1 H_{1LT}^\perp - w_1 H_{1LT}$, $w_{HH}=2w_1^2 \bar w_1 - w_1 w_2 - \frac{1}{2}w_0 \bar w_1$}\label{tab:FFunpol}
%\begin{tabular}{c|cc|c|ccc|cc|c}\hline
LT-PC & $F_{1LT}^{\cos\varphi_{LT}}$ & $F_{2LT}^{\cos\varphi_{LT}}$ &$F_{LT}^{\cos(\varphi_{LT}+\varphi)}/{8c_1^e c_2^q} $ & $F_{1LT}^{\cos(\varphi_{LT}-\varphi)}/2c_1^ec_1^q$ &$F_{2LT}^{\cos(\varphi_{LT}-\varphi)}$ &$F_{3LT}^{\cos(\varphi_{LT}-\varphi)}/4c_3^ec_3^q$ & 
$F_{1LT}^{\cos(\varphi_{LT}-2\varphi)}$ &$F_{2LT}^{\cos(\varphi_{LT}-2\varphi)}$ &$F_{LT}^{\cos(\varphi_{LT}-3\varphi)}/8c_1^ec_2^q$ \rule[-0.20cm]{0mm}{0.6cm} \\ %\hline
$\varphi$-dep. & \multicolumn{2}{c|} {$\cos\varphi_{LT}$} & $\cos(\varphi_{LT}+\varphi)$ & \multicolumn{3}{c|}{$\cos(\varphi_{LT}-\varphi)$} & \multicolumn{2}{c|}{$\cos(\varphi_{LT}-2\varphi)$}  & $\cos(\varphi_{LT}-3\varphi)$ \rule[-0.20cm]{0mm}{0.6cm} \\ %\hline
%$\theta$-dep. & $\sin\theta$ & $\sin2\theta$ & $\sin^2\theta$ & $1+\cos^2\theta$& $\sin^2\theta$ & $\cos\theta$ & $\sin\theta$ & $\sin2\theta$ & $sin^2\theta$ \rule[-0.20cm]{0mm}{0.6cm} \\ \hline
%%
twist-2 & 0 & 0 & $-\mathcal{C} [ \bar w_1 {\cal H}_{1LT}^{\perp} \bar H_1^\perp ]$ & $-\mathcal{C} [ w_1 D_{1LT}^\perp \bar D_1 ]$ & 0 
& $-\mathcal{C} [ w_1 D_{1LT}^\perp \bar D_1 ]$ & 0 & 0 & $ \mathcal{C} [ w_{hh}^{t}  H_{1LT}^\perp \bar H_1^\perp ]$  \rule[-0.20cm]{0mm}{0.6cm} \\ \hline%\hline
%%%%%%%%
LT-PV & $\tilde F_{1LT}^{\sin\varphi_{LT}}$ & $\tilde F_{2LT}^{\sin\varphi_{LT}}$ &$\tilde F_{LT}^{\sin(\varphi_{LT}+\varphi)}$ & $\tilde F_{1LT}^{\sin(\varphi_{LT}-\varphi)}/2c_1^ec_3^q$ &$\tilde F_{2LT}^{\sin(\varphi_{LT}-\varphi)}$ &$\tilde F_{3LT}^{\sin(\varphi_{LT}-\varphi)}/4c_3^ec_1^q$ & 
$\tilde F_{1LT}^{\sin(\varphi_{LT}-2\varphi)}$ &$\tilde F_{2LT}^{\sin(\varphi_{LT}-2\varphi)}$ &$\tilde F_{LT}^{\sin(\varphi_{LT}-3\varphi)}$ \rule[-0.20cm]{0mm}{0.6cm} \\ %\hline
$\varphi$-dep. & \multicolumn{2}{c|} {$\sin\varphi_{LT}$} & $\sin(\varphi_{LT}+\varphi)$ & \multicolumn{3}{c|}{$\sin(\varphi_{LT}-\varphi)$} & \multicolumn{2}{c|}{$\sin(\varphi_{LT}-2\varphi)$}  & $\sin(\varphi_{LT}-3\varphi)$ \rule[-0.20cm]{0mm}{0.6cm} \\ %\hline
%$\theta$-dep. & $\sin\theta$ & $\sin2\theta$ & $\sin^2\theta$ & $1+\cos^2\theta$& $\sin^2\theta$ & $\cos\theta$ & $\sin\theta$ & $\sin2\theta$ & $sin^2\theta$ \rule[-0.20cm]{0mm}{0.6cm} \\ \hline
%%
twist-2 & 0 & 0 & 0 & $- \mathcal{C} [ w_1 G_{1LT}^\perp \bar D_1 ] $ & 0 & $-\mathcal{C} [ w_1 G_{1LT}^\perp \bar D_1 ]$ & 0 & 0 &0 \rule[-0.20cm]{0mm}{0.6cm} \\ \hline %\hline
%%%%%%%
%\end{tabular}
%\end{table} 
%%%%%%%%%%%%%%
%%%%%%%
%\begin{table}[!ht]
%\caption{The 9 structure functions for TT-dependent cases $w_1 H_{1T}^{\perp\prime}=\frac{1}{2} w_0 \bar w_1 H_{1T}^\perp - w_1 H_{1T}$, $w_{HH}=2w_1^2 \bar w_1 - w_1 w_2 - \frac{1}{2}w_0 \bar w_1$}\label{tab:FFttpol}
%\begin{tabular}{c|cc|c|ccc|cc|c}\hline
TT-PC & $F_{1TT}^{\cos(2\varphi_{TT}-\varphi)}$ & $F_{2TT}^{\cos(2\varphi_{TT}-\varphi)}$ & $F_{TT}^{\cos2\varphi_{TT}}/{8c_1^e c_2^q} $ &
  $F_{1TT}^{\cos(2\varphi_{TT}-2\varphi)}/2c_1^ec_1^q$ & $F_{2TT}^{\cos(2\varphi_{TT}-2\varphi)}$ & $F_{3TT}^{\cos(2\varphi_{TT}-2\varphi)}/4c_3^ec_3^q$ &
  $F_{1TT}^{\cos(2\varphi_{TT}-3\varphi)}$ & $F_{2TT}^{\cos(2\varphi_{TT}-3\varphi)}$ & $F_{TT}^{\cos(2\varphi_{TT}-4\varphi)}/4c_1^ec_2^q$ \rule[-0.20cm]{0mm}{0.6cm} \\ %\hline
$\varphi$-dep. & \multicolumn{2}{c|} {$\cos(2\varphi_{TT}-\varphi)$} & $\cos2\varphi_{TT}$ & \multicolumn{3}{c|}{$\cos(2\varphi_{TT}-2\varphi)$} & \multicolumn{2}{c|}{$\cos(2\varphi_{TT}-3\varphi)$}  & $\cos(2\varphi_{TT}-4\varphi)$ \rule[-0.20cm]{0mm}{0.6cm} \\ %\hline
%$\theta$-dep. & $\sin\theta$ & $\sin2\theta$ & $\sin^2\theta$ & $1+\cos^2\theta$& $\sin^2\theta$ & $\cos\theta$ & $\sin\theta$ & $\sin2\theta$ & $sin^2\theta$ \rule[-0.20cm]{0mm}{0.6cm} \\ \hline
%%
twist-2 & 0 & 0 & $\mathcal{C} [ w_2 {\cal H}_{1TT}^{\perp} \bar H_1^\perp ]$ & $ \mathcal{C} [ w_{dd}^{tt}  D_{1TT}^\perp \bar D_1 ] $ & 0 
& $\mathcal{C} [ w_{dd}^{tt}  D_{1TT}^\perp \bar D_1 ]$ & 0 & 0 & $-\mathcal{C} [ w_{hh}^{tt}  H_{1TT}^\perp \bar H_1^\perp ]$  \rule[-0.20cm]{0mm}{0.6cm} \\ \hline%\hline
%%%%%%%%%%%%
TT-PV & $\tilde F_{1TT}^{\sin(2\varphi_{TT}-\varphi)}$ & $\tilde F_{2TT}^{\sin(2\varphi_{TT}-\varphi)}$ &$\tilde F_{TT}^{\sin2\varphi_{TT}}$ 
  & $\tilde F_{1TT}^{\sin(2\varphi_{TT}-2\varphi)}/2c_1^ec_3^q$ &$\tilde F_{2TT}^{\sin(2\varphi_{TT}-2\varphi)}$ &$\tilde F_{3TT}^{\sin(2\varphi_{TT}-2\varphi)}/4c_3^ec_1^q$ & 
     $\tilde F_{TT}^{\sin(2\varphi_{TT}-3\varphi)}$ &$\tilde F_{2TT}^{\sin(2\varphi_{TT}-3\varphi)}$ & $\tilde F_{TT}^{\sin(2\varphi_{TT}-4\varphi)}$ \rule[-0.20cm]{0mm}{0.6cm} \\ %\hline
$\varphi$-dep. & \multicolumn{2}{c|} {$\sin(2\varphi_{TT}-\varphi)$} & $\sin2\varphi_{TT}$ & \multicolumn{3}{c|}{$\sin(2\varphi_{TT}-2\varphi)$} & \multicolumn{2}{c|}{$\sin(2\varphi_{TT}-3\varphi)$}  & $\sin(2\varphi_{TT}-4\varphi)$ \rule[-0.20cm]{0mm}{0.6cm} \\ %\hline
%$\theta$-dep. & $\sin\theta$ & $\sin2\theta$ & $\sin^2\theta$ & $1+\cos^2\theta$& $\sin^2\theta$ & $\cos\theta$ & $\sin\theta$ & $\sin2\theta$ & $sin^2\theta$ \rule[-0.20cm]{0mm}{0.6cm} \\ \hline
twist-2 & 0 & 0 & 0 & $\mathcal{C} [w_{dd}^{tt} G_{1TT}^\perp \bar D_1 ] $ & 0 & $\mathcal{C} [ w_{dd}^{tt}  G_{1TT}^\perp \bar D_1 ]$ & 0 & 0 & 0 \rule[-0.20cm]{0mm}{0.6cm} \\ \hline%\hline
%%%%%%%%%%%% \\ \hline\hline
\end{tabular}
\end{table} 
%\end{sidewaystable}
%\begin{landscape}
%\end{landscape}

\end{widetext}

\subsection{Discussion about the twist-2 results}

As we mentioned earlier in this paper, the twist-2 results presented here just correspond to the results obtained 
from the intuitive parton model with FFs defined in the gauge invariant form. 
Just as for the structure functions in inclusive deep-inelastic lepton-nucleon scattering (DIS) obtained using the original intuitive parton model, 
at the LO in pQCD and twist-2, the results exhibit a number of simple regularities (symmetries) such as Callan-Gross relation. 
To see these regularities more clearly, we list the leading twist results in Table~\ref{tab:StructureFunctions}.

Indeed, from these results, we see that although there are 81 independent structure functions, a large part of them vanish at twist-2. 
Totally 27 of them are non-zero, among them 19 are parity conserved and 8 are parity violated. 
Furthermore we see following regularities. 

(1) Among the 27 non-zero structure functions, 5 with $c_1^e c_1^q$, 5 with $c_3^e c_3^q$  and 9 with $c_1^e c_2^q$ are parity even, 
4 with $c_1^e c_3^q$ and 4 with $c_3^e c_1^q$ are parity odd. 
This can be understood easily since from Eq.~(\ref{eq:leptonictensor}) we see that $c_1^e$ symbolizes the symmetric parity conserving part and $c_3^e$ 
the anti-symmetric parity violating part of the tensor. 

(2) The non-vanishing structure functions are associated with either $1+\cos^2\theta$, or $\cos\theta$ or $\sin^2\theta$. 

For those associated with $1+\cos^2\theta$ or $\cos\theta$, 5 with coefficient $c_1^e c_1^q$ and 5 with $c_3^e c_3^q$. 
They are all from ${\cal C}[D\bar D]$, i.e. fragmentations of unpolarized quark and are parity conserving. 
There are also 4 with coefficient $c_1^e c_3^q$ and 4 with $c_3^e c_1^q$. 
They are all from ${\cal C}[G\bar D]$, i.e. fragmentations of longitudinally polarized quark and unpolarized anti-quark and are parity violating. 

Those associated with $\sin^2\theta$ all have coefficient $c_1^e c_2^q$ and are from  ${\cal C}[H\bar H]$, i.e. transversely polarized quark and anti-quark. 

To understand such regularities, we recall the result for the basic weak process $e^+e^-\to Z\to q\bar q$. 
We recall that the differential cross section is~\cite{ff:theta}, 
\begin{align}
\frac{d\hat\sigma}{d\Omega}=\frac{\alpha^2}{4s}\chi\bigl[ c_1^ec_1^q(1+\cos^2\theta)+2c_3^ec_3^q\cos\theta\bigr],
\end{align}  
and the produced quark (anti-quark) is longitudinally polarized and the polarization is given by, 
\begin{align}
P_q(\theta)=-\frac{c_1^ec_3^q(1+\cos^2\theta)+2c_3^ec_1^q\cos\theta}{c_1^ec_1^q(1+\cos^2\theta)+2c_3^ec_3^q\cos\theta}.
\end{align} 
Furthermore, although the quark (anti-quark) is not transversely polarized,  their transverse spin components are correlated. 
We define, 
 \begin{align}
 c_{nn}^q=\frac{|M_{n++}|^2+|M_{n--}|^2-|M_{n+-}|^2-|M_{n-+}|^2}{|M_{n++}|^2+|M_{n--}|^2+|M_{n+-}|^2+|M_{n-+}|^2},
 \end{align}
where $M$ is the scattering amplitude, $+$ or $-$ denotes that the quark or anti-quark is in $s_n=1/2$ or $-1/2$ state. 
We obtain that, for $\vec n$ is in the normal of the production plane, 
  \begin{align}
 c_{nn}^q(\theta)=\frac{c_1^ec_2^q\sin^2\theta}{c_1^ec_1^q(1+\cos^2\theta)+2c_3^ec_3^q\cos\theta},  \label{eq:cnnq}
 \end{align}
 which is in fact also true for any transverse direction $\vec n$ if we replace $\sin^2\theta$ in the numerator by $\sin^2\theta\cos2\varphi_n$ 
 where $\varphi_n$ is the azimuthal angle between $\vec n$ and the normal of the production plane.
In terms of $y=(1+\cos\theta)/2$, we have, 
\begin{align}
&\frac{d\hat\sigma}{d\Omega} = \frac{\alpha^2}{2s}\chi T_0^q(y) , \\
&P_q(y) = T_1^q(y)/T_0^q(y),\\
& c_{nn}^q(y) = c_1^ec_2^qC(y)/2T_0^q(y),
 \end{align}
where $T_0^q(y) =  c_1^e c_1^q A(y)  - c_3^e c_3^q B(y)$ is the relative production weight for flavor $q$, 
$T_1^q(y) = -c_1^e c_3^q A(y) + c_3^e c_1^q B(y)$; 
$A(y)$, $B(y)$ and $C(y)$ are given in Sec.~\ref{Subsec:FFs_HGJ} by Eqs.~(\ref{eq:A(y)}-\ref{eq:C(y)}). 
We see clearly why we have the regularities  for the structure functions mentioned at the beginning of this point. 

(3) It is also clear that if we consider $e^+e^-\to\gamma^*\to q\bar q$, i.e. the electromagnetic process, 
we have, $T_0^{q(em)}(y) = e_q^2 A(y)$, $P_q^{(em)}=0$. 
The quark transverse spin correlation $c_{nn}^{q(em)}(y) = C(y)/2A(y)$ independent of flavor of the quark. 
In this case, we will not have ${\cal C}[G\bar D]$-terms but ${\cal C}[D\bar D]$ and ${\cal C}[H\bar H]$-terms.
 
(4) If we integrate over $p_2$, we obtain the results for the inclusive process $e^+e^-\to Z\to VX$. 
The non-vanishing structure functions are,  
\begin{align}
& z_1 F_{1U1,in} = {2c_1^e c_1^q} D_1(z_1), \\
& z_1 F_{3U1,in} = {4c_3^e c_3^q}  D_1(z_1) , \\
& z_1  \tilde F_{1L1,in} = -{2c_1^e c_3^q}  G_{1L}(z_1), \\
& z_1 \tilde F_{3L1,in} = -{4c_3^e c_1^q} G_{1L}(z_1), \\
& z_1 F_{1LL1,in} = {2c_1^e c_1^q}  D_{1LL}(z_1) , \\
& z_1 F_{3LL1,in} = {4c_3^e c_3^q}  D_{1LL}(z_1) .
\end{align}
All the others vanish at twist-2. This is consistent with the results obtained in \cite{Wei:2013csa}. 
We emphasize in particular that Callan-Gross relation in DIS now is replaced by $F_{2U1,in}=0$,  
and all the structure functions associated with the transverse spin components vanish at leading twist.

\subsection{Twist-3 contributions}\label{subsec:twist3FFs}

Among the 54 structure functions that vanish at twist-2, 36 have twist-3 contributions as the leading power contributions. 
The results are a bit lengthy so we present them as an appendix (see appendix \ref{sec:twist3FFs}). 
We see that all the 36 structure functions associated with $\sin\theta$ and $\sin2\theta$ have twist-3 contributions as leading power contributions.   
Besides others, we have $F_{1U2}^{\cos\varphi}$, $F_{2U2}^{\cos\varphi}$, $\tilde F_{1U2}^{\sin\varphi}$ and $\tilde F_{2U2}^{\sin\varphi}$ in the unpolarized part, 
also $F_{1T2}^{\sin\varphi_S}$, $F_{2T2}^{\sin\varphi_S}$, $\tilde F_{1T2}^{\cos\varphi_S}$ and $\tilde F_{2T2}^{\cos\varphi_S}$ in the vector polarization dependent part. 
This means that at the twist-3 level there should be parity conserved azimuthal asymmetry $\langle \cos\varphi \rangle_U$ and parity violated asymmetry $\langle \sin\varphi \rangle_U$ 
in the unpolarized case and parity conserved transverse polarization in the normal direction of the lepton-hadron plane and parity violated component in the plane. 
We will discuss this more in next section.

\section{Azimuthal asymmetries and hadron polarizations}\label{sec:Tpol}

\subsection{Azimuthal asymmetries}

At leading twist and for unpolarized $V$ (i.e. polarization is not measured), there is only one azimuthal asymmetry as given by Eq.~(\ref{eq:Acos2}), i.e.,   
\begin{align}
\langle \cos2\varphi\rangle_{U}^{(0)} = - \frac{ C(y) \sum_q c_1^e c_2^q 
\mathcal{C} [ w_{hh} H_1^\perp \bar H_1^\perp ] }{ \sum_q T_0^q(y) \mathcal{C} [ D_1 \bar D_1 ] }.
\end{align}
This is the only leading twist azimuthal asymmetry in the unpolarized case due to Collins effect~\cite{Collins:1992kk} and transverse spin correlation $c_{nn}^q$ 
given by Eq.~(\ref{eq:cnnq}) for $q\bar q$ produced via $e^+e^-$ annihilation.  
Here, as well as in the following of this paper, when writing the expressions for azimuthal asymmetries and/or polarizations in terms of FFs, 
to avoid confusion, we include the summation over $q$ explicitly but  
still keep the $q\leftrightarrow \bar q$ terms implicitly and omit the flavor indices for the FFs. 

If we could consider the polarization and azimuthal asymmetry simultaneously, we would have, 
\begin{align}
&\langle \cos2\varphi\rangle_{LL}^{(0)} = - \frac{ C(y) \sum_q c_1^e c_2^q 
\mathcal{C} [ w_{hh} (H_1^\perp + S_{LL} H_{1LL}^\perp) \bar H_1^\perp ] }{ \sum_q T_0^q(y) \mathcal{C} [ ( D_1 + S_{LL} D_{1LL} ) \bar D_1 ] },\\
 &\langle \sin2\varphi\rangle_{L}^{(0)} = - \frac{ \lambda C(y) \sum_q c_1^e c_2^q  \mathcal{C} [ w_{hh} H_{1L}^\perp \bar H_1^\perp ] }
 { \sum_q T_0^q(y) \mathcal{C}( D_1 -\lambda G_{1L} ) \bar D_1 }.
 \end{align}
Although it is academic since it will be very difficult to measure this asymmetry, it is interesting to see the existence of such asymmetry. 

Up to twist-3, we have another two azimuthal asymmetries in the unpolarized case, i.e., 
\begin{align}
&\langle\cos\varphi\rangle_U^{(1)}=  -\frac{8D(y)}{ z_1 z_2 Q  F_{Ut}^{(0)}} \nonumber\\ 
&~~\times \sum_q \biggl\{T_2^q(y) \Bigl(M_1 \mathcal{C} [ w_1 D^\perp z_2 \bar D_1]  + M_2  \mathcal{C} [ \bar w_1 z_1 D_1 \bar D^{\perp\prime}] \Bigr) \nonumber\\
&~~~~~ + T_4^q(y) \Bigl( M_1 \mathcal{C}[ \bar w_1 H z_2 \bar H_1^\perp] + M_2  \mathcal{C}[ w_1 z_1 H_1^\perp  \bar H^{\perp\prime}] \Bigr) \biggr\},\\ 
%%%%%%%%
& \langle\sin\varphi\rangle_U^{(1)}=  \frac{8D(y)}{ z_1 z_2 Q  F_{Ut}^{(0)}} \nonumber\\
&~~\times \sum_q \biggl\{ T_3^q(y) \Bigl( M_1 \mathcal{C} [ w_1 G^\perp z_2 \bar D_1] - M_2 \mathcal{C} [ \bar w_1 z_1 D_1 \bar G^\perp]\Bigr) \nonumber\\
&~~~~~ + 2 c_3^e c_2^q \Bigl( M_1 \mathcal{C} [ \bar w_1 E z_2 \bar H_1^\perp] - M_2 \mathcal{C} [ w_1 z_1 H_1^\perp \bar E] \Bigr)  \biggr\},
\end{align}
where $D(y) = \sqrt{y(1-y)}$, $T_2^q(y) = -c_3^e c_3^q  + c_1^e c_1^q B(y)$, $T_3^q(y) = c_3^e c_1^q  - c_1^e c_3^q B(y)$, $T_4^q(y)=4 c_1^e c_2^q B(y)$ 
and  $F_{Ut}^{(0)}$ is the twist-2 contribution to $F_{Ut}$ and is given by, 
\begin{align}
F_{Ut}^{(0)}= {4} \sum_q T_0^q(y) \mathcal{C} [ D_1 \bar D_1 ]. 
\end{align}
We see that they depend on several twist-3 FFs. 

If we consider $e^+e^-\to\gamma^*\to V\pi X$, we have, 
\begin{align}
&\langle \cos2\varphi\rangle_{U}^{(0,em)}= - \frac{ C(y)}{A(y)} \frac{\sum_q e_q^2 \mathcal{C} [w_{hh} H_1^\perp \bar H_1^\perp ] }{\sum_q e_q^2 \mathcal{C} [ D_1 \bar D_1 ] },\\
%\end{align}
%\begin{align}
&\langle\cos\varphi\rangle_U^{(1,em)} =  -\frac{2\tilde B(y)} {A(y)}\frac{1}{z_1 z_2 Q{\sum_q e_q^2 \mathcal{C} [ D_1 \bar D_1 ] }} \nonumber\\ 
&\phantom{XXXXXX} \times \sum_q e_q^2 
\Bigl\{ M_1 \mathcal{C} [ w_1 D^\perp z_2 \bar D_1+ 4 \bar w_1 H z_2 \bar H_1^\perp]  \nonumber\\
&\phantom{XXXXXXXX}+  M_2  \mathcal{C}[ \bar w_1 z_1 D_1 \bar D^{\perp\prime} + 4w_1 z_1 H_1^\perp  \bar H^{\perp\prime}]  \Bigr\}   ,
%%%%%%%%
\end{align}
and $\langle\sin\varphi\rangle_U^{(1,em)}  = 0$, where $\tilde B(y)=\sqrt{y(1-y)}B(y)$.
In this case we have a non-zero azimuthal asymmetry $\langle \cos2\varphi\rangle_{U}^{(0,em)}$ 
at leading twist due to Collins effect~\cite{Collins:1992kk} and a twist-3 asymmetry $\langle \cos\varphi\rangle_{U}^{(0,em)}$ 
similar to Cahn effect~\cite{Cahn:1978se} in deep-inelastic lepton-nucleon scattering.

\subsection{Hadron polarizations at twist-2}

The polarization is in general dependent on $\varphi$. 
Experimentally it is much easier to consider the case where $\varphi$ is integrated. 
In this case,  at the leading twist, we have, for the longitudinal polarization, 
\begin{align}
&\langle \lambda\rangle^{(0)} %=& \frac{2}{3F_{Ut}^{(0)}} \frac{4}{z_1 z_2} \sum_q T_1^q(y) \mathcal{C}[G_{1L} \bar D_1] \nonumber\\
%=& \frac{2}{3F_{Ut}^{(0)}} \frac{4}{z_1 z_2} \sum_q P_q(y)T_0^q(y) \mathcal{C}[G_{1L} \bar D_1] \nonumber \\
= \frac{2}{3} \frac{\sum_q P_q(y)T_0^q(y) \mathcal{C}[G_{1L} \bar D_1]}{\sum_q T_0^q(y) \mathcal{C}[D_1 \bar D_1]}, \label{eq:ALambdat2}\\
%%%
&\langle S_{LL}\rangle^{(0)} %=& \frac{1}{2F_{Ut}^{(0)}} \frac{4}{z_1 z_2} \sum_q T_0^q(y) \mathcal{C} [D_{1LL} \bar D_1] \nonumber\\
= \frac{1}{2} \frac{\sum_q T_0^q(y) \mathcal{C} [ D_{1LL} \bar D_1 ]}{\sum_q T_0^q(y) \mathcal{C}[D_1 \bar D_1]}. \label{eq:ASllt2}
\end{align}

For transverse dependent components w.r.t. the hadron-hadron plane,  we have, 
\begin{align}
&\langle S_T^{n}\rangle^{(0)} = \frac{2}{3} \frac{\sum_q T_0^q(y) \mathcal{C} [ w_1 D_{1T}^\perp \bar D_1 ]}{\sum_q T_0^q(y) \mathcal{C}[D_1 \bar D_1]}, \label{eq:AStnt2}\\
&\langle S_T^{t}\rangle^{(0)} = -\frac{2}{3} \frac{ \sum_q P_q(y) T_0^q(y)  \mathcal{C} [ w_1 G_{1T}^\perp \bar D_1 ]}{\sum_q T_0^q(y) \mathcal{C}[D_1 \bar D_1]},\label{eq:ASttt2}\\
%%%
&\langle S_{LT}^{n}\rangle^{(0)} = \frac{2}{3} \frac{\sum_q P_q(y) T_0^q(y)  \mathcal{C} [ w_1 G_{1LT}^\perp \bar D_1 ]}{\sum_q T_0^q(y) \mathcal{C}[D_1 \bar D_1]}, \label{eq:ASltnt2}\\
&\langle S_{LT}^{t}\rangle^{(0)} = -\frac{2}{3}\frac{\sum_q T_0^q(y)  \mathcal{C} [ w_1 D_{1LT}^\perp \bar D_1 ]}{\sum_q T_0^q(y) \mathcal{C}[D_1 \bar D_1]},\label{eq:ASlttt2}\\
%%%
&\langle S_{TT}^{nn}\rangle^{(0)} = -\frac{2}{3}\frac{\sum_q T_0^q(y)  \mathcal{C} [ w_{dd}^{tt} D_{1TT}^\perp \bar D_1 ]}{\sum_q T_0^q(y) \mathcal{C}[D_1 \bar D_1]}, \label{eq:ASttnnt2}\\
&\langle S_{TT}^{nt}\rangle^{(0)} = -\frac{2}{3}\frac{\sum_q P_q(y) T_0^q(y)  \mathcal{C} [ w_{dd}^{tt} G_{1TT}^\perp \bar D_1 ]}{\sum_q T_0^q(y) \mathcal{C}[D_1 \bar D_1]}.\label{eq:ASttntt2}
\end{align}
The transverse components w.r.t. the lepton-hadron plane are zero at the leading twist in $\varphi$ integrated case. 

If we consider $e^+e^-\to\gamma^*\to V\pi X$, i.e. annihilate via electromagnetic interaction only, we have, 
\begin{align}
&\langle S_{LL}\rangle^{(0,em)}=\frac{1}{2} \frac{ \sum_q  e_q^2 \mathcal{C} [ D_{1LL} \bar D_1 ] }{\sum_q  e_q^2 \mathcal{C} [ D_{1} \bar D_1 ]},\\
%\end{align}
%\begin{align}
&\langle S_T^{n}\rangle^{(0,em)}= \frac{2}{3}\frac{\sum_q  e_q^2 \mathcal{C} [ w_1 D_{1T}^\perp \bar D_1 ]}{\sum_q  e_q^2 \mathcal{C} [ D_{1} \bar D_1 ]}, \\
&\langle S_{LT}^{t}\rangle^{(0,em)}= \frac{2}{3}\frac{\sum_q  e_q^2 \mathcal{C} [ w_1 D_{1LT}^\perp \bar D_1 ]}{\sum_q  e_q^2 \mathcal{C} [ D_{1} \bar D_1 ]} , \\
&\langle S_{TT}^{nn}\rangle^{(0,em)}=-\frac{2}{3}\frac{\sum_q  e_q^2  \mathcal{C} [ w_{dd}^{tt} D_{1TT}^\perp \bar D_1 ]}{3\sum_q  e_q^2 \mathcal{C} [ D_{1} \bar D_1 ]}, 
\end{align}
while the parity violating components, 
\begin{align}
&\langle \lambda \rangle^{(0,em)}=\langle S_T^{t} \rangle^{(0,em)}=\langle S_{LT}^{n}\rangle^{(0,em)}= \langle S_{TT}^{nt}\rangle^{(0,em)}=0.
\end{align}
We see in particular that the $S_{LL}$-component is non-zero at leading twist also in the parity conserved case. 
Parity conserving transverse components exist due to Sivers-type FFs such as $D_{1T}^\perp$, $D_{1LT}^\perp$ and $D_{1TT}^\perp$ 
similar to the Sivers function $f_{1T}^\perp$ in three dimensional PDFs~\cite{Sivers:1989cc}. 

For the inclusive process $e^+e^-\to Z\to VX$, we have, 
\begin{align}
&\langle \lambda\rangle^{(0)}_{in}= \sum_q 2 P_q(y) T_0^q(y) G_{1L}(z_1)/\sum_q 3T_0^q(y) D_{1}(z_1), \\
&\langle S_{LL}\rangle^{(0)}_{in}= \sum_q T_0^q(y) D_{1LL}(z_1)/\sum_q 2T_0^q(y) D_{1}(z_1),
\end{align}
while all the transverse components such as $\langle S_T^{i}\rangle^{(0)}_{in}$, 
$\langle S_{LT}^{i}\rangle^{(0)}_{in}$ and $\langle S_{TT}^{ij}\rangle^{(0)}_{in}$ ($i,j=x$ or $y$) vanish at twist-2. 
We also see that $\langle S_{LL}\rangle^{(0)}_{in}$ is non-zero also in parity conserved reactions while $\langle\lambda\rangle^{(0)}_{in}$ exists only in parity violated case.

\subsection{Transverse polarizations with respect to the lepton-hadron plane at twist-3}

As mentioned in Sec.~\ref{subsec:twist3FFs}, twist-3 contribution exists only for those structure functions that are zero at twist-2. 
They are the leading power contributions for the corresponding structure functions.  
In particular we see that there is no twist-3 contribution to the transverse components w.r.t. the hadron-hadron plane discussed in last subsection.
However, for the transverse components w.r.t. the lepton-hadron plane, 4 of them,
i.e. $\langle S_T^x\rangle$, $\langle S_T^y\rangle$, $\langle S_{LT}^x\rangle$ and $\langle S_{LT}^y\rangle$ have twist-3 contributions. 
They are determined by $F_{jT2}^{\sin\varphi_S}$, $\tilde F_{jT2}^{\cos\varphi_S}$,  $\tilde F_{jLT2}^{\sin\varphi_S}$ and $F_{jLT2}^{\cos\varphi_S}$ 
given in appendix \ref{sec:twist3FFs} respectively. 
The expressions can easily be obtained by inserting these results into Eqs.~(\ref{eq:AStx}-\ref{eq:ASlty}) but are a bit lengthy so we omit them here. 
However we emphasize that if we consider $e^+e^-\to\gamma^*\to V\pi X$, the parity violating parts vanish 
and we have only the following two components,   
\begin{align}
&\langle S_T^{y}\rangle^{(1,em)} = \frac{8M_1 \tilde B(y)}{3z_1z_2Q A(y) \sum_q e_q^2 \mathcal{C} [D_1\bar D_1]} \nonumber\\
&~~~\times \sum_qe_q^2 \Bigl\{ z_2\mathcal{C} [ {\cal D}^\perp_T \bar D_1 - 2\frac{w_2}{M_1} H_T^{\perp-} \bar H_1^\perp] \nonumber\\
&~~~ - \frac{z_1M_2}{2M_1} \mathcal{C} [ {w_2}  (D_{1T}^\perp \bar D^{\perp\prime} - G_{1T}^\perp \bar G^\perp)  - 8{\mathcal H}_{1T}^{\perp} \bar H_1^{\perp\prime} ] \Bigr\}, \\
%%%%
&\langle S_{LT}^{x}\rangle^{(1,em)} = -\frac{8M_1\tilde B(y)}{3z_1z_2QA(y) \sum_q e_q^2 \mathcal{C} [D_1\bar D_1]}  \nonumber\\
&~~~\times\sum_qe_q^2 \Bigl\{ z_2\mathcal{C} [ {\cal D}^\perp_{LT} \bar D_1 - 2 \frac{w_2}{M_1} H_{LT}^{\perp+} \bar H_1^\perp ] \nonumber\\
&~~~+ \frac{z_1M_2}{2M_1} \mathcal{C} [ {w_2} (D_{1LT}^\perp \bar D^{\perp\prime} + G_{1LT}^\perp \bar G^\perp )
-8 {\mathcal H}_{1LT}^\perp \bar H_1^{\perp\prime} ] \Bigr\}.
\end{align}

It is also interesting to see that these transverse components are defined w.r.t. the lepton-hadron plane and exist also in the inclusive process. 
For $e^+e^-\to Z\to VX$, we have, 
\begin{align}
&\langle S_T^{x}\rangle^{(1)}_{in}=-\frac{8M_1D(y)}{3z_1Q} \frac{\sum_q T_3^q(y) G_T}{\sum_q T_0^q(y) D_1}, \label{eq:AStxt3}\\
&\langle S_T^{y}\rangle^{(1)}_{in}=\frac{8M_1D(y)}{3z_1Q} \frac{\sum_q T_2^q(y) D_T}{\sum_q T_0^q(y) D_1}, \label{eq:AStyt3}\\
&\langle S_{LT}^{x}\rangle^{(1)}_{in}=-\frac{8M_1D(y)}{3z_1Q} \frac{\sum_q T_2^q(y) D_{LT}}{\sum_q T_0^q(y) D_1}, \label{eq:ASltxt3}\\
&\langle S_{LT}^{y}\rangle^{(1)}_{in}=\frac{8M_1D(y)}{3z_1Q} \frac{\sum_q T_3^q(y) G_{LT}}{\sum_q T_0^q(y) D_1}. \label{eq:ASltyt3}
\end{align}
We recall that $\langle S_T^{y}\rangle$ is $P$-even and naive $T$-odd, $\langle S_T^{x}\rangle$ is $P$-odd and naive $T$-even 
and $\langle S_{LT}^{y}\rangle$ is $P$-odd and naive $T$-odd. 
Neither of these three can exist in deep-inelastic scattering such as $e^-N\to e^-X$.  
The only existing one is $\langle S_{LT}^{x}\rangle$ which is both $P$- and $T$-even. 
We see also from table~\ref{tab:TMDFFChiralTime} whether the corresponding FFs are $T$-odd or $T$-even which is 
consistent with the structure functions and/or the polarizations.  

For $e^+e^-\to\gamma^*\to V X$, we have, 
\begin{align}
&\langle S_T^{y}\rangle^{(1,em)}_{in} = \frac{8M_1\tilde B(y)}{3z_1QA(y)} \frac{\sum_q e_q^2 D_T}{\sum_q e_q^2 D_1}, \\
&\langle S_{LT}^{x}\rangle^{(1,em)}_{in} = -\frac{8M_1\tilde B(y)}{3z_1QA(y)} \frac{\sum_q e_q^2 D_{LT}}{\sum_q e_q^2 D_1},
\end{align}
and other two parity violating components are zero.

\section{Summary and discussion} \label{sec:Summary}

Three parts have been presented in this paper: A summary of results of a general decomposition of the quark-quark correlator that leads to the operator definition of TMD FFs, 
a general kinematical analysis for $e^+e^-\to V\pi X$ and a complete twist-3 calculation based on the partonic picture at leading order in pQCD.  
We summarize the main results in the following.

(1) We presented the results of general decomposition of quark-quark correlator for fragmentation of quark to spin-1 hadron. 
The correlator is expressed as a sum of a spin independent, a vector polarization dependent and a tensor polarization dependent part. 
Formally, the spin independent part is identical to that for spin-0 hadrons, 
the vector polarization dependent part is the same as that for spin-1/2 hadrons, while the tensor polarization dependent part is novel for spin-1 hadrons. 
The decomposition leads to totally 72 TMD FFs, 8 for spin independent, 24 for the vector polarization dependent  
and the other 40 for the tensor polarization dependent part. 
Among them, 18 contribute at leading twist, 36 at twist-3 and the other 18 at twist-4; 
half of them (36) are $T$-odd, the other half are $T$-even;  also half are $\chi$-odd and the other half are $\chi$-even. 

(2) These TMD FFs are used in describing the semi-inclusive high energy reaction (see e.g. \cite{Wei:2014pma}). 
We note that usually for a complete description of a semi-inclusive reaction, the quark-quark correlator is not sufficient. 
One usually needs quark-$j$-gluon-quark correlator, too ($j=1,2,...$ represents the number of gluons). 
They contribute at higher twist starting at twist-$(j+2)$.   
For example, to make a complete calculation up to twist-3, besides the quark-quark correlator, one needs the quark-gluon-quark correlator. 
These contributions should be taken into account simultaneously. 
It is also important to note that because of the QCD equation of motion, they are often not independent 
and relationships obtained from QCD equation of motion should be used.  

(3) We presented also the results for a general kinematic analysis for $e^+e^-\to V \pi X$. 
This process is in general described by 81 structure functions, 42 are parity conserving and 39 are parity violating. 
The azimuthal asymmetries and hadron polarizations are in general coupled with each other and are described by the corresponding structure functions.  
In practice, it is much simpler to study the azimuthal asymmetries in the unpolarized case and hadron polarizations averaged over the azimuthal angle $\varphi$. 
For unpolarized hadrons, there are 4 azimuthal asymmetries i.e. 
$\langle\cos\varphi\rangle_U$,  $\langle\sin\varphi\rangle_U$,  $\langle\cos2\varphi\rangle_U$,  and $\langle\sin2\varphi\rangle_U$. 
The two $\cos$-asymmetries are parity conserving while the two $\sin$-asymmetries are parity violating.

(4) The hadron polarizations are most conveniently studied in the helicity Gottfried-Jackson frame. 
Here, we have two longitudinal components $\langle\lambda\rangle$ and $\langle S_{LL}\rangle$ defined in the helicity basis,  
and 6 transverse components that can be defined either w.r.t. the lepton-hadron plane, i.e., 
$\langle S_{T}^x\rangle$, $\langle S_{T}^y\rangle$, $\langle S_{LT}^x\rangle$, 
$\langle S_{LT}^y\rangle$, $\langle S_{TT}^{xx}\rangle$ and $\langle S_{TT}^{xy}\rangle$, 
or w.r.t the hadron-hadron plane i.e. $\langle S_{T}^n\rangle$, $\langle S_{T}^t\rangle$, $\langle S_{LT}^n\rangle$, 
$\langle S_{LT}^t\rangle$, $\langle S_{TT}^{nn}\rangle$ and $\langle S_{TT}^{nt}\rangle$. 
In the case of averaging over $\varphi$, they correspond to different structure functions as 
given by Eqs.~(\ref{eq:AStx}-\ref{eq:ASttxy}) and Eqs.~(\ref{eq:AStn}-\ref{eq:ASttnt}) respectively. 
Half of them are parity conserving while the other half are parity violating. 

(5) The results obtained in partonic picture at LO pQCD up to twist-3 are also presented in terms of the gauge invariant FFs. 
These results show that at leading twist there are 27 non-vanishing structure functions 19 correspond to parity conserving 
and 8 are parity violating. We have also 36 structure functions that have twist-3 as leading power contributions. 

(6) For unpolarized hadrons, there is only one azimuthal asymmetry $\langle\cos2\varphi\rangle$ at leading twist due to 
Collins effect\cite{Collins:1992kk} in fragmentation and transverse spin correlation $c_{nn}^q$ given by Eq.~(\ref{eq:cnnq}) in $e^+e^-$-annihilations,
and two twist-3 asymmetries $\langle\cos\varphi\rangle$ and $\langle\sin\varphi\rangle$, the former is similar to the Cahn effect~\cite{Cahn:1978se} 
in DIS and the latter exists only in parity violating reactions. 

(7) Longitudinal components of hadron polarization $\langle \lambda\rangle$ and $\langle S_{LL}\rangle$ exist at leading twist 
as given by Eqs.~(\ref{eq:ALambdat2}-\ref{eq:ASllt2}). 
While the former depends on the initial polarization $P_q$ of the quark produced at the $e^+e^-$ annihilation vertex and exists only 
in weak interaction processes, the latter is independent of $P_q$ and exists also in electromagnetic processes. 

(8) Transverse components $\langle S_{T}^n\rangle$, $\langle S_{T}^t\rangle$, $\langle S_{LT}^n\rangle$, $\langle S_{LT}^t\rangle$, 
$\langle S_{TT}^{nn}\rangle$ and $\langle S_{TT}^{nt}\rangle$ w.r.t. the hadron-hadron plane exist at leading twist given by Eqs.~(\ref{eq:AStnt2}-\ref{eq:ASttntt2}). 
Among them $\langle S_{T}^n\rangle$, $\langle S_{LT}^t\rangle$ and $\langle S_{TT}^{nn}\rangle$ are parity conserving and 
$\langle S_{T}^t\rangle$, $\langle S_{LT}^n\rangle$ and $\langle S_{TT}^{nt}\rangle$ are parity violating.
 
(9) There are also twist-3 transverse components $\langle S_{T}^x\rangle$, $\langle S_{T}^y\rangle$, $\langle S_{LT}^x\rangle$, 
$\langle S_{LT}^y\rangle$, $\langle S_{TT}^{xx}\rangle$ and $\langle S_{TT}^{xy}\rangle$ w.r.t. lepton-hadron plane.
They are determined by the corresponding twist-3 FFs as given by Eqs.~(\ref{eq:AStxt3}-\ref{eq:ASltyt3}).
Similarly, $\langle S_{T}^y\rangle$, $\langle S_{LT}^x\rangle$ and $\langle S_{TT}^{xx}\rangle$ are parity conserving and 
$\langle S_{T}^x\rangle$, $\langle S_{LT}^y\rangle$ and $\langle S_{TT}^{xy}\rangle$ are parity violating.

(10) For inclusive reaction $e^+e^-\to VX$, we can only study $z$-dependence. 
Kinematically, the hadronic tensor and/or cross section take the same form as that of the inclusive reaction 
$e^+e^-\to V\pi X$ averaged over $\varphi$. 
We have two longitudinal components of polarization i.e. $\langle\lambda\rangle$ and $\langle S_{LL}\rangle$ at leading twist. 
In particular we have 4 transverse components $\langle S_{T}^x\rangle$, $\langle S_{T}^y\rangle$, $\langle S_{LT}^x\rangle$, $\langle S_{LT}^y\rangle$ at twist-3. 
Three of them are either $T$-odd or $P$-odd and do not exist in deep-inelastic scattering such as $e^-h\to e^-X$. 
The only both $P$ and $T$-even one is $\langle S_{LT}^x\rangle$. \\

Finally, we would like to emphasize in particular that, in experiments, different components of the (vector) polarizations of octet hyperons such as $\Lambda$, $\Sigma^\pm$ and $\Xi^{0,-}$ 
and those of the tensor polarizations of vector mesons such as $\rho$ and $K^*$ can be measured in a conceptually simple way. 
Polarizations of these hyperons  can be measured by studying the angular distributions of the decay products of their spin self analyzing parity violating decays. 
All the five independent components of the tensor polarization, $S_{LL}$, $S_{LT}^{x}$, $S_{LT}^{y}$, $S_{TT}^{xx}$ and $S_{TT}^{xy}$, 
of these vector mesons can also be measured via the angular distributions in their strong decays into two pseudoscalar mesons~\cite{Bacchetta:2000jk}. 
Such measurements have also been carried out in past in different high energy reactions. 
Transverse polarizations of different hyperons have been observed in unpolarized hadron-hadron, 
hadron-nucleus collisions~\cite{tran-exp}, in $e^+e^-$-annihilations~\cite{Althoff:1984iz} and lepton-hadron reactions~\cite{Adamovich:1994gy}
that correspond to the Sivers type FF $D_{1T}^\perp$ and higher twist addenda to it.  
We see in particular that in experiments with $e^+e^-$ annihilation at high energies where FFs can be best studied, 
measurements have been carried out e.g. at LEP on longitudinal polarization of $\Lambda$ 
hyperon production~\cite{Buskulic:1996vb,Ackerstaff:1997nh} by ALEPH and OPAL collaborations, 
and also on the spin alignment $\rho_{00}=(1-2S_{LL})/3$ for vector mesons such as $K^*$, $\rho$ and so on~\cite{Ackerstaff:1997kj,Ackerstaff:1997kd,Abreu:1997wd}. 
Results for $z$ dependences have been obtained in both cases. 
Even non-diagonal components (corresponds to higher twist contributions only) have also been measured~\cite{Ackerstaff:1997kj,Ackerstaff:1997kd,Abreu:1997wd}.
The data available are definitely still far from enough to limit the precise forms of the FFs involved. 
They have however provided important hints for the corresponding components and have attracted much attention theoretically.
Many phenomenological model studies have been carried out in last few years~\cite{Gustafson:1992iq,Boros:1998kc,Liu:2000fi,Liu:2001yt,Liang:2002ub,Xu:2002hz,
Ma:1998pd,Ma:1999gj,Ma:1999wp,Ma:1999hi,Ma:2000uu,Ma:2000cg,Chi:2013hka,Ellis:2002zv,Anselmino:1997ui,Anselmino:1998jv,Anselmino:1999cg,Xu:2001hz,Xu:2003fq}.

Recent measurements have been carried out on azimuthal asymmetries for two hadron production 
by BELLE, BARBAR and BES III Collaborations~\cite{Abe:2005zx,Seidl:2008xc, Vossen:2011fk,TheBABAR:2013yha,Ablikim:2015pta}. 
They provide useful constraints on Collins function~\cite{Anselmino:2007fs,Anselmino:2013vqa}.
Presently, related measurements can be and are being carried out e.g in $pp$-collisions by STAR at RHIC,   
and in the existing $e^+e^-$ colliders such as BELLE at KEK and BES at BEPC~\cite{current}.  
They can certainly also be studied in future $e^+e^-$ colliders at high energies,  
electron-ion colliders discussed in the community~\cite{future}. 
We would in particular like to note that usually the production rates of vector mesons are much higher than hyperons in high energy reactions. 
Hence, we expect that studies of vector meson tensor polarization might provide us a more sensitive window 
to study polarization effects in fragmentation process in particular and to develop QCD theory in general.  \\

\section*{Acknowledgements}
We thank Yu-kun Song and Long Chen for helpful discussions.
This work was supported in part by the National Natural Science Foundation of China
(No. 11375104),  the Major State Basic Research Development Program in China (No. 2014CB845406) 
and the CAS Center for Excellence in Particle Physics (CCEPP).

\begin{widetext}
\begin{appendix}

\section{Fragmentation functions defined via quark-quark correlator}\label{FFs}

We make a full list of the TMD FFs defined via quark-quark correlator in this appendix. 

\subsection{The spin independent part}
The general decomposition of the spin independent part of the quark-quark correlator is given by,
\begin{align}
& z\Xi^{U(0)}(z,k_{F\perp};p) = ME(z,k_{F\perp}), \label{eq:XiUS}\\
& z\tilde\Xi^{U(0)}(z,k_{F\perp};p) =0, \label{eq:XiPS}\\
& z\Xi_\alpha^{U(0)}(z,k_{F\perp};p) = p^+ \bar n_\alpha D_1(z,k_{F\perp})+ k_{F\perp\alpha} D^\perp(z,k_{F\perp}) + \frac{M^2}{p^+}n_\alpha D_3(z,k_{F\perp}),\label{eq:XiUV}\\
 & z\tilde\Xi_\alpha^{U(0)}(z,k_{F\perp};p) = -\tilde k_{F\perp\alpha} G^\perp(z,k_{F\perp}), \label{eq:XiUAV}\\
& z\Xi_{\rho\alpha}^{U(0)}(z,k_{F\perp} ;p) = -\frac{p^+}{M} \bar n_{[\rho}\tilde k_{F\perp\alpha]} H_1^\perp(z,k_{F\perp}) + M\varepsilon_{\perp\rho\alpha} H(z,k_{F\perp}) 
- \frac{M}{p^+} n_{[\rho}\tilde k_{F\perp\alpha]} H_3^\perp(z,k_{F\perp}).  \label{eq:XiUT}
\end{align}
Here, we note in particular that, compared with the corresponding $\bar n$ component, 
the $n_\perp$ and $n$ components are suppressed by $M/p^+$ and $(M/p^+)^2$ and contribute at twist-3 and twist-4 respectively.
If we integrate over $d^2k_{F\perp}$, terms with $k_{F\perp}$ odd Lorentz structures vanish and we obtain,
\begin{align}
& z\Xi^{U(0)}(z;p) = ME(z), %\\
 & z\tilde\Xi^{U(0)}(z;p) =0, && \label{eq:XiUPS1}\\
& z\Xi_\alpha^{U(0)}(z;p) = p^+ \bar n_\alpha D_1(z) + \frac{M^2}{p^+}n_\alpha D_3(z),%\\
 & z\tilde\Xi_\alpha^{U(0)}(z;p) = 0, && \label{eq:XiUAV1}\\
& z\Xi_{\rho\alpha}^{U(0)}(z;p) = M\varepsilon_{\perp\rho\alpha} H(z), &&& \label{eq:XiUT1}
\end{align}
where the one-dimensional FF is just equal to the corresponding three-dimensional one integrated over $d^2k_{F\perp}$ such as,
\begin{align} 
D_1(z)=\int \frac{d^2k_{F\perp}}{(2\pi)^2}D_1(z,k_{F\perp})
=z\sum_X \int \frac{d\xi^-}{2\pi}  e^{-ip^+\xi^-/z}  
 \langle p,S;X|\bar\psi(\xi^-) \mathcal{L}(\xi^-;\infty) |0\rangle \frac{\gamma^+}{4} \langle 0|\mathcal{L}^\dag (0;\infty)\psi(0) |p,S;X\rangle. \label{eq:D1z}
\end{align}
The factor $z$ before $\Xi^{(0)}$ on the left-hand-side of Eqs.~(\ref{eq:XiUS}-\ref{eq:XiUT}) is needed so that 
$D_1(z)$ obtained this way is the number density for a quark fragmentation into a specified hadron. 
However,  when polarization is involved, we note the difference: While for phenomenologically defined  $D_1(z)$, a sum over spin of $h$ and an average 
over the spin of quark is understood, for $D_1(z)$ defined via quark-quark correlator as given by Eq.~(\ref{eq:D1z}), we have an average over hadron spin and 
a sum over quark spin.  Hence $D_1(z)$ is identical in the two cases only for spin $1/2$ hadrons. 

\subsection{Vector polarization dependent part} \label{FFs:Vector}

We build the $S$-dependent basic Lorentz covariants with the corresponding properties under space reflection as demanded 
and obtain the general decomposition of the $S$-dependent part of the quark-quark correlator as,
%\begin{widetext}
\begin{align}
& z\Xi^{V(0)}(z,k_{F\perp};p,S) = (\tilde k_{F\perp} \cdot S_T) E_T^\perp(z,k_{F\perp}), \label{eq:XiVS}\\
& z\tilde \Xi^{V(0)}(z,k_{F\perp};p,S) = M \Bigl[ \lambda E_L(z,k_{F\perp}) + \frac{k_{F\perp} \cdot S_T}{M} E^{\prime\perp}_T(z,k_{F\perp}) \Bigr],  \label{eq:XiVPS}\\
& z\Xi_\alpha^{V(0)}(z,k_{F\perp};p,S) = p^+ \bar n_\alpha \frac{\tilde k_{F\perp} \cdot S_T}{M} D_{1T}^\perp(z,k_{F\perp}) 
 - M\tilde S_{T\alpha} D_T(z,k_{F\perp}) \nonumber\\
& \hspace{2cm} - \tilde k_{F\perp\alpha} \Bigl[ \lambda D_L^\perp(z,k_{F\perp}) + \frac{k_{F\perp} \cdot S_T}{M}D_T^{\perp}(z,k_{F\perp}) \Bigr] 
 + \frac{M}{p^+}n_\alpha (\tilde k_{F\perp} \cdot S_T) D_{3T}^\perp(z,k_{F\perp}),  \label{eq:XiVV}\\
& z\tilde\Xi_\alpha^{V(0)}(z,k_{F\perp};p,S) = p^+ \bar n_\alpha \Bigl[ \lambda G_{1L}(z,k_{F\perp}) + \frac{k_{F\perp} \cdot S_T}{M} G_{1T}^\perp(z,k_{F\perp}) \Bigr] 
- MS_{T\alpha} G_T(z,k_{F\perp}) \nonumber \\
& \hspace{2cm} - k_{F\perp\alpha} \Bigl[ \lambda G_{L}^\perp(z,k_{F\perp}) + \frac{k_{F\perp} \cdot S_T}{M} G_{T}^{\perp}(z,k_{F\perp}) \Bigr] 
+ \frac{M^2}{p^+}n_\alpha \Bigl[ \lambda G_{3L}(z,k_{F\perp}) + \frac{k_{F\perp} \cdot S_T}{M} G_{3T}^\perp(z,k_{F\perp}) \Bigr],  \label{eq:XiVAV}\\
& z\Xi_{\rho\alpha}^{V(0)}(z,k_{F\perp};p,S) = p^+ \bar n_{[\rho}S_{T\alpha]} H_{1T}(z,k_{F\perp}) + \frac{p^+}{M} \bar n_{[\rho}k_{F\perp\alpha]} \Bigl[ \lambda H_{1L}^\perp(z,k_{F\perp}) 
+ \frac{k_{F\perp} \cdot S_T}{M} H_{1T}^\perp(z,k_{F\perp}) \Bigr] \nonumber\\
& \hspace{2cm} + k_{F\perp[\rho}S_{T\alpha]} H_T^\perp(z,k_{F\perp}) + M \bar n_{[\rho}n_{\alpha]} \Bigl[ \lambda H_L(z,k_{F\perp}) 
+ \frac{k_{F\perp} \cdot S_T}{M} H_T^{\prime\perp}(z,k_{F\perp}) \Bigr] \nonumber\\
& \hspace{2cm} + \frac{M^2}{p^+} n_{[\rho}S_{T\alpha]} H_{3T}(z,k_{F\perp}) 
+ \frac{M}{p^+} n_{[\rho}k_{F\perp\alpha]} \Bigl[ \lambda H_{3L}^\perp(z,k_{F\perp}) + \frac{k_{F\perp} \cdot S_T}{M} H_{3T}^\perp(z,k_{F\perp}) \Bigr].  \label{eq:XiVT}
\end{align}
%\end{widetext}
%where $\varepsilon_\perp^{k_{F\perp}S_T}\equiv \varepsilon_{\perp\rho\sigma}k_{F\perp}^\rho S_T^\sigma$. 
%
If we integrate over $d^2k_{F\perp}$, only 8 terms survive, i.e.,
\begin{align}
z\Xi^{V(0)}(z;p,S) =& 0,  \label{eq:XiVS1}\\
z\tilde \Xi^{V(0)}(z;p,S) =& \lambda M E_L(z),  \label{eq:XiVPS1}\\
 z\Xi_\alpha^{V(0)}(z;p,S) = & -M \tilde S_{T\alpha} D_{T}(z),  \label{eq:XiVV1}\\
z\tilde\Xi_\alpha^{V(0)}(z;p,S) = & \lambda p^+ \bar n_\alpha  G_{1L}(z)  - MS_{T\alpha} G_T(z) + \lambda \frac{M^2}{p^+}n_\alpha  G_{3L}(z),  \label{eq:XiVAV1}\\
z\Xi_{\rho\alpha}^{V(0)}(z;p,S) = & p^+ \bar n_{[\rho}S_{T\alpha]} H_{1T}(z) - \lambda M \bar n_{[\rho}n_{\alpha]}  H_L(z) + \frac{M^2}{p^+}  n_{[\rho}S_{T\alpha]} H_{3T}(z),  \label{eq:XiVT1}
\end{align}
where the one-dimensional FF in the longitudinally polarized case is just equal to the corresponding three-dimensional FF integrated over $d^2k_{F\perp}$, 
while in the transversely polarized case, we have,   
\begin{align}
& K_{T}(z) = \int \frac{d^2k_{F\perp}}{(2\pi)^2} {\cal K}_T^\perp(z,k_{F\perp}), 
& {\cal K}_T^\perp(z,k_{F\perp}) \equiv  K_T(z,k_{F\perp}) + \frac{k_{F\perp}^2}{2M^2} K_T^{\perp}(z,k_{F\perp}), \label{eq:calK}
\end{align}
for the transverse polarization dependent FFs such as $K_T=D_T$, $G_T$, $H_{1T}$ or $H_{3T}$, and similar for the $S_{LT}$-dependent part in the following. 

%\begin{align}
%& D_{T}(z) = \int \frac{d^2k_{F\perp}}{(2\pi)^2} \Bigl[ D_T(z,k_{F\perp}) + \frac{k_{F\perp}^2}{2M^2} D_T^{\perp}(z,k_{F\perp}) \Bigr], \\
%& G_T(z) = \int \frac{d^2k_{F\perp}}{(2\pi)^2} \Bigl[ G_T(z,k_{F\perp}) + \frac{k_{F\perp}^2}{2M^2} G_T^{\perp}(z,k_{F\perp}) \Bigr],\\
%& H_{1T}(z) = \int \frac{d^2k_{F\perp}}{(2\pi)^2} \Bigl[ H_{1T}(z,k_{F\perp}) + \frac{k_{F\perp}^2}{2M^2} H_{1T}^{\perp}(z,k_{F\perp}) \Bigr],\\
%& H_{3T}(z) = \int \frac{d^2k_{F\perp}}{(2\pi)^2} \Bigl[ H_{3T}(z,k_{F\perp}) + \frac{k_{F\perp}^2}{2M^2} H_{3T}^{\perp}(z,k_{F\perp}) \Bigr].
%\end{align}

\subsection{Tensor polarization dependent part} \label{FFs:tensor}

The most general decomposition for the tensor polarization dependent part is given by,
\begin{align}
& z\Xi^{T(0)}(z,k_{F\perp};p,S) =M \Bigl[ S_{LL}E_{LL}(z,k_{F\perp}) 
+ \frac{k_{F\perp} \cdot S_{LT}}{M} E_{LT}^\perp(z,k_{F\perp}) 
%+ \frac{k_{F\perp} \cdot S_{TT} \cdot k_{F\perp}}{M^2} E_{TT}^{\perp}(z,k_{F\perp})\Bigr],  \label{eq:XiTS} \\
+ \frac{S_{TT}^{k_Fk_F}}{M^2} E_{TT}^{\perp}(z,k_{F\perp})\Bigr],  \label{eq:XiTS} \\
& z\tilde \Xi^{T(0)}(z,k_{F\perp};p,S) =M \Bigl[ \frac{\tilde k_{F\perp} \cdot S_{LT}}{M} E_{LT}^{\prime\perp} (z,k_{F\perp}) 
+ \frac{S_{TT}^{\tilde k_{F} k_{F}}}{M^2} E_{TT}^{\prime\perp}(z,k_{F\perp}) \Bigr],  \label{eq:XiTPS}\\
& z\Xi_\alpha^{T(0)}(z,k_{F\perp};p,S) = p^+ \bar n_\alpha \Bigl[ S_{LL} D_{1LL}(z,k_{F\perp}) 
+ \frac{k_{F\perp} \cdot S_{LT}}{M}D_{1LT}^\perp (z,k_{F\perp}) + \frac{S_{TT}^{k_Fk_F}}{M^2} D_{1TT}^\perp(z,k_{F\perp}) \Bigr] \nonumber\\
%\frac{k_{F\perp} \cdot S_{TT} \cdot k_{F\perp}}{M^2} D_{1TT}^\perp(z,k_{F\perp}) \Bigr] \nonumber\\
 & \hspace{2cm} + M S_{LT\alpha} D_{LT}(z,k_{F\perp}) +S_{TT\alpha}^{k_F} D_{TT}^{\prime\perp}(z,k_{F\perp}) \nonumber\\
 % k_{F\perp}^\sigma S_{TT\sigma\alpha} D_{TT}^{\prime\perp}(z,k_{F\perp}) \nonumber\\
 & \hspace{2cm} + k_{F\perp\alpha} \Bigl[ S_{LL}D_{LL}^\perp(z,k_{F\perp}) + \frac{k_{F\perp} \cdot S_{LT}}{M}D_{LT}^\perp (z,k_{F\perp}) 
 + \frac{S_{TT}^{k_Fk_F}}{M^2} D_{TT}^\perp(z,k_{F\perp}) \Bigr] \nonumber\\
 %+ \frac{k_{F\perp} \cdot S_{TT} \cdot k_{F\perp}}{M^2} D_{TT}^\perp(z,k_{F\perp}) \Bigr] \nonumber\\
 & \hspace{2cm} + \frac{M^2}{p^+}n_\alpha \Bigl[ S_{LL} D_{3LL}(z,k_{F\perp}) + \frac{k_{F\perp} \cdot S_{LT}}{M}D_{3LT}^\perp (z,k_{F\perp}) 
% + \frac{k_{F\perp} \cdot S_{TT} \cdot k_{F\perp}}{M^2} D_{3TT}^\perp(z,k_{F\perp}) \Bigr],  \label{eq:XiTV}\\
 + \frac{S_{TT}^{k_Fk_F}}{M^2} D_{3TT}^\perp(z,k_{F\perp}) \Bigr],  \label{eq:XiTV}\\
& z\tilde \Xi_\alpha^{T(0)}(z,k_{F\perp};p,S) = p^+ \bar n_\alpha \Bigl[ \frac{\tilde k_{F\perp} \cdot S_{LT}}{M}G_{1LT}^\perp(z,k_{F\perp}) 
  + \frac{S_{TT}^{\tilde k_Fk_{F}}}{M^2}G_{1TT}^\perp(z,k_{F\perp}) \Bigr] \nonumber\\
%  + \frac{\tilde k_{F\perp} \cdot S_{TT}^{k_{F\perp}}}{M^2}G_{1TT}^\perp(z,k_{F\perp}) \Bigr] \nonumber\\
 & \hspace{2cm} - M\tilde S_{LT\alpha} G_{LT}(z,k_{F\perp}) 
 - \tilde S_{TT\alpha}^{k_{F}} G_{TT}^{\prime\perp}(z,k_{F\perp}) \nonumber\\
& \hspace{2cm} - \tilde k_{F\perp\alpha} \Bigl[ S_{LL}G_{LL}^\perp(z,k_{F\perp})  
  + \frac{k_{F\perp} \cdot S_{LT}}{M}G_{LT}^\perp (z,k_{F\perp}) + \frac{S_{TT}^{k_Fk_F}}{M^2} G_{TT}^\perp(z,k_{F\perp}) \Bigr] \nonumber\\
 & \hspace{2cm} + \frac{M^2}{p^+}n_\alpha \Bigl[ \frac{ \tilde k_{F\perp} \cdot S_{LT} }{M}G_{3LT}^\perp(z,k_{F\perp}) 
 + \frac{S_{TT}^{\tilde k_Fk_F}}{M^2} G_{3TT}^\perp(z,k_{F\perp}) \Bigr]  , \label{eq:XiTAV}\\
& z\Xi_{\rho\alpha}^{T(0)}(z,k_{F\perp} ;p,S) = -p^+ \bar n_{[\rho} \tilde S_{LT\alpha]} H_{1LT}(z,k_{F\perp}) 
 - \frac{p^+}{M} \bar n_{[\rho} \tilde S_{TT\alpha]}^{k_{F}} H_{1TT}^{\prime\perp}(z,k_{F\perp}) \nonumber\\
 & \hspace{2cm} -\frac{p^+}{M} \bar n_{[\rho} \tilde k_{F\perp\alpha]} \Bigl[ S_{LL} H_{1LL}^\perp(z,k_{F\perp}) 
+ \frac{k_{F\perp} \cdot S_{LT}}{M}H_{1LT}^\perp (z,k_{F\perp}) + \frac{S_{TT}^{k_Fk_F}}{M^2} H_{1TT}^\perp(z,k_{F\perp}) \Bigr] \nonumber\\
 & \hspace{2cm} + M\varepsilon_{\perp\rho\alpha} \Bigl[ S_{LL} H_{LL}(z,k_{F\perp}) + \frac{k_{F\perp} \cdot S_{LT}}{M}H_{LT}^\perp (z,k_{F\perp}) 
 + \frac{S_{TT}^{k_Fk_F}}{M^2} H_{TT}^\perp(z,k_{F\perp}) \Bigr] \nonumber\\
 & \hspace{2cm} + \bar n_{[\rho}n_{\alpha]} \Bigl[ (\tilde k_{F\perp} \cdot S_{LT}) H_{LT}^{\prime\perp}(z,k_{F\perp}) 
 + \frac{S_{TT}^{\tilde k_Fk_F}}{M}H_{TT}^{\prime\perp}(z,k_{F\perp}) \Bigr] \nonumber\\
 & \hspace{2cm} - \frac{M}{p^+} n_{[\rho} \tilde k_{F\perp\alpha]} \Bigl[ S_{LL} H_{3LL}^\perp(z,k_{F\perp}) 
 + \frac{k_{F\perp} \cdot S_{LT}}{M}H_{3LT}^\perp (z,k_{F\perp}) + \frac{S_{TT}^{k_Fk_F}}{M^2} H_{3TT}^\perp(z,k_{F\perp}) \Bigr] \nonumber\\
 & \hspace{2cm} - \frac{M}{p^+} n_{[\rho} \Bigl[ M \tilde S_{LT\alpha]} H_{3LT}(z,k_{F\perp}) 
 + \tilde S_{TT\alpha]}^{k_{F}}H_{3TT}^{\prime\perp}(z,k_{F\perp}) \Bigr].  \label{eq:XiTT}
\end{align}
We integrate over $d^2k_{F\perp}$ and obtain, 
\begin{align}
z\Xi^{T(0)}(z;p,S) = & M S_{LL}E_{LL}(z),  \label{eq:XiTS1}\\
z\tilde\Xi^{T(0)}(z;p,S) =& 0,  \label{eq:XiTPS1} \\
 z\Xi_\alpha^{T(0)}(z;p,S) = & p^+ \bar n_\alpha S_{LL} D_{1LL}(z)+ MS_{LT\alpha} D_{LT}(z) +   \frac{M^2}{p^+}n_\alpha S_{LL} D_{3LL}(z),  \label{eq:XiTV1}\\
z\tilde\Xi_\alpha^{T(0)}(z;p,S) = & -M \tilde S_{LT\alpha} G_{LT} (z), \label{eq:XiTAV1}\\
z\Xi_{\rho\alpha}^{T(0)}(z ;p,S) =& -p^+ \bar n_{[\rho} \tilde S_{LT\alpha]} H_{1LT}(z) 
+ M\varepsilon_{\perp\rho\alpha} S_{LL} H_{LL}(z)  - \frac{M^2}{p^+}  n_{[\rho} \tilde S_{LT\alpha]} H_{3LT}(z).  \label{eq:XiTT1}
\end{align}
%We have totally 8 terms, 4 of them are $S_{LL}$-dependent and the other 4 are $S_{LT}$-dependent.  
%They have exact one to one correspondence to the unpolarized and $S_T$-dependent parts. 
Again, the 4 $S_{LL}$-dependent one-dimensional FFs  are just equal to the corresponding three-dimensional FFs integrated over $d^2k_{F\perp}$, 
while the 4 $S_{LT}$-dependent FFs are given by Eq.~(\ref{eq:calK}) for $K_T=D_{LT}$, $G_{LT}$, $H_{1LT}$ and $H_{3LT}$.
%\begin{align}
%& D_{LT}(z) = \int \frac{d^2k_{F\perp}}{(2\pi)^2} \Bigl[ D_{LT}(z,k_{F\perp}) + \frac{k_{F\perp}^2}{2M^2} D_{LT}^{\perp}(z,k_{F\perp}) \Bigr], \\
%& G_{LT}(z) = \int \frac{d^2k_{F\perp}}{(2\pi)^2} \Bigl[ G_{LT}(z,k_{F\perp}) + \frac{k_{F\perp}^2}{2M^2} G_{LT}^{\perp}(z,k_{F\perp}) \Bigr],\\
%& H_{1LT}(z) = \int \frac{d^2k_{F\perp}}{(2\pi)^2} \Bigl[ H_{1LT}(z,k_{F\perp}) + \frac{k_{F\perp}^2}{2M^2} H_{1LT}^{\perp}(z,k_{F\perp}) \Bigr],\\
%& H_{3LT}(z) = \int \frac{d^2k_{F\perp}}{(2\pi)^2} \Bigl[ H_{3LT}(z,k_{F\perp}) + \frac{k_{F\perp}^2}{2M^2} H_{3LT}^{\perp}(z,k_{F\perp}) \Bigr].
%\end{align}

We list those twist-2 FFs in table \ref{tab:TMDFF1}, and those twist-3 FFs in table \ref{tab:TMDFF2}. 
The twist-4 FFs have the same structure of those at twist-2, so we do not make a separate table.
We also list them according to chiral and time-reversal properties in table \ref{tab:TMDFFChiralTime}.  
%
%\end{widetext}

\begin{table}[!ht]
\caption{The 18 leading twist components of the FFs for quark fragments to spin-1 hadrons}\label{tab:TMDFF1}
\begin{tabular}{p{1.8cm}l@{\hspace{0.6cm}}lp{2.8cm}l} \hline
\begin{minipage}[c]{1.5cm}quark\\ polarization \end{minipage}  & \begin{minipage}[c]{1.5cm} hadron \\ polarization \end{minipage}  
 & TMD FFs  & integrated over $\vec k_{F\perp}$ &  name  \rule[-0.39cm]{0mm}{0.95cm} \\[0.12cm]  \hline 
 & \hspace{0.5cm} $U$ & $D_1(z,k_{F\perp})$ &  \hspace{0.5cm} $D_1(z)$  & number density \\ [-0.0cm]\cline{2-5}
     & \hspace{0.5cm} $T$ & $D_{1T}^\perp(z,k_{F\perp})$ & \hspace{0.6cm} $\times$ &  \\ [0.1cm] \cline{2-5}
\hspace{0.5cm} $U$ & \hspace{0.5cm} $LL$ & $D_{1LL}(z,k_{F\perp})$    & \hspace{0.35cm} $D_{1LL}(z)$  & spin alignment \\ [-0.0cm]
     & \hspace{0.5cm} $LT$ & $D_{1LT}^\perp(z,k_{F\perp})$  & \hspace{0.6cm} $\times$ &  \\ [0.1cm]
     & \hspace{0.5cm} $TT$ & $D_{1TT}^\perp(z,k_{F\perp})$  & \hspace{0.6cm} $\times$ &  \\ [0.1cm] \hline
\vspace{0.7cm} \hspace{0.5cm} $L$ & \hspace{0.5cm} $L$  & $G_{1L}(z,k_{F\perp})$ & \hspace{0.5cm} $G_{1L}(z)$ & spin transfer (longitudinal) \\ [-0.8cm]
 & \hspace{0.5cm} $T$ & $G_{1T}^\perp(z,k_{F\perp})$ & \hspace{0.6cm} $\times$ &  \\ [0.1cm] \cline{2-5}
 & \hspace{0.5cm} $LT$  & $G_{1LT}^\perp(z,k_{F\perp})$   & \hspace{0.6cm} $\times$ &  \\ [-0.0cm]
 & \hspace{0.5cm} $TT$ & $G_{1TT}^\perp(z,k_{F\perp})$  & \hspace{0.6cm} $\times$ &  \\ [0.1cm] \hline
 & \hspace{0.5cm} $U$ & $H_{1}^\perp(z,k_{F\perp})$ &  \hspace{0.6cm} $\times$ & Collins function \\[0.1cm]\cline{2-5}
 & \hspace{0.4cm} $T(\parallel)$ & $H_{1T}(z,k_{F\perp})$ & \vspace{0.15cm} \hspace{0.5cm} $H_{1T}(z)$ & \vspace{-0.2cm} spin transfer (transverse)  \\[0.1cm]
 & \hspace{0.4cm} $T(\perp)$ & $H_{1T}^\perp(z,k_{F\perp})$ &  &   \\[0.1cm]
 \hspace{0.5cm} $T$ & \hspace{0.5cm} $L$ & $H_{1L}^\perp(z,k_{F\perp})$ & \hspace{0.6cm} $\times$ &  \\[0.1cm] \cline{2-5}
 & \hspace{0.5cm} $LL$ & $H_{1LL}^\perp(z,k_{F\perp})$  & \hspace{0.6cm} $\times$ &  \\[-0.0cm]
 & \hspace{0.5cm} $LT$ & $H_{1LT}(z,k_{F\perp}),~H_{1LT}^\perp(z,k_{F\perp})$  & \vspace{-0.2cm} \hspace{0.35cm} $H_{1LT}(z)$ & \vspace{-0.1cm}   \\[0.2cm]
 & \hspace{0.5cm} $TT$ & $H_{1TT}^\perp(z,k_{F\perp}),~H_{1TT}^{\prime\perp}(z,k_{F\perp})$  & \hspace{0.6cm}$\times$, $\times$ &   \\[0.1cm]\hline
\end{tabular}
\end{table}

\begin{table}[!ht]
\caption{The 36 twist-3 components of the FFs for quark fragments to spin-1 hadrons}\label{tab:TMDFF2}
\begin{tabular}{p{1.8cm}l@{\hspace{0.6cm}}lp{2.8cm}l} \hline
\begin{minipage}[c]{1.5cm}quark\\ polarization \end{minipage}  & \begin{minipage}[c]{1.5cm} hadron \\ polarization \end{minipage}  
 & TMD FFs  & \hspace{0.2cm} integrated over $\vec k_{F\perp}$   \rule[-0.39cm]{0mm}{0.95cm} \\[0.12cm]  \hline 
\vspace{1.3cm} \hspace{0.5cm} $U$  & \hspace{0.5cm} $U$ & $E(z,k_{F\perp})$, $D^\perp(z,k_{F\perp})$ &  \hspace{0.6cm} $E(z)$, $\times$   \\ [-1.3cm]\cline{2-5}
       & \hspace{0.5cm} $L$ & $D_{L}^\perp(z,k_{F\perp})$ & \hspace{0.6cm} $\times$   \\ [0.1cm] 
       & \hspace{0.5cm} $T$ & $E_{T}^\perp(z,k_{F\perp})$, $D_T(z,k_{F\perp})$, $D_T^{\perp}(z,k_{F\perp})$ & \hspace{0.6cm} $\times$, $D_T(z)$   \\ [0.1cm] \cline{2-5}
      & \hspace{0.5cm} $LL$ & $E_{LL}(z,k_{F\perp})$, $D_{LL}^\perp(z,k_{F\perp})$    & \hspace{0.6cm} $E_{LL}(z)$, $\times$    \\ [0.0cm]
      & \hspace{0.5cm} $LT$ & $E_{LT}^\perp(z,k_{F\perp})$, $D_{LT}(z,k_{F\perp})$, $D_{LT}^\perp(z,k_{F\perp})$  & \hspace{0.6cm} $\times$, $D_{LT}(z)$   \\ [0.1cm]
      & \hspace{0.5cm} $TT$ & $E_{TT}^\perp(z,k_{F\perp})$, $D_{TT}^\perp(z,k_{F\perp})$, $D_{TT}^{\prime\perp}(z,k_{F\perp})$  & \hspace{0.6cm} $\times$, $\times$, $\times$   \\ [0.1cm] \hline
\vspace{1.3cm} \hspace{0.5cm} $L$ & \hspace{0.5cm} $U$  & $G^\perp(z,k_{F\perp})$ & \hspace{0.6cm} $\times$  \\ [-1.3cm] \cline{2-5}
 & \hspace{0.5cm} $L$ & $E_L(z,k_{F\perp})$, $G_L^\perp(z,k_{F\perp})$ & \hspace{0.6cm} $E_L(z)$, $\times$   \\ [0.1cm]
 & \hspace{0.5cm} $T$ & $E_T^{\prime\perp}(z,k_{F\perp})$, $G_T(z,k_{F\perp})$, $G_T^\perp(z,k_{F\perp})$ & \hspace{0.6cm} $\times$, $G_T(z)$   \\ [0.1cm] \cline{2-5}
& \hspace{0.5cm} $LL$  & $G_{LL}^\perp(z,k_{F\perp})$   & \hspace{0.6cm} $\times$   \\ [-0.0cm]
& \hspace{0.5cm} $LT$  & $E_{LT}^{\prime\perp}(z,k_{F\perp})$, $G_{LT}(z,k_{F\perp})$, $G_{LT}^{\perp}(z,k_{F\perp})$   & \hspace{0.6cm} $\times$, $G_{LT}(z)$   \\ [-0.0cm]
 & \hspace{0.5cm} $TT$ & $E_{TT}^{\prime\perp}(z,k_{F\perp})$, $G_{TT}^{\perp}(z,k_{F\perp})$, $G_{TT}^{\prime\perp}(z,k_{F\perp})$  & \hspace{0.6cm} $\times$, $\times$, $\times$   \\ [0.1cm] \hline
& \hspace{0.5cm} $U$ & $H(z,k_{F\perp})$ &  \hspace{0.6cm} $H(z)$  \\[0.1cm]\cline{2-5}
& \hspace{0.5cm} $L$ & $H_L(z,k_{F\perp})$ &  \hspace{0.6cm} $H_L(z)$  \\[0.1cm]
 & \hspace{0.4cm} $T(\parallel)$ & $H_{T}^\perp(z,k_{F\perp})$ & \hspace{0.6cm} $\times$  \vspace{-0.0cm} \\[0.1cm]
\hspace{0.5cm} $T$ & \hspace{0.4cm} $T(\perp)$ & $H_{T}^{\prime\perp}(z,k_{F\perp})$ & \hspace{0.6cm} $\times$   \\[0.1cm] \cline{2-5}
 & \hspace{0.5cm} $LL$ & $H_{LL}(z,k_{F\perp})$  & \hspace{0.6cm} $H_{LL}(z)$  \\[-0.0cm]
 & \hspace{0.5cm} $LT$ & $H_{LT}^\perp(z,k_{F\perp})$, $H_{LT}^{\prime\perp}(z,k_{F\perp})$  & \hspace{0.6cm} $\times$, $\times$  \vspace{-0.1cm}   \\[0.2cm]
 & \hspace{0.5cm} $TT$ & $H_{TT}^\perp(z,k_{F\perp})$, $H_{TT}^{\prime\perp}(z,k_{F\perp})$  & \hspace{0.6cm}$\times$, $\times$ &   \\[0.1cm]\hline
\end{tabular}
\end{table}

\begin{table}[!ht]
\caption{Chiral and time reversal properties of TMD FFs from quark-quark correlator}\label{tab:TMDFFChiralTime}
\begin{tabular}{@{}ccllll@{}}
\hline
\multirow{2}{*}{\begin{tabular}[c]{@{}c@{}}quark\\ polarization\end{tabular}} & \multirow{2}{*}{\begin{tabular}[c]{@{}c@{}}hadron\\ polarization\end{tabular}}~~~ & \multicolumn{2}{l}{chiral-even} & \multicolumn{2}{l}{chiral-odd} \\ \cline{3-6} 
& & T-even          & T-odd         & T-even         & T-odd         \\ \hline
\multirow{6}{*}{$U$} & $U$  &  $D_1$,~ $D^\perp$,~ $D_3$~ &  & $E$~ &  \\ \cline{2-6}
& $L$ &  & $D_L^\perp$~ &  & \\
& $T$  & & $D_{1T}^\perp$,~ $D_T$,~ $D_T^{\perp}$,~ $D_{3T}^\perp$~ &  & $E_T^\perp$~ \\ \cline{2-6}
& $LL$ & $D_{1LL}$,~ $D_{LL}^\perp$,~ $D_{3LL}$~ & & $E_{LL}$~ & \\
& $LT$ & $D_{1LT}^\perp$,~ $D_{LT}$,~ $D_{LT}^\perp$,~ $D_{3LT}^\perp$~ &  & $E_{LT}^\perp$~ & \\
& $TT$ & $D_{1TT}^\perp$,~ $D_{TT}^\perp$,~ $D_{TT}^{\prime\perp}$,~ $D_{3TT}^{\perp}$~~~ & &$E_{TT}^\perp$~ &\\ \hline
\multirow{6}{*}{$L$}& $U$& & $G^\perp$~ &  &   \\ \cline{2-6}
& $L$ & $G_{1L}$,~ $G_L^\perp$,~ $G_{3L}$~ &  &  & $E_L$~  \\
& $T$ & $G_{1T}^\perp$,~ $G_T$,~ $G_T^\perp$,~ $G_{3T}^\perp$~ & &  & $E_T^{\prime\perp}$~  \\ \cline{2-6}
& $LL$ & & $G_{LL}^\perp$~ &  & \\
& $LT$ & & $G_{1LT}^\perp$,~ $G_{LT}$,~ $G_{LT}^\perp$,~ $G_{3LT}^\perp$~ & $E_{LT}^{\prime\perp}$~ & \\
& $TT$ & & $G_{1TT}^\perp$,~ $G_{TT}^\perp$,~ $G_{TT}^{\prime\perp}$,~ $G_{3TT}^\perp$~~~~~ & $E_{TT}^{\prime\perp}$~ & \\ \hline
\multirow{7}{*}{$T$} & $U$ & & & & $H_1^\perp$,~ $H$,~ $H_3^\perp$~ \\ \cline{2-6}
& $L$ & & & $H_{1L}^\perp$,~ $H_L$,~ $H_{3L}^\perp$ & \\
& $T(\parallel)$  &  &  & $H_{1T}$,~ $H_T^\perp$,~ $H_{3T}$ &  \\
& $T(\perp)$ &  & & $H_{1T}^\perp$,~ $H_T^{\prime\perp}$,~ $H_{3T}^\perp$~~~ &  \\ \cline{2-6}
& $LL$ & & &  & $H_{1LL}^\perp$,~ $H_{LL}$,~ $H_{3LL}^\perp$~ \\
& $LT$ &  & & & $H_{1LT}$,~ $H_{1LT}^\perp$,~ $H_{LT}^\perp$,~ $H_{LT}^{\prime\perp}$,~ $H_{3LT}$,~ $H_{3LT}^\perp$~ \\
& $TT$ &  & & & $H_{1TT}^\perp$,~ $H_{1TT}^{\prime\perp}$,~ $H_{TT}^\perp$,~ $H_{TT}^{\prime\perp}$,~ $H_{3TT}^\perp$,~ $H_{3TT}^{\prime\perp}$~ \\ \hline 
\end{tabular}
\end{table}

\subsection{Twist-3 FFs defined via quark-gluon-quark correlator}\label{FFs:qgq}

Twist-3 components are the leading twist contributions that we obtain from $\hat\Xi^{(1)}_\rho$. 
There has to be one $\bar n$ involved in the basic Lorentz covariants and the other(s) are from the transverse components. 
Since the $\bar n$ component of gluon field goes into the gauge link, we only have the other three components for $D_\rho$ 
thus no $\bar n_\rho$-component exists in the basic Lorentz covariants. 
We therefore do not have twist-3 contributions from $\Xi_\rho^{(1)}$ or $\tilde\Xi_\rho^{(1)}$. 
The twist-3 contributions are obtained from $\Xi^{(1)}_{\rho\alpha}$,  $\tilde\Xi^{(1)}_{\rho\alpha}$ and $\Xi^{(1)}_{\rho\alpha\beta}$ and are given in the following.

For the unpolarized part, we have,
\begin{align}
z \Xi^{U(1)}_{\rho\alpha}&(z,k_{F\perp};p) = - p^+\bar{n}_\alpha  k_{F\perp\rho} D_d^\perp(z,k_{F\perp}) + \cdots\ ,  \label{eq:Xi1UV}\\
z \tilde \Xi^{U(1)}_{\rho\alpha}&(z,k_{F\perp};p)= -i p^+ \bar{n}_\alpha  \tilde k_{F\perp\rho} G_d^{\perp}(z,k_{F\perp}) + \cdots\ ,  \label{eq:Xi1UAV}\\
z \Xi^{U(1)}_{\rho\alpha\beta}&(z,k_{F\perp};p)= -p^+\Bigl[ M \varepsilon_{\perp\rho[\alpha}\bar{n}_{\beta]}  H_d(z,k_{F\perp}) - \frac{1}{M} \tilde k_{F\perp\rho} k_{F\perp[\alpha}\bar{n}_{\beta]} H_d^\perp(z,k_{F\perp})\Bigr] +\cdots \   \label{eq:Xi1UT}.
\end{align}
%%%%%%
For the vector polarization dependent part, we have, 
%\begin{widetext}
\begin{align}
z \Xi^{V(1)}_{\rho\alpha}(z,k_{F\perp};p,S) = & p^+\bar{n}_\alpha \Bigl\{ M \tilde S_{T\rho} D_{dT}(z,k_{F\perp})  
+ \tilde k_{F\perp\rho} \Bigl[ \lambda D_{dL}^\perp(z,k_{F\perp}) + \frac{k_{F\perp} \cdot S_T}{M} D_{dT}^\perp(z,k_{F\perp}) \Bigr]\Bigr\}+\cdots\  , \label{eq:Xi1VV}\\
%%%
z \tilde \Xi^{V(1)}_{\rho\alpha}(z,k_{F\perp};p,S) = & - i p^+ \bar{n}_\alpha\Bigl\{ M S_{T\rho} G_{dT}(z,k_{F\perp}) 
+ k_{\perp\rho} \Bigl[ \lambda G_{dL}^\perp(z,k_{F\perp}) + \frac{k_{F\perp} \cdot S_T}{M} G_{dT}^\perp(z,k_{F\perp}) \Bigr] \Bigr\}+\cdots \ ,\label{eq:Xi1VAV}\\
z \Xi^{V(1)}_{\rho\alpha\beta}(z,k_{F\perp};p,S) = & p^+ \Bigl\{ 
\lambda \Bigl[ M g_{\perp\rho[\alpha} \bar{n}_{\beta]} H_{dL}(z,k_{F\perp}) + \frac{1}{M} k_{F\perp\rho} k_{F\perp[\alpha} \bar{n}_{\beta]} H_{dL}^\perp(z,k_{F\perp}) \Bigr]  \nonumber\\
&- (\tilde k_{F\perp}\cdot S_T) \Bigl[ \varepsilon_{\perp\rho[\alpha}\bar{n}_{\beta]} H_{dT}^{\perp}(z,k_{F\perp}) 
 - \frac{1}{M^2} \tilde k_{F\perp\rho} k_{F\perp[\alpha} \bar{n}_{\beta]} H_{dT}^{\perp\prime}(z,k_{F\perp}) \Bigr]  \nonumber\\
&+ (k_{F\perp}\cdot S_T) \Bigl[ g_{\perp\rho[\alpha} \bar{n}_{\beta]} H_{dT}^{\prime\perp}(z,k_{F\perp}) 
 + \frac{1}{M^2} k_{F\perp\rho} k_{F\perp[\alpha}\bar{n}_{\beta]} H_{dT}^{\prime\perp\prime}(z,k_{F\perp}) \Bigr]  \Bigr\}+\cdots \  . \label{eq:Xi1VT}
\end{align}
%%%%%%%
For tensor polarization dependent part, we have,
\begin{align}
z \Xi^{T(1)}_{\rho\alpha}(z,k_{F\perp};p,S) =& -p^+\bar{n}_\alpha \Bigl[ k_{F\perp\rho} S_{LL} D_{dLL}^\perp(z,k_{F\perp}) 
   + M S_{LT\rho} D_{dLT}(z,k_{F\perp}) + k_{F\perp\rho} \frac{k_{F\perp} \cdot S_{LT}}{M} D_{dLT}^\perp(z,k_{F\perp}) \nonumber\\
&+ S_{TT\rho}^{ k_F} D_{dTT}^{\prime\perp}(z,k_{F\perp})+ k_{F\perp\rho} \frac{S_{TT}^{k_Fk_F}}{M^2} D_{dTT}^{\perp}(z,k_{F\perp})\Bigr]+\cdots\ , \label{eq:Xi1TV}\\
%%%
z \tilde \Xi^{(1)}_{\rho\alpha} (z,k_{F\perp};p,S) = & -i p^+ \bar{n}_\alpha\Bigl[ \tilde k_{F\perp\rho} S_{LL} G_{dLL}^\perp(z,k_{F\perp}) 
+ M \tilde S_{LT\rho} G_{dLT}(z,k_{F\perp}) + \frac{1}{M} \tilde k_{F\perp\rho} k_{F\perp} \cdot S_{LT} G_{dLT}^\perp(z,k_{F\perp}) \nonumber\\
&+ \tilde S_{TT\rho}^{k_{F}} G_{dTT}^{\prime \perp}(z,k_{F\perp}) 
  + \tilde k_{F\perp\rho} \frac{S_{TT}^{k_Fk_F}}{M^2} G_{dTT}^{\perp}(z,k_{F\perp}) \Bigr]+\cdots\ , \label{eq:Xi1TAV}\\
%%%%%%
z \Xi^{T(1)}_{\rho\alpha\beta} (z,k_{F\perp};p,S) = & p^+ \Bigl\{ S_{LL} \Bigl[ M \varepsilon_{\perp\rho[\alpha} \bar{n}_{\beta]} H_{dLL}(z,k_{F\perp})
- \frac{1}{M} \tilde k_{F\perp\rho} k_{F\perp[\alpha} \bar{n}_{\beta]} H_{dLL}^\perp(z,k_{F\perp})\Bigr]  \nonumber\\
& (k_{F\perp} \cdot {S}_{LT}) \Bigl[ \varepsilon_{\perp\rho[\alpha}\bar{n}_{\beta]} H_{dLT}^\perp(z,k_{F\perp}) 
  - \frac{1}{M^2} \tilde k_{F\perp\rho} k_{F\perp[\alpha} \bar{n}_{\beta]}H_{dLT}^{\perp\prime}(z,k_{F\perp}) \Bigr] \nonumber\\
&- (\tilde k_{F\perp} \cdot S_{LT}) \Bigl[ g_{\perp\rho[\alpha} \bar{n}_{\beta]}H_{dLT}^{\prime\perp}(z,k_{F\perp}) 
  + \frac{1}{M^2} k_{F\perp\rho} k_{F\perp[\alpha} \bar{n}_{\beta]}H_{dLT}^{\prime\perp\prime}(z,k_{F\perp}) \Bigr] \nonumber\\
&+ \frac{S_{TT}^{k_Fk_F}}{M} \Bigl[ \varepsilon_{\perp\rho[\alpha}\bar{n}_{\beta]} H_{dTT}^{\perp}(z,k_{F\perp}) 
- \frac{1}{M^2} \tilde k_{F\perp\rho} k_{F\perp[\alpha} \bar{n}_{\beta]}  H_{dTT}^{\perp\prime}(z,k_{F\perp}) \Bigr] \nonumber\\
&+\frac{S_{TT}^{\tilde k_Fk_F}}{M} \Bigl[ g_{\perp\rho[\alpha} \bar{n}_{\beta]} H_{dTT}^{\prime\perp}(z,k_{F\perp}) 
+ \frac{1}{M^2} {k_{F\perp\rho} k_{F\perp[\alpha} \bar{n}_{\beta]} }H_{dTT}^{\prime\perp\prime}(z,k_{F\perp}) \Bigr]  \Bigr\}+\cdots\ . \label{eq:Xi1TT}
\end{align}
Here, we use a subscript $d$ to specify that they are defined via quark-gluon-quark correlator. 
A prime in the superscript before the $\perp$ denotes different polarization situation, that after the $\perp$ specifies 
different FF for the same polarization situation. 
We see that we have totally 36 FFs at twist-3 defined via quark-gluon-quark correlator.  
This is just the same as what we obtained from the quark-quark correlator. 
Among them, 18 are $\chi$-even and the other 18 are $\chi$-odd; 
4 contribute to unpolarized part, 12 to vector polarized part and 20 to the tensor polarized part. 
We note in particular that the hermiticity in this case does not demand that the FFs defined via quark-gluon-quark correlator are real. 
They can have both real and imaginary parts.

\section{Twist-3 contributions to the hadronic tensor}\label{sec:twist3Ws}

In the two-hadron-collinear frame, the twist-3 contributions to other parts of the hadronic tensor besides $W_{\mu\nu}^{(1)U}$ and $W_{\mu\nu}^{(1)LL}$ 
given by Eqs.~(\ref{eq:W1U}) and (\ref{eq:W1LL}) are given by,   
\begin{align}
W_{\mu\nu}^{(1)L}  = &\frac{4\lambda}{z_1z_2} \int\frac{d^2k_\perp}{(2\pi)^2} \frac{d^2k^\prime_\perp}{(2\pi)^2} \delta^2(k_\perp + k^\prime_\perp - q_\perp) 
\biggl\{  \frac{1}{p_1^+} \Bigl[ -\omega_{\mu\nu}(\tilde k)   D_L^\perp + \tilde \omega_{\mu\nu}(k) G_L^\perp \Bigr] \bar D_1
 + \frac{2M_1 c_2^q}{M_2 p_1^+} \Bigl[ 2 (\tilde k^\prime_{n} - \tilde k^\prime_{\bar n})_{\{\mu\nu\}} H_{L}  + i (\tilde k^\prime_{n} - \tilde k^\prime_{\bar n})_{[\mu\nu]} E_{L} \Bigr] \bar H_1^\perp \nonumber\\
%%%%%%%%%%%%%%%%%
&+ \frac{1}{p_2^-} G_{1L} \Bigl[ \omega_{\mu\nu}(k^\prime) \bar D^\perp + \tilde \omega_{\mu\nu}(\tilde k^\prime) \bar G^\perp \Bigr] 
 - \frac{2M_2 c_2^q}{M_1p_2^-} H_{1L}^\perp \Bigl[ 2 (\tilde k_{n} - \tilde k_{\bar n})_{\{\mu\nu\}} \bar H + i (\tilde k_{n} - \tilde k_{\bar n})_{[\mu\nu]} \bar E \Bigr]  \nonumber\\
%%%%%%%%%%%%%%%%%
& - \frac{\sqrt{2}}{Q} \Bigl[ \tilde \omega_{\mu\nu}(k^\prime, k)   G_{1L}  \bar D_1
+ \frac{4c_2^q}{M_1M_2} %\omega_{\mu\nu}^{(9)}(k,k^\prime)  
\bigl( k_\perp^2 \tilde k^\prime_{\bar n} + \tilde k_\perp \cdot k_\perp^\prime k^\prime_{n} + k_\perp^\prime \cdot k_\perp k^\prime_{n}\bigr)_{\{\mu\nu\}} 
H_{1L}^\perp \bar H_1^\perp \Bigr] \biggr\}, \label{eq:W1L}\\
%\end{align}
%\begin{align}
%
W_{\mu\nu}^{(1)T}  = & \frac{4}{ z_1z_2} \int\frac{d^2k_\perp}{(2\pi)^2} \frac{d^2k^\prime_\perp}{(2\pi)^2} \delta^2(k_\perp + k^\prime_\perp - q_\perp)   %\nonumber\\
%& \times
\biggl\{   \frac{k_\perp \cdot S_T}{M_1}  \Bigl( \frac{1}{p_1^+}  \Bigl[ \omega_{\mu\nu}(-\tilde k, \tilde k)  D_{T}^\perp +  \tilde \omega_{\mu\nu}(k) G_{T}^\perp \Bigr]  \bar D_1
+ \frac{1}{p_2^-} G_{1T} \Bigl[   \tilde \omega_{\mu\nu}(k^\prime) \bar D^\perp + \tilde \omega_{\mu\nu}(\tilde k^\prime) \bar G^\perp  \Bigr]  \nonumber\\ 
&\phantom{XXXXXX}+ \frac{2M_1 c_2^q}{M_2 p_1^+}  \Bigl[  2 (\tilde k^\prime_{n} - \tilde k^\prime_{\bar n})_{\{\mu\nu\}} H_{T}^{\prime\perp} +  i (\tilde k^\prime_{n} - \tilde k^\prime_{\bar n})_{[\mu\nu]}  E_{T}^{\prime\perp} \Bigr] \bar H_1^\perp
   - \frac{2M_2 c_2^q}{M_1p_2^-}  H_{1T}^\perp \Bigl[  2 (\tilde k_{n} - \tilde k_{\bar n})_{\{\mu\nu\}}  \bar H  +  i (\tilde k_{n} - \tilde k_{\bar n})_{[\mu\nu]} \bar E \Bigr]   \nonumber\\
&\phantom{XXXXXX} - \frac{\sqrt{2}}{Q}  \Bigl[ \tilde\omega_{\mu\nu}(k^\prime, k) G_{1T}^\perp  \bar D_1  + \frac{4c_2^q}{M_1M_2} %\omega_{\mu\nu}^{(9)}(k,k^\prime) 
\bigl( k_\perp^2 \tilde k^\prime_{\bar n} + \tilde k_\perp \cdot k_\perp^\prime k^\prime_{n} + k_\perp^\prime \cdot k_\perp k^\prime_{n}\bigr)_{\{\mu\nu\}} H_{1T}^\perp \bar H_1^\perp \Bigr] \Bigr) \nonumber\\
%%%%%%%%
+& \frac{\tilde k_\perp \cdot S_T}{M_1} \Bigl(    
 \frac{1}{p_2^-}  D_{1T}^\perp  \Bigl[ -\omega_{\mu\nu}(k^\prime)  \bar D^\perp - \tilde \omega_{\mu\nu}(\tilde k^\prime)  \bar G^\perp\Bigr] 
 + \frac{2M_1 c_2^q}{M_2 p_1^+}  \Bigl[ 2 ( k^\prime_{n} - k^\prime_{\bar n})_{\{\mu\nu\}} H_T^\perp + i (k^\prime_{n} - k^\prime_{\bar n})_{[\mu\nu]} E_T^\perp  \Bigr]  \bar H_1^\perp 
 +  \frac{\sqrt{2}}{Q} \omega_{\mu\nu}(k^\prime, k)  D_{1T}^\perp \bar D_1 \Bigr) \nonumber\\
 %%%%%%%%%%%%%
 & - \frac{2M_2 c_2^q}{p_2^-}  H_{1T} \Bigl[
2 (\tilde S_{n} - \tilde S_{\bar n})_{\{\mu\nu\}} \bar H + i (\tilde S_{n} - \tilde S_{\bar n})_{[\mu\nu]} \bar E \Bigr]  
+ \frac{M_1}{p_1^+} \Bigl[ - \omega_{\mu\nu}(\tilde S) D_T + \tilde \omega_{\mu\nu}(S) G_T \Bigr] \bar D_1 \nonumber\\
& - \frac{4\sqrt{2}c_2^q}{M_2 Q} %\omega_{\mu\nu}^{(8)}(k,k^\prime,S) 
 \Bigl[  -\tilde k'_\perp\cdot S_\perp  (k_{\bar n}+ k'_n)_{\{\mu\nu\}} 
+ \tilde k'_\perp \cdot k_\perp S_{\bar n\{\mu\nu\}} 
+ k_\perp \cdot S_\perp \tilde k'_{\bar n\{\mu\nu\}} + k_\perp^\prime \cdot S_\perp  \tilde k'_{n\{\mu\nu\}} \Bigr]
H_{1T} \bar H_1^\perp \biggr\}, \label{eq:W1T}
\end{align}

\begin{align}
W_{\mu\nu}^{(1)LT} & = \frac{4}{ z_1z_2} \int\frac{d^2k_\perp}{(2\pi)^2} \frac{d^2k^\prime_\perp}{(2\pi)^2} \delta^2(k_\perp + k^\prime_\perp - q_\perp) %\nonumber\\
%\times&
 \biggl\{ \frac{k_\perp \cdot S_{LT}}{M_1} \Bigl(
 \frac{1}{p_1^+} \Bigl[  \omega_{\mu\nu}(k) D_{LT}^\perp
+ \tilde \omega_{\mu\nu}(\tilde k) G_{LT}^\perp \Bigr] \bar D_1
 - \frac{1}{p_2^-}  \Bigl[ \omega_{\mu\nu}(k^\prime)  D_{1LT}^\perp \bar D^\perp + \tilde \omega_{\mu\nu}(\tilde k^\prime) D_{1LT}^\perp \bar G^\perp\Bigr]  \nonumber\\
%---------------------------------
&\hspace{15mm} +\frac{2M_1 c_2^q}{M_2 p_1^+}  \Bigl[ 2 (k^\prime_{n} - k^\prime_{\bar n})_{\{\mu\nu\}}  H_{LT}^\perp
 + i (k^\prime_{n} - k^\prime_{\bar n})_{[\mu\nu]} E_{LT}^\perp \Bigr] \bar H_1^\perp 
- \frac{2M_2 c_2^q}{M_1 p_2^-} \Bigl[ 2 (k_{n} - k_{\bar n})_{\{\mu\nu\}} H_{1LT}^\perp \bar H + i (k_{n} - k_{\bar n})_{[\mu\nu]} H_{1LT}^\perp \bar E \Bigr]  \nonumber\\
%----------------------------------
& \hspace{15mm} +\frac{\sqrt{2}}{Q} 
 \Bigl[ \omega_{\mu\nu}(k^\prime, k)  D_{1LT}^\perp \bar D_1
  - \frac{4}{M_1 M_2} \omega_{\mu\nu}^{(n)}(k,k^\prime) H_{1LT}^\perp \bar H_1^\perp \Bigr] \Bigr) \nonumber\\
  %%%%%%%%%%%%%%%%%%
  %%%%%%%%%%%%%%%%%%
  +& \frac{\tilde k_\perp \cdot S_{LT}}{M_1} \Bigl( \frac{1}{p_2^-}  G_{1LT}^\perp \Bigl[ \tilde \omega_{\mu\nu}(k^\prime) \bar D^\perp + \omega_{\mu\nu}(\tilde k^\prime) \bar G^\perp \Bigr]
+\frac{2 M_1 c_2^q}{M_2 p_1^+}  \Bigl[ 2 (\tilde k^\prime_{n} - \tilde k^\prime_{\bar n})_{\{\mu\nu\}} H_{LT}^{\prime\perp} + i (\tilde k^\prime_{n} - \tilde k^\prime_{\bar n})_{[\mu\nu]} E_{LT}^{\prime\perp}  \Bigr] \bar H_1^\perp
- \frac{\sqrt{2}}{Q} \tilde \omega_{\mu\nu}(k^\prime, k) G_{1LT}^\perp \bar D_1 \Bigr) \nonumber\\
%%%%%%%%%%%%%%%%%%%
%%%%%%%%%%%%%%%%%%%
+& \frac{M_1}{p_1^+} \Bigl[ \omega_{\mu\nu}(S_{LT})  D_{LT} + \tilde \omega_{\mu\nu}(\tilde S_{LT}) G_{LT}  \Bigr] \bar D_1 
- \frac{2M_2 c_2^q}{p_2^-} \Bigl[ 2 (S_{LTn} - S_{LT\bar n})_{\{\mu\nu\}} H_{1LT} \bar H + i (S_{LTn} - S_{LT\bar n})_{[\mu\nu]} H_{1LT} \bar E \Bigr]  \nonumber\\
%-------------------------------------
-& \frac{4\sqrt{2}c_2^q}{M_2 Q} %\omega_{\mu\nu}^{(10)}(k,k^\prime,S_{LT}) 
\bigl(  k_\perp\cdot S_{LT} k^\prime_{\bar n} 
- k'_\perp\cdot S_{LT} k_{\bar n} + k_\perp \cdot k_\perp^\prime S_{LT\bar n} + k_\perp^{\prime2} S_{LTn}\bigr)_{\{\mu\nu\}}  H_{1LT} \bar H_1^\perp
 \biggr\}, \label{eq:W1LT}
 \end{align}
 
\begin{align}
W_{\mu\nu}^{(1)TT} & =\frac{4}{ z_1z_2} \int\frac{d^2k_\perp}{(2\pi)^2} \frac{d^2k^\prime_\perp}{(2\pi)^2} \delta^2(k_\perp + k^\prime_\perp - q_\perp) %\nonumber\\
 %\times & 
 \biggl\{ \frac{S_{TT}^{kk}}{M_1^2} \Bigl(
   \frac{1}{p_1^+} \Bigl[ \omega_{\mu\nu}(k) D_{TT}^\perp + \tilde \omega_{\mu\nu}(\tilde k) G_{TT}^\perp \Bigr] \bar D_1
-\frac{1}{p_2^-}  \Bigl[ \omega_{\mu\nu}(k^\prime) D_{1TT}^\perp \bar D^\perp + \tilde \omega_{\mu\nu}(\tilde k^\prime)  D_{1TT}^\perp \bar G^\perp \Bigr] \nonumber\\
%-------------------------------------------
 &~~~~ +\frac{2M_1 c_2^q}{M_2 p_1^+} \Bigl[ 2 (k^\prime_{n} - k^\prime_{\bar n})_{\{\mu\nu\}} H_{TT}^\perp + i (k^\prime_{n} - k^\prime_{\bar n})_{[\mu\nu]} E_{TT}^\perp \Bigr] \bar H_1^\perp
- \frac{2M_2 c_2^q}{M_1p_2^-} \Bigl[ 2 (k_{n} - k_{\bar n})_{\{\mu\nu\}} H_{1TT}^\perp \bar H + i (k_{n} - k_{\bar n})_{[\mu\nu]} S_{TT}^{kk} H_{1TT}^\perp \bar E \Bigr]  \nonumber\\
%-------------------------------------------
&~~~~ + \frac{\sqrt{2}}{Q} \Bigl[ \omega_{\mu\nu}(k^\prime, k) D_{1TT}^\perp \bar D_1 - \frac{4}{M_1M_2} \omega_{\mu\nu}^{(n)}(k,k^\prime) H_{1TT}^\perp \bar H_1^\perp \Bigr] \Bigr) \nonumber\\
%%%%%%%%%%%%%%%%%%%%%
%%%%%%%%%%%%%%%%%%%%%
+& \frac{S_{TT}^{k\tilde k}}{M_1^2} \Bigl(
\frac{1}{p_2^-}  G_{1TT}^\perp \Bigl[ \tilde \omega_{\mu\nu}(k^\prime) \bar D^\perp + \omega_{\mu\nu}(\tilde k^\prime) \bar G^\perp \Bigr]
 +\frac{2M_1 c_2^q}{M_2 p_1^+} \Bigl[ 2 (\tilde k^\prime_{n} - \tilde k^\prime_{\bar n})_{\{\mu\nu\}} H_{TT}^{\prime\perp} 
 + i (\tilde k^\prime_{n} - \tilde k^\prime_{\bar n})_{[\mu\nu]} E_{TT}^{\prime\perp}  \Bigr] \bar H_1^\perp
 - \frac{\sqrt{2}}{Q} \tilde \omega_{\mu\nu}(k^\prime, k)  G_{1TT}^\perp \bar D_1 \Bigr) \nonumber\\
%%%%%%%%%%%%%%%%%%%%%%
%%%%%%%%%%%%%%%%%%%%%%
+& \frac{1}{p_1^+} \Bigl[ \omega_{\mu\nu}(S_{TT}^{k}) D_{TT}^{\prime\perp}
 + \tilde \omega_{\mu\nu}(\tilde S_{TT}^k) G_{TT}^{\prime\perp} \Bigr] \bar D_1
- \frac{2M_2 c_2^q}{p_2^-M_1} \Bigl[ 2 (S_{TTn}^k - S_{TT\bar n}^k)_{\{\mu\nu\}} H_{1TT}^{\prime\perp} \bar H 
+ i (S_{TTn}^k - S_{TT\bar n}^k)_{[\mu\nu]} H_{1TT}^{\prime\perp} \bar E \Bigr]  \nonumber\\
%------------------------------------------
-& \frac{4\sqrt{2}c_2^qM_1^2}{M_2 Q} %\omega_{\mu\nu}^{(10)}(k,k^\prime,S_{TT}^{k}) 
\bigl(  S_{TT}^{kk} k^\prime_{\bar n} 
- S_{TT}^{kk'} k_{\bar n} + k_\perp \cdot k_\perp^\prime S^k_{TT\bar n} + k_\perp^{\prime2} S^k_{TTn}\bigr)_{\{\mu\nu\}}  H_{1TT}^{\prime\perp} \bar H_1^\perp \biggr\}. \label{eq:W1TT}
\end{align}

Transforming them into the Helicity-GJ-frame, we obtain from Eqs.~(\ref{eq:W1L}-\ref{eq:W1TT}) the contributions at twist-3 
and they take exactly the same form as given in these equations. 
However, we obtain also additional twist-3 contributions from the twist-2 parts given by Eqs.~(\ref{eq:W0U}-\ref{eq:W0TT}). 
The corresponding terms for the unpolarized part is given by Eq. (\ref{eq:dW1U}). 
Other parts are given in the following.

\begin{align}
\delta W^{(1)L}_{\mu\nu} = \frac{4\sqrt{2}}{z_1z_2 Q}& \int\frac{d^2k_\perp}{(2\pi)^2} \frac{d^2k^\prime_\perp}{(2\pi)^2}  \delta^2(k_\perp + k^\prime_\perp - q_\perp) %\nonumber\\
%& \times 
\lambda \Bigl\{
 \bigl( c_3^q q_{\bar n\{\mu\nu\}} - ic_1^q \tilde q_{\bar n[\mu\nu]} \bigr)  G_{1L}  \bar D_1
 + \frac{4c_2^q}{M_1M_2} k_\perp \cdot (q_\perp - \tilde k_\perp^\prime) k^\prime_{\bar n\{\mu\nu\}} H_{1L}^\perp \bar H_1^\perp \Bigr\}, \label{eq:dW1L}\\
%\end{align}
%\begin{align}
\delta W^{(1)T}_{\mu\nu} = \frac{4\sqrt{2}}{z_1z_2 Q}& \int\frac{d^2k_\perp}{(2\pi)^2} \frac{d^2k^\prime_\perp}{(2\pi)^2} \delta^2(k_\perp + k^\prime_\perp - q_\perp) %\nonumber\\
%\times & 
\Bigl\{ \frac{k_\perp \cdot S_\perp}{M_1} \Bigl[ \bigl( c_3^q q_{\bar n\{\mu\nu\}} - ic_1^q \tilde q_{\bar n[\mu\nu]} \bigl) G_{1T}^\perp \bar D_1 
+ \frac{4c_2^q}{M_1M_2} k_\perp \cdot (q_\perp - \tilde k_\perp^\prime) k^\prime_{\bar n\{\mu\nu\}} H_{1T}^\perp \bar H_1^\perp \Bigr] \nonumber\\
& - \frac{\tilde k_\perp \cdot S_\perp}{M_1} \bigl( c_1^q q_{\bar n\{\mu\nu\}} - ic_3^q \tilde q_{\bar n[\mu\nu]} \bigr)  D_{1T}^\perp \bar D_1
+ \frac{4c_2^q}{M_2} \bigl( k_\perp \cdot S_\perp \tilde k^\prime_{\bar n}+ k_\perp \cdot \tilde k_\perp^\prime S_{\bar n} - \tilde k_\perp^\prime \cdot S_\perp k_{\bar n} \bigr)_{\{\mu\nu\}}  H_{1T} \bar H_1^\perp \Bigr\}, \label{eq:dW1T} \\
%\end{align}
%\begin{align}
\delta W^{(1)LL}_{\mu\nu} = \frac{4\sqrt{2}}{z_1z_2 Q} & \int\frac{d^2k_\perp}{(2\pi)^2} \frac{d^2k^\prime_\perp}{(2\pi)^2} \delta^2(k_\perp + k^\prime_\perp - q_\perp) %\nonumber\\
%\times & 
S_{LL} \Bigl\{ -\bigl( c_1^q q_{\bar n\{\mu\nu\}} - ic_3^q \tilde q_{\bar n[\mu\nu]} \bigr) D_{1LL} \bar D_1
+ \frac{4c_2^q}{M_1M_2} \bigl(  k_\perp^2 k^\prime_{\bar n} + k_\perp^{\prime2} k_{\bar n} \bigr)_{\{\mu\nu\}} H_{1LL}^\perp \bar H_1^\perp \Bigr\}, \label{eq:dW1LL}\\
%\end{align}
%\begin{align}
\delta W^{(1)LT}_{\mu\nu} = \frac{4\sqrt{2}}{z_1z_2 Q} & \int\frac{d^2k_\perp}{(2\pi)^2} \frac{d^2k^\prime_\perp}{(2\pi)^2} \delta^2(k_\perp + k^\prime_\perp - q_\perp) %\nonumber\\
%\times & 
\Bigl\{ \frac{k_\perp \cdot S_{LT}}{M_1} \Bigl[-\bigl( c_1^q q_{\bar n\{\mu\nu\}} - ic_3^q \tilde q_{\bar n[\mu\nu]} \bigr) D_{1LT}^\perp \bar D_1 
+ \frac{4c_2^q}{M_1M_2} \bigl( k_\perp^2 k^\prime_{\bar n} + k_\perp^{\prime2} k_{\bar n} \bigr)_{\{\mu\nu\}} H_{1LT}^\perp \bar H_1^\perp \Bigr] \nonumber\\
& + \frac{\tilde k_\perp \cdot S_{LT}}{M_1} \bigl( c_3^q q_{\bar n\{\mu\nu\}} - ic_1^q \tilde q_{\bar n[\mu\nu]} \bigr) G_{1LT}^\perp \bar D_1 
+ \frac{4c_2^q}{M_2} \bigl( q_\perp \cdot k_\perp^\prime S_{LT\bar n} + k_\perp \cdot S_{LT} k^\prime_{\bar n} - k_\perp^\prime \cdot S_{LT} k_{\bar n} \bigr)_{\{\mu\nu\}} H_{1LT} \bar H_1^\perp \Bigr\}, \label{eq:dW1LT}\\
%\end{align}
%\begin{align}
\delta W^{(1)TT}_{\mu\nu} = \frac{4\sqrt{2}}{z_1z_2 Q} & \int\frac{d^2k_\perp}{(2\pi)^2} \frac{d^2k^\prime_\perp}{(2\pi)^2} \delta^2(k_\perp + k^\prime_\perp - q_\perp) %\nonumber\\
%\times & 
\Bigl\{ \frac{S_{TT}^{kk}}{M_1^2} \Bigl[-\bigl( c_1^q q_{\bar n\{\mu\nu\}} - ic_3^q \tilde q_{\bar n[\mu\nu]} \bigr) D_{1TT}^\perp \bar D_1 
+ \bigl( k_\perp^2 k^\prime_{\bar n} + k_\perp^{\prime2} k_{\bar n} \bigr)_{\{\mu\nu\}} H_{1TT}^\perp \bar H_1^\perp \Bigr] \nonumber\\
& + \frac{S_{TT}^{k\tilde k}}{M_1^2} \bigl( c_3^q q_{\bar n\{\mu\nu\}} - ic_1^q \tilde q_{\bar n[\mu\nu]} \bigr) G_{1LT}^\perp \bar D_1 
+ \frac{4c_2^q}{M_1M_2} \bigl( q_\perp \cdot k_\perp^\prime S^k_{TT\bar n} + S_{TT}^{kk} k^\prime_{\bar n} - S_{TT}^{kk^\prime} k_{\bar n} \bigr)_{\{\mu\nu\}} H_{1TT} \bar H_1^\perp \Bigr\}. \label{eq:dW1TT}
\end{align}

\section{Twist-3 contributions to the structure functions}\label{sec:twist3FFs}

In the partonic picture at the LO pQCD, 36 of the 81 structure functions for $e^+e^-\to V\pi X$ have twist-3 contributions. 
We list the results in this appendix in the following.

\begin{align}
F_{1U2}^{\cos\varphi} =& \frac{8 c_3^e c_3^q}{z_1z_2Q} \mathcal{C} [ M_1 w_1 D^\perp z_2 \bar D_1 + M_2 \bar w_1 z_1 D_1 \bar D^{\perp\prime} ], \label{eq:F1U2cos}\\
F_{2U2}^{\cos\varphi} =& \frac{4 c_1^e}{z_1z_2Q}\Bigl\{ c_1^q \mathcal{C} [  M_1 w_1 D^\perp z_2 \bar D_1 + M_2 \bar w_1 z_1 D_1 \bar D^{\perp\prime} ] 
+ 4 c_2^q  \mathcal{C} [M_1 \bar w_1 H z_2 \bar H_1^\perp - M_2 w_1 z_1 H_1^\perp \bar H^{\perp\prime}  ]\Bigr\}, \label{eq:F2U2cos} \\
\tilde F_{1U2}^{\sin\varphi} =& \frac{8c_3^e}{z_1z_2Q} \Bigl\{ c_1^q \mathcal{C} [  ( M_1 w_1 G^\perp z_2 \bar D_1 - M_2 \bar w_1 z_1 D_1 \bar G^\perp ) ]
+ 2 c_2^q \mathcal{C} [( M_1 \bar w_1 E z_2 \bar H_1^\perp - M_2 w_1 z_1 H_1^\perp \bar E ) ]\Bigr\}, \label{eq:tF1U2sin} \\
\tilde F_{2U2}^{\sin\varphi} =& \frac{4 c_1^e c_3^q}{z_1z_2Q} \mathcal{C} [ M_1 w_1 G^\perp z_2 \bar D_1 - M_2 \bar w_1 z_1 D_1 \bar G^\perp ], \label{eq:tF2U2sin} \\
%
%%%%%%%%%%%%%%%%%%%%%%%%%%
%%%%%%%%%%%%%%%%%%%%%%%%%%
\tilde F_{1L2}^{\cos\varphi} =& \frac{8 c_3^e}{z_1z_2Q}\Bigl\{ c_1^q\mathcal{C} [ M_1 w_1 G_L^\perp z_2 \bar D_1 - M_2 \bar w_1 z_1 G_{1L} \bar D^{\perp\prime}]
+ 2 c_2^q \mathcal{C} [ -M_1 \bar w_1 E_L z_2 \bar H_1^\perp + M_2 w_1 z_1 H_{1L}^\perp \bar E  ]\Bigr\}, \\
\tilde F_{2L2}^{\cos\varphi} =& \frac{4 c_1^e c_3^q}{z_1z_2Q} \mathcal{C} [ M_1 w_1 G_L^\perp z_2 \bar D_1 - M_2 \bar w_1 z_1 G_{1L} \bar D^{\perp\prime} ], \\
F_{1L2}^{\sin\varphi} =& \frac{8 c_3^e c_3^q}{z_1z_2Q} \mathcal{C} [ - M_1 w_1 D_L^\perp z_2 \bar D_1 + M_2 \bar w_1 z_1 G_{1L} \bar G^\perp ], \\
F_{2L2}^{\sin\varphi} =& \frac{4 c_1^e}{z_1z_2Q}\Bigl\{ c_1^q \mathcal{C} [  -M_1 w_1 D_L^\perp z_2 \bar D_1 + M_2 \bar w_1 z_1 G_{1L} \bar G^\perp ]
 + 4 c_2^q \mathcal{C} [  M_1 \bar w_1 H_L z_2 \bar H_1^\perp - M_2 w_1 z_1 H_{1L}^\perp \bar H^{\perp\prime} ]\Bigr\}, \\
%%%%%%%%%%%%%%%%%%%%%%%%%%
%%%%%%%%%%%%%%%%%%%%%%%%%%
\tilde F_{1T2}^{\cos\varphi_S} =& \frac{4 c_3^e}{z_1z_2Q}  \Bigl\{ c_1^q \mathcal{C} [ 2M_1 {\cal G}^\perp_T z_2 \bar D_1 
  + M_2 w_2 z_1 (G_{1T}^\perp \bar D^{\perp\prime}+ D_{1T}^\perp \bar G^\perp) ]  %\nonumber\\ &
+ c_2^q \mathcal{C} [ -2M_1 w_2 E_T^{\perp-} z_2 \bar H_1^\perp + M_2 z_1 {\mathcal H}_{1T}^{\perp} \bar E ] \Bigr\}, \label{eq:tF1T2cos}\\
\tilde F_{2T2}^{\cos\varphi_S} =& \frac{4 c_1^e c_3^q}{z_1z_2Q} \mathcal{C} [ M_1 {\cal G}^\perp_T z_2 \bar D_1 
  + M_2 \frac{w_2}{2} z_1 (G_{1T}^\perp \bar D^{\perp\prime}+ D_{1T}^\perp \bar G^\perp) ],\label{eq:tF2T2cos}\\
F_{1T2}^{\sin\varphi_S} =& \frac{8 c_3^e c_3^q}{z_1z_2Q} \mathcal{C} [ -M_1 {\cal D}^\perp_T z_2 \bar D_1 +
 M_2 \frac{w_2}{2} z_1 (D_{1T}^\perp \bar D^{\perp\prime} -  G_{1T}^\perp \bar G^\perp) ], \label{eq:F1T2sin}\\
F_{2T2}^{\sin\varphi_S} =& \frac{4 c_1^e}{z_1z_2Q}  
\Bigl\{ c_1^q \mathcal{C}[ -M_1 {\cal D}^\perp_T z_2 \bar D_1 +M_2 \frac{w_2}{2} z_1 (D_{1T}^\perp \bar D^{\perp\prime} - G_{1T}^\perp \bar G^\perp) ] %\nonumber\\ & 
 +4 c_2^q \mathcal{C} [ M_1 \frac{w_2}{2} H_T^{\perp-} z_2 \bar H_1^\perp - M_2 z_1 {\mathcal H}_{1T}^{\perp} \bar H_1^{\perp\prime} ] \Bigr\}, \label{eq:F2T2sin}\\
%%%%%%%%%%%%%%%%%%%%%%%%%%
%%%%%%%%%%%%%%%%%%%%%%%%%%
\tilde F_{1T2}^{\cos(\varphi_S-2\varphi)} =& \frac{8 c_3^e}{z_1z_2Q}  
\Bigl\{ c_1^q \mathcal{C} [ M_1 w_3 G_T^\perp z_2 \bar D_1 - M_2 w_4 z_1 (G_{1T}^\perp \bar D^{\perp\prime} - D_{1T}^\perp \bar G^\perp) ] %\nonumber\\&
+ 2 c_2^q \mathcal{C} [ -M_1 w_4 E_T^{\perp+} z_2 \bar H_1^\perp + M_2 w_3 z_1 H_{1T}^\perp \bar E ] \Bigr\}, \\
\tilde F_{2T2}^{\cos(\varphi_S-2\varphi)} =& \frac{4 c_1^e c_3^q}{z_1z_2Q} 
\mathcal{C} [ M_1 w_3 G_T^\perp z_2 \bar D_1 - M_2 w_4 z_1 (G_{1T}^\perp \bar D^{\perp\prime}- D_{1T}^\perp \bar G^\perp) ], \\
F_{1T2}^{\sin(\varphi_S-2\varphi)} =& \frac{8 c_3^e c_3^q}{z_1z_2Q} 
\mathcal{C} [ M_1 w_3 D_T^\perp z_2 \bar D_1 - M_2 w_4 z_1 (D_{1T}^\perp \bar D^{\perp\prime} + G_{1T}^\perp \bar G^\perp) ], \\
F_{2T2}^{\sin(\varphi_S-2\varphi)} =& \frac{4 c_1^e}{z_1z_2Q}  
\Bigl\{ c_1^q \mathcal{C}[ M_1 w_3 D_T^\perp z_2 \bar D_1 - M_2 w_4 z_1 (D_{1T}^\perp \bar D^{\perp\prime} + G_{1T}^\perp \bar G^\perp) ] % \nonumber\\ &
+ 4 c_2^q \mathcal{C}[ -M_1 w_4 H_T^{\perp+} z_2 \bar H_1^\perp + M_2 w_3 z_1 H_{1T}^\perp \bar H^{\perp\prime} ] \Bigr\}, \\
%%%%%%%%%%%%%%%%%%%%%%%%%%
%%%%%%%%%%%%%%%%%%%%%%%%%%
F_{1LL2}^{\cos\varphi} =& \frac{8 c_3^e c_3^q}{z_1z_2Q} \mathcal{C} [ M_1 w_1 D_{LL}^\perp z_2 \bar D_1 + M_2 \bar w_1 z_1 D_{1LL} \bar D^{\perp\prime} ], \\
F_{2LL2}^{\cos\varphi} =& \frac{4 c_1^e}{z_1z_2Q} \Bigl\{ c_1^q \mathcal{C} [  M_1 w_1 D_{LL}^\perp z_2 \bar D_1 + M_2 \bar w_1 z_1 D_{1LL} \bar D^{\perp\prime} ] 
+ 4 c_2^q \mathcal{C} [ M_1 \bar w_1 H_{LL} z_2 \bar H_1^\perp - M_2 w_1 z_1 H_{1LL}^\perp \bar H^{\perp\prime} ]\Bigr\}, \\
\tilde F_{2LL2}^{\sin\varphi} =& \frac{4 c_1^e c_3^q}{z_1z_2Q} \mathcal{C} [ M_1 w_1 G_{LL}^\perp z_2 \bar D_1 - M_2 \bar w_1 z_1 D_{1LL} \bar G^\perp ], \\
\tilde F_{1LL2}^{\sin\varphi} =& \frac{8 c_3^e}{z_1z_2Q}
\Bigl\{ c_1^q \mathcal{C} [  M_1 w_1 G_{LL}^\perp z_2 \bar D_1 - M_2 \bar w_1 z_1 D_{1LL} \bar G^\perp ] 
+ 2 c_2^q \mathcal{C} [ M_1 \bar w_1 E_{LL} z_2 \bar H_1^\perp - M_2 w_1 z_1 H_{1LL}^\perp \bar E ]\Bigr\}, \\
%%%%%%%%%%%%%%%%%%%%%%%%%%
%%%%%%%%%%%%%%%%%%%%%%%%%%
F_{1LT2}^{\cos\varphi_{LT}} =& \frac{8 c_3^ec_3^q}{z_1z_2Q} 
\mathcal{C} [ M_1 {\cal D}^\perp_{LT} z_2 \bar D_1 - M_2 \frac{w_2}{2} z_1 (D_{1LT}^\perp \bar D^{\perp\prime} + G_{1LT}^\perp \bar G^\perp) ], \\
F_{2LT2}^{\cos\varphi_{LT}} =& \frac{4 c_1^e}{z_1z_2Q} 
\Bigl\{ c_1^q \mathcal{C} [ M_1 {\cal D}^\perp_{LT} z_2 \bar D_1 - M_2 \frac{w_2}{2} z_1 (D_{1LT}^\perp \bar D^{\perp\prime} + G_{1LT}^\perp \bar G^\perp) ] %\nonumber\\
%&
-4 c_2^q \mathcal{C} [ M_1 \frac{w_2}{2} H_{LT}^{\perp+} z_2 \bar H_1^\perp + M_2 z_1 {\mathcal H}_{1LT}^\perp \bar H_1^{\perp\prime} ] \Bigr\}, \\
\tilde F_{1LT2}^{\sin\varphi_{LT}} =& \frac{8 c_3^e}{z_1z_2Q} 
\Bigl\{ c_1^q \mathcal{C} [ M_1 {\cal G}^\perp_{LT} z_2 \bar D_1 - M_2 \frac{w_2}{2} z_1 (G_{1LT}^\perp \bar D^{\perp\prime} - D_{1LT}^\perp \bar G^\perp) ] %\nonumber\\ & 
+ c_2^q \mathcal{C} [ -M_1 w_2 E_{LT}^{\perp+} z_2 \bar H_1^\perp - 2M_2 z_1 {\mathcal H}_{1LT}^\perp \bar E ]\Bigr\}, \\
\tilde F_{2LT2}^{\sin\varphi_{LT}} =& \frac{4 c_1^e c_3^q}{z_1z_2Q} 
\mathcal{C} [ M_1 {\cal G}^\perp_{LT} z_2 \bar D_1 - M_2 \frac{w_2}{2} z_1 (G_{1LT}^\perp \bar D^{\perp\prime} - D_{1LT}^\perp \bar G^\perp) ], \\
%%%%
F_{1LT2}^{\cos(\varphi_{LT}-2\varphi)} =& \frac{8 c_3^e c_3^q}{z_1z_2Q} 
\mathcal{C} [ M_1 w_3 D_{LT}^\perp z_2 \bar D_1 + M_2 w_4 z_1 (D_{1LT}^\perp \bar D^{\perp\prime} - G_{1LT}^\perp \bar G^\perp) ], \\
F_{2LT2}^{\cos(\varphi_{LT}-2\varphi)} =& \frac{4 c_1^e}{z_1z_2Q} 
\Bigl\{ c_1^q\mathcal{C} [ M_1 w_3 D_{LT}^\perp z_2 \bar D_1 + M_2 w_4 z_1 (D_{1LT}^\perp \bar D^{\perp\prime} - G_{1LT}^\perp \bar G^\perp) ] %\nonumber\\ & 
+ 4 c_2^q\mathcal{C} [ M_1 w_4 H_{LT}^{\perp-} z_2 \bar H_1^\perp - M_2 w_3 z_1 H_{1LT}^\perp \bar H^{\perp\prime} ]\Bigr\}, \\
\tilde F_{1LT2}^{\sin(\varphi_{LT}-2\varphi)} =& \frac{8 c_3^e}{z_1z_2Q} 
\Bigl\{ c_1^q \mathcal{C} [ -M_1 w_3 G_{LT}^\perp z_2 \bar D_1 + M_2 w_4 z_1 (G_{1LT}^\perp \bar D^{\perp\prime} +  D_{1LT}^\perp \bar G^\perp )] %\nonumber\\ & 
-2 c_2^q\mathcal{C} [ M_1 w_4 E_{LT}^{\perp-} z_2 \bar H_1^\perp - M_2 w_3 z_1 H_{1LT}^\perp \bar E ]\Bigr\}, \\
\tilde F_{2LT2}^{\sin(\varphi_{LT}-2\varphi)} =& \frac{4 c_1^e c_3^q}{z_1z_2Q} 
\mathcal{C} [ -M_1 w_3 G_{LT}^\perp z_2 \bar D_1 + M_2 w_4 z_1 (G_{1LT}^\perp \bar D^{\perp\prime} + D_{1LT}^\perp \bar G^\perp) ], \\
%%%%%%%%%%%%%%%%%%%%%%%%%%
%%%%%%%%%%%%%%%%%%%%%%%%%%
F_{1TT2}^{\cos(2\varphi_{TT}-\varphi)} =& \frac{8 c_3^e c_3^q}{z_1z_2Q} 
\mathcal{C} [ -M_1 w_1 {\cal D}^\perp_{TT} z_2 \bar D_1 - M_2 z_1(w_3 \bar w_1 D_{1TT}^\perp \bar D^{\perp\prime} - w_5 G_{1TT} ^\perp \bar G^\perp) ], \\
F_{2TT2}^{\cos(2\varphi_{TT}-\varphi)} =& \frac{4 c_1^e}{z_1z_2Q} 
\Bigl\{ c_1^q \mathcal{C} [ -M_1 w_1 {\cal D}^\perp_{TT} z_2 \bar D_1 - M_2 z_1 (w_3 \bar w_1 D_{1TT}^\perp \bar D^{\perp\prime} - w_5 G_{1TT} ^\perp \bar G^\perp) ] \nonumber\\
& + 4 c_2^q \mathcal{C} [ M_1 (w_6 H_{TT}^\perp z_2 \bar H_1^\perp - w_8 H_{TT}^{\prime\perp} \bar H_1^\perp) + M_2 w_1 z_1 {\cal H}_{1TT}^{\perp} \bar H^{\perp\prime} ]\Bigr\}, \\
\tilde F_{1TT2}^{\sin(2\varphi_{TT}-\varphi)} =& \frac{8 c_3^e}{z_1z_2Q}
\Bigl\{ c_1^q \mathcal{C} [ -M_1 w_1 {\cal G}^\perp_{TT} z_2 \bar D_1 + M_2 z_1( w_6  G_{1TT}^\perp \bar D^{\perp\prime} + w_3 \bar w_1 D_{1TT}^\perp \bar G^\perp) ]  \nonumber\\
& + 2 c_2^q \mathcal{C} [ M_1 z_2 (w_6 E_{TT}^\perp  \bar H_1^\perp - w_8 E_{TT}^{\prime\perp}  \bar H_1^\perp) + M_2 w_1 z_1 {\cal H}_{1TT}^{\perp} \bar E ] \Bigr\}, \\
\tilde F_{2TT2}^{\sin(2\varphi_{TT}-\varphi)} =& \frac{4 c_1^e c_3^q}{z_1z_2Q} 
\mathcal{C} [ -M_1 w_1 {\cal G}^\perp_{TT} z_2 \bar D_1 + M_2 z_1 (w_6  G_{1TT}^\perp \bar D^{\perp\prime} + w_3 \bar w_1 D_{1TT}^\perp \bar G^\perp) ], \\
%%%%%
F_{1TT2}^{\cos(2\varphi_{TT}-3\varphi)} =& \frac{8 c_3^e c_3^q}{z_1z_2Q} 
\mathcal{C} [ M_1 w_9 D_{TT}^\perp z_2 \bar D_1 + M_2 \frac{w_7}{2} z_1 (D_{1TT}^\perp \bar D^{\perp\prime} -  G_{1TT}^\perp \bar G^\perp) ], \\
F_{2TT2}^{\cos(2\varphi_{TT}-3\varphi)} =& \frac{4 c_1^e}{z_1z_2Q} 
\Bigl\{ c_1^q \mathcal{C} [ M_1 w_9 D_{TT}^\perp z_2 \bar D_1 + M_2 \frac{w_7}{2} z_1 (D_{1TT}^\perp \bar D^{\perp\prime} - G_{1TT}^\perp \bar G^\perp) ] 
+ 4c_2^q \mathcal{C} [ M_1 \frac{w_7}{2} H_{TT}^{\perp-} z_2 \bar H_1^\perp - M_2 w_9 z_1 H_{1TT}^\perp \bar H^{\perp\prime} ]\Bigr\}, \\
\tilde F_{1TT2}^{\sin(2\varphi_{TT}-3\varphi)} =& \frac{8 c_3^e}{z_1z_2Q} 
\Bigl\{ c_1^q \mathcal{C} [ -M_1 w_9 G_{TT}^\perp z_2 \bar D_1 + M_2 \frac{w_7}{2} z_1 (G_{1TT}^\perp \bar D^{\perp\prime} +  D_{1TT}^\perp \bar G^\perp) ] 
- c_2^q \mathcal{C} [ M_1 w_7 E_{TT}^{\perp-} z_2 \bar H_1^\perp - 2M_2 w_9 z_1 H_{1TT}^\perp \bar E ]\Bigr\}, \\
\tilde F_{2TT2}^{\sin(2\varphi_{TT}-3\varphi)} =& \frac{4 c_1^e c_3^q}{z_1z_2Q} \mathcal{C} [ -M_1 w_9 G_{TT}^\perp z_2 \bar D_1 + M_2 \frac{w_7}{2} z_1 D_{1TT}^\perp \bar G^\perp ].
\end{align}
Here, just as for the $S_T$- and $S_{LT}$-dependent FFs given by Eq.~(\ref{eq:calK}), for $S_{TT}$-dependent $K$,  
we define, 
%\begin{align}
%&{\cal K}^\perp_\sigma=K_{\sigma} + \frac{k_\perp^2}{2M_1^2} K_{\sigma}^\perp, \end{align}
%for $K=E$ or $D$ and $\sigma=T$ or  $LT$, and   
\begin{align}
&{\cal K}^\perp_{TT}(z,k_\perp)=K_{TT}^{\prime\perp}(z,k_\perp) + \frac{k_\perp^2}{2M_1^2} K_{TT}^\perp(z,k_\perp), 
\end{align}
for $K=D, G$ or $H$. Also, 
$K_\sigma^{\perp\pm} =K_\sigma^\perp \pm K_\sigma^{\prime\perp}$, 
for all different $K$'s and polarization $\sigma$'s, and for the leading twist involved combinations,
\begin{align}
& \bar D^{\perp\prime} = z_2 \bar D_1 - \bar D^\perp,  &
& \bar H^{\perp\prime} = \bar H - \bar w_0 z_2 \bar H_1^\perp. &
\end{align}
Besides the $w$'s given by Eqs.~(\ref{eq:w0}-\ref{eq:w2}) and in the text in Sec. \ref{sec:StructureFunctions}, 
we have also introduced the scalar weights defined as,   
\begin{align}
%& w_{hh} = 2w_1\bar w_1-w_0,\\
%& w_{hh}^t = 2w_1^2\bar w_1-w_0(w_1+\frac{1}{2}\bar w_1),\\
%& w_{tt}^{tt} = w_0w_2-4w_0w_1\bar w_1+4w_1^2w_2+8w_1^3\bar w_1,\\
& w_3 = \frac{1}{2}w_0-w_1^2, &
& w_4 = \frac{1}{2}w_2-w_1\bar w_1,&
& w_5 = w_1w_2-w_0w_2+\frac{1}{2}w_0w_1, & &\\
& w_6 = w_1w_2-\frac{1}{2}w_0\bar w_1, &
& w_7 = 4w_1^2\bar w_1-2w_1w_2-w_0\bar w_1, &
& w_8 = 4w_1^2\bar w_1-2w_1w_2-\frac{1}{2}w_0\bar w_1,&
& w_9 = (2w_1^2-\frac{3}{2}w_0)w_1.
\end{align}
They are all scalar functions of $k_\perp$, $k'_\perp$ and $p_{2T}$.

\end{appendix}

\end{widetext}

%\include{references}
%\end{document}


\begin{thebibliography}{0}
\bibitem{ffTR} One can similarly define whether a basic Lorentz tensor $h^{\mu\nu}$ is time reversal $T$-even, 
$h_{\mu\nu}(q^{\cal T},p_1^{\cal T},S^{\cal T},p_2^{\cal T})=h^{\mu\nu}(q,p_1,S,p_2)$, or $T$-odd, 
$h_{\mu\nu}(q^{\cal T},p_1^{\cal T},S^{\cal T},p_2^{\cal T})=-h^{\mu\nu}(q,p_1,S,p_2)$, 
where $a^{\cal T}$ denotes the $a$ under time reversal. 
We note that for $a=p$, $S_{LL}$, $S_{LT}$ or $S_{TT}$, $a^{\cal T}=a^{\cal P}$,  
while for the polarization vector $S$, $S^{\cal T}=-S^{\cal P}$. 
Hence for the symmetric tensors, the unpolarized, $S_{LL}$, $S_{LT}$ and $S_{TT}$-dependent parts have the same $P$ and $T$ behavior, 
while the $S$-dependent part have different $P$ and $T$ behaviors. 
For the anti-symmetric parts, the unpolarized, $S_{LL}$, $S_{LT}$ and $S_{TT}$-dependent parts have different $P$ and $T$ behaviors, 
while the $S$-dependent part have the same $P$ and $T$ behaviors.
\bibitem{ff:theta} The angle $\theta$ here is the angle between the incident electron and the produced quark for $e^+e^-\to q\bar q$. 
Up to $1/Q$ it is the same as that defined in the Helicity-GJ-frame for $e^+e^-\to V\pi X$.
\end{thebibliography}
\end{document}